\documentclass[12pt]{article}
\usepackage{amsmath}
\usepackage{ulem}
\usepackage{graphicx}
\usepackage{cite,./mcite}
\usepackage{amssymb}

        \newdimen\eqskip
        \newdimen\txtskip
        \baselineskip=\txtskip
        \newdimen\mysep                
        \newdimen\hmysep
\numberwithin{equation}{section}        
\begin{document}

  \newcommand{\ccaption}[2]{
    \begin{center}
    \parbox{0.9\textwidth}{
      \caption[#1]{\small{{#2}}}
      }
    \end{center}
    }
\newcommand{\BS}{\bigskip}
\def    \be             {\begin{equation}}
\def    \ee             {\end{equation}}
\def    \ba             {\begin{eqnarray}}
\def    \ea             {\end{eqnarray}}
\def    \nn             {\nonumber}
\def    \=              {\;=\;}
\def    \ret            {\\[\eqskip]}
\def    \ie             {{\em i.e.\/} }
\def    \eg             {{\em e.g.\/} }
\def \lsim{\lesssim}
\def \gsim{\gtrsim}
\def    \ev            {\mbox{$\mathrm{eV}$}}
\def    \kev            {\mbox{$\mathrm{keV}$}}
\def    \mev            {\mbox{$\mathrm{MeV}$}}
\def    \gev            {\mbox{$\mathrm{GeV}$}}
\def	\tev		{\mbox{$\mathrm{TeV}$}}
\def    \gr            {\mbox{$\mathrm{gr}$}}
\def    \fm            {\mbox{$\mathrm{fm}$}}
\def    \cm            {\mbox{$\mathrm{cm}$}}
\def    \km            {\mbox{$\mathrm{km}$}}
\def    \isec            {\mbox{$\mathrm{s}^{-1}$}}
\def    \yr            {\mbox{$\mathrm{yr}$}}
\def    \pt             {\mbox{$p_{\mathrm T}$}}
\def    \et             {\mbox{$E_{\mathrm T}$}}
\def    \ifb            {\mbox{$\mathrm{fb}^{-1}$}}
\def\tesla{\ifmmode \mathrm{Tesla} \else Tesla \fi}
\def\gauss{\ifmmode \mathrm{G} \else G \fi}

\def    \as             {\ifmmode \alpha_s \else $\alpha_s$ \fi}
\def \mpl {\ifmmode M_{\mathrm{Pl}} \else $M_{\mathrm{Pl}}$ \fi}
\def \mplsq {\ifmmode M^2_{\mathrm{Pl}} \else $M^2_{\mathrm{Pl}}$ \fi}
\def\rem           {\ifmmode R_{\mathrm{EM}} \else $R_{\mathrm{EM}}$ \fi}
\def\remsq           {\ifmmode R^2_{\mathrm{EM}} \else $R^2_{\mathrm{EM}}$ \fi}
\def\remf         {\ifmmode R_{\mathrm{EM,4}} \else $R_{\mathrm{EM,4}}$ \fi}
\def\Rs {R}
\def\ktd {{\tilde k}_D}
\def\kd {k_D}
\def\khd{{\hat k}_D}
\def\arad{a}
\def\rson{R_B}
\def\hf{{1\over 2}}
\def\Mnr{M_{\rm NR}}
\def\Eprod{E_i}
\def\pprod{p_i}
\def\Mprod{M_i}
\def\Rprod{R_i}
\def\bhm{{\hat b}_{min}}
\def\bhmq{{\hat b}_{min,q}}
\def\cac{c_{ac,p}}
\def\cacm{c_{ac,M}}
\def\cel{c_{sc}}
\def\rhat{{\hat r}}
\def\bhat{{\hat b}}
\def\rshat{{\sqrt{\hat s}}}
\def\sun{{\odot}}
\def\Msun{M_\odot}
\def\gauss{\mathrm{G}}
\begin{titlepage}
\nopagebreak
{\begin{flushright}
{\begin{minipage}{7cm}
    CERN-PH-TH/2008-025 
\end{minipage}}
\end{flushright}}
\vfill
\begin{center}
{\bf\sc\LARGE Astrophysical implications of  \\[.5cm] hypothetical
  stable TeV-scale black holes} 
\end{center}
\vfill
\begin{center}
{\large 
Steven B. Giddings$^{a,}$\footnote{giddings@physics.ucsb.edu}
 and Michelangelo
 L.~Mangano$^{b,}$\footnote{michelangelo.mangano@cern.ch}
}
\end{center}

\vskip 0.5cm

{\it  $^{a}$ Department of Physics, 
University of California, Santa Barbara, CA 93106}

{\it $^{b}$ PH-TH, CERN, Geneva, Switzerland
}
\vfill
\begin{abstract}
We analyze macroscopic effects of TeV-scale black holes, such as could
possibly be produced at the LHC, in what is regarded as an extremely
hypothetical scenario in which they are stable and, if trapped inside
Earth, begin to accrete matter.  We examine a wide variety of
TeV-scale gravity scenarios, basing the resulting accretion models on
first-principles, basic, and well-tested physical laws.  These
scenarios fall into two classes, depending on whether accretion could
have any macroscopic effect on the Earth at times shorter than the
Sun's natural lifetime.  We argue that cases with such effect at
shorter times than the solar lifetime are ruled out, since in these
scenarios black holes produced by cosmic rays impinging on much denser
white dwarfs and neutron stars would then catalyze their decay on
timescales incompatible with their known lifetimes.  We also comment
on relevant lifetimes for astronomical objects that capture primordial
black holes.  In short, this study finds no basis for concerns that
TeV-scale black holes from the LHC could pose a risk to Earth on time
scales shorter than the Earth's natural lifetime. Indeed, conservative
arguments based on detailed calculations and the best-available
scientific knowledge, including solid astronomical data, conclude,
from multiple perspectives, that there is no risk of any significance
whatsoever from such black holes.
 \end{abstract}
CERN-PH-TH/2008-025\hfill \\  
\vfill
\end{titlepage}
\tableofcontents
\newpage
\section{Introduction}
One of the most spectacular manifestations of nature realizing certain extra-dimensional 
scenarios\cite{ArkaniHamed:1998rs,Antoniadis:1998ig,RSi,GKP}
could be the production of microscopic black holes
at the LHC~\cite{Giddings:2001bu,Dimopoulos:2001hw}.\footnote{For
reviews with more references see
\cite{Giddings:2001ih,Giddings:2007nr}.} These are expected to undergo
prompt, quasi-thermal, Hawking~\cite{Hawking:1974sw} decay to large
multiplicities of elementary particles, leading to very characteristic
final states. It has been suggested~\cite{Unruh:2004zk,Vilkovisky:2005db},
however, that black hole decay via Hawking radiation may not be a
universal feature, and could for example depend on the details of the
Planck-scale degrees of freedom.  While this suggestion is not based
on any complete 
microphysical picture, and furthermore appears contradictory to basic 
quantum-mechanical principles\footnote{See, for example,
\cite{'tHooft:1996tq}.},  it does raise a possible question about stability of 
microscopic black holes that might be produced at the LHC, in
TeV-scale gravity scenarios.  This in turn has led some to express
concern about the fate of their evolution: could their accretion pose
any threat to the Earth? This is the question addressed in this paper.

The structure of this work, and a succinct summary of our findings,
are outlined here. We begin our work by
reviewing what are widely regarded as quite convincing arguments for
the robustness of the exceedingly rapid decay of microscopic black
holes. We then discuss the arguments asserting that macroscopic
consequences of new and unknown particles can be ruled out by the lack
of evidence for their effects from production by cosmic rays
hitting the surface of the Earth or other astronomical bodies. We
argue here that charged black holes will lose enough energy to stop
when traversing the Earth or the Sun, via standard electromagnetic
processes. Since black holes would be typically
produced by the collision of quark
pairs, whether in cosmic-ray interactions or at the LHC, they would often be
initially charged. To the extent that no mechanism leads to their
neutralization, the cosmic-ray based argument for their being harmless
is therefore robust. Their neutralization through the Schwinger mechanism
proceeds  according to quantum principles like those underlying Hawking
radiation.  There is therefore no concrete framework where
neutralization occurs without Hawking decay taking place as well,
leading to a likely contradiction in assuming that stable black
holes must be neutral.  We nonetheless make the hypothesis that 
this odd
situation could occur, and analyze the possible effects of such neutral and stable black holes,
beginning with a review of some essential features of gravity and black
holes in $D> 4$ dimensions, including both large- and
warped-extra-dimensions models. 

Next, we develop the formalism to describe the evolution of such black
holes trapped inside the Earth or inside dense objects such as white
dwarfs and neutron stars.  We introduce and discuss accretion
scenarios that apply within nuclear, atomic and macroscopic matter,
evaluating the time scales corresponding to various phases in the
evolution of a growing black hole.  We establish upper and lower
limits to the rate at which accretion can take place, building on very
basic principles such as conservation laws and classical and quantum
dynamics. This is possible since, in order for accretion to become
macroscopic, it is necessary that a black hole exerts its influence
over distance scales much larger than its event horizon. Such black
holes are very small, and their accretion power, if limited to
absorbing particles that have impact parameters of the order of the
Schwarzschild radius, is typically insufficient to cause macroscopic
growth. At large distances the physical processes become independent of the
short-distance properties of the black hole, which only acts through
its attractive potential, and as a mass sink. In this long-distance domain,
classical or quantum dynamics are well-tested, making the
study of accretion possible, independently of our detailed knowledge
or ignorance about the physics inside the black hole.

We investigate two concrete scenarios, for various configurations of
the extra dimensions: growth of hypothetical LHC-produced stable black
holes within the Earth, and growth of such black holes produced by
cosmic rays impinging on dense stars, such as white dwarfs or neutron
stars.  In both cases, we make conservative assumptions, namely
fastest possible growth in Earth, and slowest possible growth in a
dense star.  The first scenario shows that, if the radius of crossover
from higher-dimensional gravity to four-dimensional gravity is less
than about 200 \AA, the natural lifetime of the solar system is 
too short to allow significant growth of stable black holes that may
be captured inside the Earth.  In contrast, in a scenario where the crossover
radius exceeds the 200 \AA\ scale, accretion times could be shorter than
the solar time scale.  In this case, however, examination of the
latter dense-star scenario then produces an  argument that, given observational data
setting the lifetimes of such objects at a billion years or more, such
stable black holes cannot in fact exist for a crossover radius
greater than $\simeq15$\AA.

These arguments thus conclude with the exclusion of any relevant consequences
for Earth due to the evolution of black holes possibly produced by the
LHC.

While the main line of the argument is relatively straightforward, being
based on well-established macroscopic properties of matter, several
issues require an in-depth study in order to provide a robust basis
to our conclusions. When studying accretion, for example, we need to
consider the possibility of effects such as the Eddington limit, where radiation emitted
during the infall of matter would slow down the accretion via radiative
pressure, constraining the accretion rate of a black hole and thus
possibly slowing down its growth inside Earth or a dense star. The estimates of the
number of black holes produced by cosmic rays need to rely on a solid
understanding of the latter's spectra and composition. Here we
therefore consider the worse-case scenarios where, contrary to
mounting experimental evidence, the cosmic primaries are dominantly
heavy nuclei rather than protons. The loss of acceptance for the
highest-energy cosmic ray flux hitting dense stars, due to the intense
surface magnetic fields, is also an important element that we
analyze. Finally, the slow-down and stopping of relativistic black
holes inside dense stars requires a detailed study of energy loss in
gravitational scattering and absorption processes.

We note that our bounds are conservative.  In particular, at each
point where we have encountered an uncertainty, we have replaced it by
a conservative or ``worst case" assumption.  For this reason, our
bounds can likely be improved through further work removing such
uncertainties. Furthermore, we do not exclude that there could be
other independent arguments, based for example on astrophysical or
cosmological production of microscopic black holes, leading to
the exclusion of stable black holes, or of their macroscopic effects.

In outline, the next section summarizes existing arguments (and their
extension) against risk from TeV-scale black holes, namely the
robustness of quantum black hole decay, and constraints from cosmic
rays impinging on objects in the solar system.  Section three then
summarizes aspects of TeV-scale gravity scenarios and the
corresponding black holes.  Section four discusses black hole
accretion in Earth.  Section five gives an in-depth analysis of the
question of stopping of neutral black holes, via gravitational
interactions.  Section six describes production of black holes by
cosmic rays impinging on white dwarfs.  Section seven describes the
resulting accretion of a white dwarf, and thus derives constraints on hypothetical
TeV-scale scenarios that otherwise might have been of concern.  Section eight
describes similar constraints arising from black hole production on,
and accretion of, neutron stars.  A summary and conclusions appear in
section eight.

Many of the technical details are treated in a series of  Appendices. Appendix~\ref{app:bondi} summarizes
macroscopic Bondi accretion, which is the canonical framework to deal
with the flow of matter into a black hole. For its use in our context,
Bondi accretion is generalized here to higher-dimensional
gravitational fields.  Appendix~\ref{app:rerad} discusses the effects
of radiation emitted during accretion, and  examines the radiative
transport conditions relevant to the question of an Eddington limit
for the growth of microscopic black holes. In
Appendix~\ref{app:scattering} we derive the main equations governing
the particle scattering and capture in the field of a $D$-dimensional
black hole,  and Appendix~\ref{app:accretion} proves few facts required
by the study of stopping by white dwarfs. 
Appendix~\ref{app:prod} reviews the formalism for the
calculation of black hole production at the LHC and in the collisions
of cosmic rays, including the case of high-energy neutrinos.  Here we
discuss the impact of different assumptions about the composition of
the high-energy cosmic rays, we evaluate the production rates under
the various assumptions, and establish the most conservative lower
limits on such rates. We also discuss the production properties of
black holes produced by cosmic rays, to serve as initial conditions
for their slow-down  in dense stars. In Appendix~\ref{app:boundBH}
we calculate the probability that black holes produced at the LHC are
trapped in the Earth's gravitational field. We argue that,
independently of our general conclusion that trapped black holes would
not accrete macroscopically, the velocity spectrum of black holes
produced during the LHC lifetime in allowed extra-dimensional
scenarios is such that they would  typically escape the Earth's
attraction. Appendix~\ref{app:synch} discusses the impact of magnetic
fields on the penetration of cosmic rays down to the surface of dense
stars.  Appendix~\ref{app:background} addresses the production of
black holes by cosmic rays hitting ``background'' objects, such as
neutron star companions, the interstellar medium, or even dark matter.
Here we discuss
the conditions for the companion of a neutron star in X-ray binary
systems to efficiently act as a beam-dump for cosmic rays, leading to
the production of black holes and their capture in the neutron star,
unimpeded by the presence of magnetic fields.  A final Appendix
contains useful conversion factors and physical constants that are
used throughout the paper.

\section{Assessment of pre-existing arguments}

We begin by assessing the strengths and weaknesses of previously
existing arguments regarding risk from black hole production.

\subsection{Instability of microscopic black holes}
\label{sec:instab}
One of the most profound steps so far towards the yet-incomplete unification
of quantum mechanical and gravitational principles was Hawking's
discovery \cite{Hawking:1974sw} that black holes evaporate.  This
provided an important missing link in the pre-existing laws of black hole
thermodynamics, by explicitly calculating the temperature that
characterizes a radiating black hole.

While Hawking's result has become nearly universally accepted, it is certainly
true that elements of the original derivation of black hole radiance
rely on assumptions that are apparently not valid.  Notable among
these is the use of modes of ultra-planckian frequencies at
intermediate steps in the derivation.  This
naturally raises the question of the robustness of the result.

Belief in the robustness of Hawking's prediction of nearly thermal
evaporation has been boosted by arguments for the result which have
now been produced from several different directions.  These
derivations have the virtue of either facing head-on the issue of the
transplanckian modes, or being independent of them, and the basic effect has survived a number of important consistency checks.

One early approach relying on the trace anomaly and avoiding explicit
reference to transplanckian modes was pioneered by Christensen and
Fulling\cite{Christensen:1977jc}.  In this approach, the stress tensor
describing the Hawking radiation is found by combining the known trace
anomaly in two dimensions, and the constraint that the stress tensor
be conserved.  This approach has been used to give explicit models of
evaporating black holes \cite{Callan:1992rs}, and has also recently
been generalized to higher dimensions in \cite{Robinson:2005pd} and a
number of followup works.

Another approach is to modify the short-distance physics to remove the
offending transplanckian modes.  One summary of such a prescription
appears in \cite{Jacobson:2003vx}, where it is argued that despite a
different origin for the modes, this approach supports the statement that
Hawking radiation persists.  This
approach is also related to analog gravity models.\footnote{For a
review, see \cite{Barcelo:2005fc}.}  These are models based on
condensed matter systems, where a horizon for sound waves occurs, and
exhibits a precise analog of the Hawking effect.  For example, the
abstract of a recent overview talk\cite{Unruh} on the subject of
Hawking radiation states ``analog models of gravity have given us a
clue that despite the shaky derivation, the effect is almost certainly
right."  These models illustrate various detailed features of the
corresponding radiation in the analog systems.

One of the puzzles of Hawking radiation stems from the fact that it
appears to lead to loss of quantum coherence, when followed to the end
state of evaporation \cite{Hawking:1976ra}, and this has led to the
``black hole information paradox."\footnote{For reviews, see
\cite{Giddings:1994pj,Strominger:1994tn,Giddings:1995gd}.}  Many
workers feel that the resolution will be that there are subtle
corrections to Hawking's thermal spectrum, that lead to unitary
evolution.  Thus while very few question that black holes Hawking
evaporate, it is clear that there are detailed aspects of the
evaporation process that we do not understand.

Indeed, the basic result, that black holes evaporate,  appears quite robust
from a very general perspective.  Nature is quantum-mechanical, and
basic quantum-mechanical principles dictate that any allowed decay
will occur.  Thus, stable (or nearly-stable) objects must be
``protected" by conservation laws, examples being baryon number,
lepton number, {\it etc.}  There is no such conserved charge carried
by a black hole that is not carried by ordinary matter, if a black hole can be
produced in collisions of partons at the LHC.  Thus by basic quantum
principles such a heavy black hole should decay into light, ordinary matter,
 and the only question is
the time scale.  Since such a black hole can have mass at most around ten
times the higher-dimensional Planck mass, $M_D\sim 1~\tev$, the only
relevant dimensionful parameter is the corresponding time scale,
$t_D\sim 1/M_D \sim 10^{-27}s$, and there are no other small dimensionless
parameters to suppress decay.  Thus, on very general grounds
such black holes are expected to
be extremely short-lived, as is indeed predicted by the more detailed
calculations of Hawking and successors.

Despite these very strong arguments for black hole decay, the possibility
of manufacturing microscopic black holes on Earth suggests
that one conduct an independent check of their benign nature.  For that
reason, this paper will test the hypothesis that the statements of
this subsection are false, by investigating possible consequences of
hypothetical black holes that do not undergo Hawking decay.  

\subsection{Cosmic ray collisions on Earth}
\label{sec:crEarth}

Cosmic rays hit Earth with energies known to exceed $10^{20}\ev$,
corresponding to center-of-mass (CM) energies exceeding $100\,\tev$.
Thus, anything that can be made with Earth-based accelerators up to
this energy is already being made by nature. This argument can be used
to effectively rule out the existence of particles
predicted by some speculative scenarios, or to exclude possible
macroscopic consequences of high energy collisions, as discussed for
example in~\cite{Busza:1999pr,Iliopoulos}.  

This argument  requires
however more attention if the large momentum imparted to the produced
particles by the cosmic-ray kinematics has an impact on how they
evolve after production. 
Consider the cosmic-ray collision producing the black hole (or any
other particle), at the parton level. $E_1$ is the energy of the parton
inside the cosmic ray nucleon, and $E_2 < m_p$ the energy of the
parton inside the atmospheric nucleon, where $m_p$ is the nucleon
mass. To produce a particle of mass
$M$, we need $E_1 \geq M^2/2E_2$, and therefore the minimum energy of
the produced black hole in the Earth rest frame is given by
$E_1>M^2/2m_p$, or, for the maximum value of $M=14$~TeV allowed at the LHC,
\be\label{BHmom}
E_1 \sim p_{BH} >  10^8 \gev\ \; ,
\ee
 where $p_{BH}$ is the momentum of the resulting black hole.
Any argument that such black holes, should they be stable, would
undergo macroscopic accretion, must be based on the proof that, while
traversing the Earth, they
slow down enough to get trapped
by the Earth's gravitational field.
Most such TeV-scale black holes should initially have color or
electric charge, since the incident partons (quarks, gluons) are
charged.  In the usual picture, where Hawking radiation is
present, these rapidly discharge through the Schwinger
mechanism~\cite{Schwinger:1951nm}  of
particle-antiparticle pair creation in an intense (chromo-)electric
field, or through breaking/absorbing a QCD string. 
The timescale for the neutralization is proportional to the black hole
Schwarzschild radius $R$.  While there is no known
example of a consistent microphysics such that Schwinger discharge occurs
and Hawking radiation does not, one can point out one difference
between them.  Specifically, the Schwinger effect can be described in
terms of pair production in the gauge field outside the horizon, but
Hawking radiation is a trans-horizon effect.  In the unlikely event
that our understanding of the horizon misses some critical element
forbidding black hole decay (an assumption that, in our view, appears
to contradict the basic quantum principles outlined above), one might
imagine that Schwinger discharge nevertheless takes place.  (One could
parameterize  such a scenario by imposing rather artificial boundary
conditions at the horizon.)

Thus, we will consider collisions on Earth in two contexts -- those
which produce hypothetical stable charged black holes, and those which
produce hypothetical stable black holes that rapidly neutralize.

Passage of a high-energy charged particle through matter leads to
well-understood energy loss~\cite{Yao:2006px,Groom}. This is due to
long-range electromagnetic effects, that have nothing to do with the
microphysics associated to the particle itself. Therefore, a muon, or
a black hole with the electric charge and the mass of a muon, would be
subject to the same energy loss through radiative processes as they
move through matter. Since at high energy the radiative properties are
mostly determined by the value of the Lorentz $\gamma$ factor of the
charged particle, we describe the energy-loss properties of a
singly-charged black hole by rescaling the energy loss of muons to the
equivalent $\gamma$ value.  For relativistic velocities, below the
threshold for $e^+e^-$ pair production in the field of nuclei, the
energy loss is described by the Bethe-Bloch equation. The energy
loss in this regime depends on the velocity, with a slow growth
proportional to $\log\gamma$. The stopping power for black holes is
therefore similar to that of muons, of the order of 2~MeV~cm$^2$/g up
to $\gamma\sim 10^3$~\cite{Yao:2006px,Groom}.  For the average
composition of Earth, this means an energy loss of about 11~MeV/cm.
Above $\gamma\sim 10^3$ (which is the case at the time of production
for black holes of mass larger than 1~TeV) pair production,
bremstrahlung and nuclear dissociation appear. These grow
approximately linearly with energy (or $\gamma$), with an energy loss
of the order of 60~MeV/cm at $\gamma\sim 10^4$, for the average Earth
density.  With Earth-like densities, the distance scale necessary to
slow down from the production energy $E\sim M^2/m_p$ to $\gamma\sim
10^3$ is thus of order $M/(6 \kev)$~cm, or, for $M\sim 14$~TeV,
more than $10^4$~km, larger than the radius of the Earth. The subsequent
slow-down below $\gamma\sim 10^3$ takes place with the constant energy
loss of $1~\gev$/m, corresponding to $\gsim10^4$~km. The stopping
distance grows with $M$, and a more careful estimate shows the Earth
provides enough stopping power for black holes with unit electric
charge up to a mass of the order of 7~\tev.
For larger masses, one can appeal to the stopping power of
the Sun.  With a core density of about 150~gr/cm$^3$, over a radius of
the order of 0.2 solar radii, namely $1.4\times 10^5$~km, the Sun
column density can stop black holes of mass well in excess of 100~TeV.
Notice also that our estimates are based on a black hole with unit
electric charge. Any minor addition to $Q=1$, due for example to
accretion of a net charge, would further increase the stopping power.
The continued health of the Sun on multi-billion year time scales --
during which many such black holes would have been produced and
stopped -- thus apparently immediately rules out any risk from charged
TeV-scale black holes.

The above arguments apply to any elementary charged particle. In
particular, they apply to possible magnetic monopoles. The stopping
power for monopoles moving through matter is about 100 times larger
than that of a charged particle~\cite{Giacomelli:2005xz}, 
since their magnetic charge is at
least $1/\alpha$ times greater than the electron electric charge. This
means that the Earth itself can stop magnetic monopoles of masses much
larger than those that can be produced at the LHC.

Let us next turn to the hypothesis that such black holes are stable,
but rapidly neutralize via Schwinger production or another mechanism.
In this case, the only interactions of a black hole with matter are
through its gravitational field.  We will give a more careful
treatment of such interactions below, but a rough cross-section is the
geometric size, $\sigma\sim 1/\tev^2$.  Again using the average
density of Earth, we find of order one scattering event expected in
its transit for this cross section.  The  maximum  momentum loss in a
collision with a parton of energy $E_p$ 
is of size $\sim
\gamma^2 E_p$.  This is achieved when particles are scattered with significant momentum transfer in the black hole frame, for example 
in a head-on elastic collision, where
the target particle is significantly deflected. Since $\gamma\sim M/E_p$,
this would  correspond to a loss of a fraction of order 1 of the
initial momentum (\ref{BHmom}).  However, this is clearly not necessarily typical:
due to the short-distance character of the $D$-dimensional
gravitational potential, the impact parameter range for hard elastic
scattering is restricted, and typically we shall either have a
small-angle scattering, or the target particle can be captured by the
black hole, leading to a large inelasticity and to a  reduced
velocity loss.  We shall discuss these issues in more detail in
Section~\ref{app:cr}.  The more careful analysis presented there
also indicates that the
stopping power of the Sun is likewise insufficient.   

To summarize, hypothetical stable charged black holes should stop in
the Earth for masses up to about 7~TeV, and in the Sun if heavier.
The multi-billion year longevity of Earth and Sun apparently provides a good safety
guarantee.  We have no concrete example of a consistent microphysics
such that black holes neutralize via Schwinger discharge but don't
Hawking radiate, but our present state of knowledge of quantum black
hole processes doesn't strictly  rule out such a possibility.  Thus,
we will seek alternative bounds on such a scenario, which will also
serve the purpose of improving the stringency of the bounds for
charged black holes. 

\section{Essentials of higher-dimensional gravity}

In this section we quickly review some features of higher-dimensional scenarios realizing TeV-scale gravity, and of the black holes that exist in these scenarios.  A brief overview discussing more aspects of these
scenarios and black hole production in them is \cite{Giddings:2001ih}.

\subsection{TeV-scale gravity scenarios}
\label{sec:tevgrav}
The basic idea of TeV-scale gravity is that either via large extra
dimensions\cite{ArkaniHamed:1998rs,Antoniadis:1998ig} or large warping\cite{RSi}, the true Planck scale is lowered to the
vicinity of a $\tev$.  
To summarize, let the $D$-dimensional action be
\begin{equation}
S=\frac{1}{ 8\pi G_D} \int d^D x \sqrt{-g} ~\frac{1}{2} {\cal R} + \int 
d^D x \sqrt{-g} {\cal L}\  ,
\label{Eact}
\end{equation}
where $G_D$ is the $D$-dimensional gravitational constant, $\cal R$ is the Ricci scalar, and $\cal L$ is the matter lagrangian.
We consider a general compact metric $g_{mn}(y)$, possibly together with warp factor $A(y)$,
\begin{equation}
ds^2=e^{2A(y)}dx^\mu dx_\mu + g_{mn}(y) dy^m dy^n \ .\label{warpmet}
\end{equation} 
Here the non-compact coordinates are $x^\mu$, and the standard model fields are typically taken to lie on a brane spanning these dimensions.
Then the relation between the higher-dimensional Planck
mass,\footnote{In this paper, we use the conventions of the ``Extra
  dimensions'' minireview by Giudice and Wells, in the  {\it
    Particle Data Book}\cite{Yao:2006px}.}  
\be
M_D^{D-2}={(2\pi)^{D-4} \over 8 \pi G_D }\ 
\label{Mpnorm}
\ee
and the four-dimensional Planck mass, defined via the four-dimensional
gravitational action  
\be
S_4= {M_4^2 \over 2} \int d^4 x \sqrt{-g_4(x)} {\cal R}_4\ ,
\ee
is
\be
{M_4^2\over M_D^2} = M_D^{D-4} \int {d^{D-4}y \over (2\pi)^{D-4}}
\sqrt{g_{D-4}} 
e^{2A}\equiv M_D^{D-4} {V_w\over (2\pi)^{D-4}}\  .\label{mplanck}
\ee
This equation defines the ``warped volume'' $V_w$.

Current lower bounds~\cite{Yao:2006px} on $M_D$ are around $1 ~\tev$.  In
order for $M_D$ to be in the \tev\ vicinity, one must have a large
warped volume.  This can be achieved by large volume and moderate
warping, by large warping, or by some combination of the two.  A
simplified version of the relationship (\ref{mplanck}), assuming the
scale of all extra dimensions is set by a radius $R_D$, is 
\be\label{scalerat}
{M_4^2\over M_D^2} \approx (M_DR_D)^{D-4} e^{2\Delta A}\ , 
\ee 
where $\Delta A$ is a measure of the relative difference in warping
between the region of maximal warp factor and the region in which
standard-model physics resides.

One can solve for the characteristic $D$-dependent size of the extra
dimensions, 
\be \label{rddef}
R_D \approx M_D^{-1} \left( {e^{-2\Delta A} M_4^2\over
  M_D^2}\right)^{1/(D-4)}\ . 
\ee
Of course, larger warping and fixed $M_D$ means that $R_D$ is smaller for a given $D$.
In particular, for unwarped scenarios $R_5$ is macroscopic and thus
ruled out, but with sufficient warping one finds viable scenarios,
such as that of \cite{RSi}. 

For $M_D = 1 ~\tev$ and with no warping, we find the following radii: 
\ba \label{radii}
&&R_D = 4.8\times 10^{-2} \cm\ , \quad \mbox{for } D=6 \\
&&R_D= 3.6\times 10^{-7} \cm\ , \quad \mbox{for } D=7 \\
&&R_D = 9.7\times 10^{-10}\cm\ , \quad  \mbox{for } D=8 \\
&&R_D = 2.8\times 10^{-11} \cm\ , \quad \mbox{for } D=9 \\
&&R_D= 2.7\times 10^{-12}\cm\ , \quad \mbox{for } D=10 \\ 
\label{radiii}&&
R_D= 4.9\times 10^{-13}\cm\ , \quad \mbox{for } D=11 \ .
\ea
For higher $M_D$ and/or in the presence of warping, these numbers
should be multiplied by a factor 
\be
\left({M_D\over \tev}\right)^{(2-D)/(D-4)} e^{-2\Delta A/(D-4)}\ .
\ee

\subsection{Higher-dimensional black holes}

\subsubsection{Schwarzschild solution}

We next turn to properties of black holes in these scenarios.  We begin with the D-dimensional Schwarzschild solution with mass $M$, which takes the form 
\be\label{Schw}
ds^2 = -\left[ 1- \left({R(M)\over r}\right)^{D-3}\right] dt^2
+{1\over 1- \left({R(M)\over r}\right)^{D-3}}dr^2 + r^2 d\Omega^2\ . 
\ee
Here the  Schwarzschild radius $R(M)$ is  
\be\label{schwarz}
R(M) = {1\over M_D} \left({k_DM\over M_D}\right)^{1/(D-3)}\ , 
\ee
where the constant $k_D$ is defined as
\be
k_D= {2(2\pi)^{D-4}\over (D-2) \Omega_{D-2}}\ ,
\ee
and where $\Omega_{D-2}$ is the volume of the unit $D-2$ sphere,
\be
\Omega_{D-2} = {2\pi^{(D-1)/2}\over \Gamma[(D-1)/2]}\ .
\ee
One can likewise write down the higher-dimensional version of the
Kerr solution \cite{MyPe}.

\subsubsection{Black holes in standard compactifications}

Nonrotating black holes with radius much less than the curvature scales and the sizes of
the extra dimensions are well approximated by the Schwarzschild
solution (\ref{Schw}).  When the Schwarzschild radius reaches a size
comparable to that of an extra dimension, one expects an inverse
Gregory-Laflamme\cite{GrLa} transition to the lower-dimensional black
hole, extended over the extra dimension.  In this case the solution
will be given by the lower-dimensional version of the solution (\ref{Schw}), with
trivial dependence on the compact coordinates.

For the purposes of computing forces due to a black hole, we will need
its gravitational potential.  In the weak-field regime this is given
by
\be\label{Phidef}
\phi = -(g_{00} +1)/2\ ,
\ee
leading to the force on a mass $m$:
\be \label{eq:Dforce}
F_G(r) = -\frac{\ktd}{M_D^{D-2}} \frac{Mm}{r^{D-2}} \; ,
\ee
where
\be
\ktd = (D-3)k_D/2 = {D-3\over D-2}{(2\pi)^{D-4}\over \Omega_{D-2}} \ .
\ee
The attractive gravitational force matches between the lower and
higher-dimensional expressions in the region $r\sim R_D$.
Specifically, by equating the D-dimensional force law to that of 4D,
we find that the forces match at a crossover radius
\be\label{fcross}
R_C= (8\pi \ktd)^{1/(D-4)} R_D\ ,
\ee
whose values are of size $5-6~R_D$ in cases of interest.

\subsubsection{Black holes in warped compactifications}
\label{sec:bhwarp}
As an example of a broad class of warped compactification scenarios,
consider a metric of the form 
\be\label{wcspec} 
ds^2 = dy^2 + e^{2y/R_D} dx_4^2 + R_D^2 ds_X^2\ ,
\ee 
where $e^{2y/R_D}\, dx_4^2$ describes the $D=4$ part of the
metric.
Here $R_D$ is both the characteristic curvature radius associated with
the warping, and the radius of the remaining extra dimensions,
compactified on some compact manifold $X$, whose metric $ds_X^2$ we
have taken to have radius ${\cal O}(1)$.  The coordinate $y$ is taken
to have range $(0,L)$.  The form (\ref{wcspec}) is representative of
many known examples of warped compactifications, such as the truncated
$AdS_5$ solutions of \cite{RSi} and the warped flux compactifications
of \cite{GKP,KKLT}, although more generally one expects for example
the metric of $X$ to vary with $y$.

For these compactifications, a black hole whose radius satisfies $R\ll
R_D$ is well-described as the D-dimensional Schwarzschild
solution (\ref{Schw}).  On the other hand, a sufficiently large black
hole is expected to be described by a four-dimensional Schwarzschild
solution, represented by (\ref{wcspec}) with $dx_4^2$ replaced by the
four-dimensional Schwarzschild metric.  
Between these extremes the
solutions are not known.

However, approximate forms for the linearized gravitational potential,
appropriate to describing the weak field regime of a concentrated mass
such as a black hole, have been derived in \cite{Froiss}.  In the
region $R_D\lsim r\lsim L$, one finds the linearized perturbation
of $dx_4^2$ given by 
\be\label{wpert} \phi = {\khd M
\over M_D^{D-2} } {e^{-j_Dr/R_D -(D-1)y/R_D}\over R_Dr^{D-4}} \ .  \ee
Here $j_D= j_{(D-3)/2,1}$ is the first zero of the relevant Bessel
function, and $\khd$ is a constant.  Thus, the radial gravitational
force at $y=0$ is 
\be\label{wforce} F_{D,w} = -{\khd M \over M_D^{D-2}} {e^{-j_D
r/R_D} \over R_D r^{D-4}} \left({j_D\over R_D} + {D-4\over r}\right) \ .
\ee

An important question is at what specific radius $R_C$ does the
gravitational force from (\ref{wpert}) match onto that for
four-dimensions.  This can be found by equating $F_{D,w}$ to
$-G_4M/r^2$.  
This yields the relation
\be\label{capprox}
{R_C\over R_D} = {1\over j_D} \left\{\ln\left[{M_4^2\over M_D^2(R_CM_D)^{D-4}}\right] + 2\ln\left({R_C\over R_D}\right) + \ln\left(8\pi {\hat k}_D\left[j_D + (D-4) {R_D\over R_C}\right]\right)\right\}\ .
\ee
From this, we see that if $R_C$ is significantly smaller than the radius given by the unwarped version of (\ref{scalerat}), this correspondingly increases the size of the region between $R_D$ and $R_C$ where warping is significant.

\subsubsection{A general perspective}
\label{genpers}

We close this section by outlining a broader perspective on TeV-scale gravity and black holes.  Note that a general feature of the above discussion is that  below the value $R_C$ the potential (\ref{Phidef}) crosses over from the four-dimensional form to one that grows more rapidly as $r$ decreases.  The flat and warped cases give both power law and exponential growth laws.  Moreover, the gravitational  potential is  proportional to the mass.  In order to have a TeV-scale model, note that the potential corresponding to a TeV-scale mass should reach the value $\phi\sim 1$ by the time $r$ reaches the value $r\sim 1/\tev$.  We assume that while the gravitational potential is modified at short distances, the dynamics of other forces is four-dimensional, as in brane-world models, in order to agree with experiment. 

 Moreover, as we will see, many features of the accretion process only depend on the long range potential, since it is at such long scales that gravity begins to compete with other effects.  
From this general perspective, a very general definition of a black hole is as an object with such a long-range gravitational potential, and which is allowed to accumulate mass at $\phi\sim1$. 

One could postulate more general forms for the potential, but clearly
the cases we have described 
are representative of a very wide class of potentials that become
strong at the TeV scale.

\section{Black hole accretion in Earth}
\label{sec:Earthacc}
\subsection{Accretion basics}
Our interest is in accretion of black holes trapped
 inside
 astronomical bodies such as planets, stars, neutron stars, {\it etc.}  
One parameter governing this accretion is the 
effective capture radius $r_c(M)$ of the black hole, for a given mass.  This is the radius out to which the
gravitational field of the black hole succeeds in attracting matter that
will eventually be absorbed.  The other parameter is the flux of mass towards the black hole, $F$.  Specifically, the black hole mass grows as
\be\label{genevol}
{dM\over dt} = \pi r_c^2 F\ .
\ee
This formula neglects reradiation of incident energy, which is discussed in Appendix~\ref{app:rerad}, and, if present, can lower the growth rate.
The flux can arise either from the motion of the black hole relative to the body, or from the motion of the constituents of the body relative to the black hole.  In the case where the dominant effect is the velocity $v$ of the black hole, we have
\be
F=\rho v
\ee
where $\rho$ is the mass density near the capture radius. 
This produces an evolution equation
\be \label{eq:defevol}
\frac{dM}{dt} = \pi \rho \, v \, r_c^2(M) \; ,
\ee

The capture radius $r_c$ is frequently different from the 
Schwarzschild radius $R$, 
and depends on the size and state of motion of the black hole, as well
as on properties of the surrounding medium.   For example, free
particles with velocity $v$ with respect to a black hole have capture
radii  
\be
r_c\approx R/v
\ee
Black holes
whose production is accessible at the LHC have an initial  radius of
the order \tev$^{-1}$. As they absorb matter, their physical and
capture radii grow.\footnote{They also may have significant initial angular momentum.  However, as they absorb matter, with negligible average angular momentum, they become increasingly well-approximated as non-spinning black holes.}  In atomic matter, there are three possible domains where the
properties of black hole interactions with matter vary.  
The first phase is that where $r_c$ 
is smaller than the nucleon size, $r_N \sim 1~\fm$. A
second phase is that where
$r_N \lsim r_c \lsim \arad$ (where $\arad\sim 1$~\AA\ is the atomic
radius). The third phase is $r_c>\arad$. 
Similar phases are present for growth inside a white dwarf, but in the case of growth
inside nuclear matter of a neutron star, the two latter phases are replaced by a single
phase with $r_c \gsim r_N$. The details of the evolution during these phases
will vary, depending on where $r_c$ is relative to $R_D$ and $R_C$, the distances characterizing crossover from the $D$-dimensional force law to that of four-dimensions.

This section will discuss evolution, first at the atomic level, and
then, for larger-scale black holes, from macroscopic matter.  
The latter is described by Bondi evolution; we also briefly discuss, and argue against, the presence of an Eddington limit, which would be relevant if emitted radiation were sufficient to slow accretion.  
 Our goal will be 
to estimate, under the most pessimistic assumptions, namely of fastest
possible growth, the timescale required for accretion of black holes
to macroscopic size. 
In subsequent sections, we will also perform similar calculations for accretion of white dwarfs and of neutron stars.

\subsection{Subatomic accretion in Earth}

\subsubsection{Competition with electromagnetic binding}

As is described in Appendix~\ref{app:boundBH}, most LHC-produced black
holes would be produced with large velocity as compared to the Earth's
escape velocity, $v_E=11$ km/s, due to imbalanced kinematics of the
initial-state partons, initial and final state radiation, etc.
However, those that are downward directed will accrete matter and slow
down while passing through the Earth.  Appendix~\ref{app:boundBH} estimates this effect based on
closely related calculations for black holes created by cosmic rays in section~\ref{app:cr}.  For present purposes, we will simply make the most conservative assumption that some of these black holes do become gravitationally bound to Earth.  Given the escape velocity and that the minimum mass of such a black hole would be ${\cal O}(\tev)$, typical kinetic and gravitational potential energies would thus be $\gsim{\cal O}(\kev)$.  This means that on the occasions where a black hole and nucleus bind, the black hole's energy overcomes the (atomic) binding energy of the nucleus to the surrounding material, and the combined system continues to fall.  Thus the black hole's motion should initially be dictated by the net gravitational field of the Earth.  

In a collision of the black hole with a nucleus, binding depends on the size of the impact parameter $b$; we will largely neglect separate capture of electrons since their capture rates are much smaller due to their smaller masses and higher velocities.  The black hole's effects are significant at impact parameters where its gravity competes with the electromagnetic binding forces of the surrounding medium.  The latter are estimated by noting that if a nucleus is displaced from its equilibrium position by a small displacement $d$, 
one
will find a restoring force of the form
\be\label{emforce}
F_E(d) = - K d
\ee
for some constant $K$. This is justified, for example, by considering
the force acting on an ion\footnote{Here we account for the fact
that the inner electrons are typically 
strongly bound to the nucleus, and so move with it.} of charge $Z'$
as it is displaced by a distance $d$ from the center of
charge of its electron cloud:
\be \label{eq:ecloud}
F_E(d)\simeq - \alpha \; \frac{Z'}{d^2} \; Z' \frac{d^3}{a^3} = - \alpha 
\frac{Z^{\prime 2}}{a^3} d\ ,
\ee
where we assumed the electron charge to be uniformly distributed in
the atomic volume.
The black hole will exert a competing gravitational force which is maximum at the point of closest approach.  
For a
 $D$-dimensional force law (\ref{eq:Dforce}), and with the nucleus displaced by $d$ towards
 the black hole, it is
\be
F_G(d) = -{\ktd Mm\over M_D^{D-2} (b-d)^{D-2}}
\ee
where the nuclear mass is given in terms of the mass number and proton mass as $m\simeq
Am_p$.  The nucleus can become bound to the black hole if this force
dominates $F_E$ for all $d$ over the range $(0,b)$.  This amounts to the
condition that, for all $d$,
\be
{\ktd Mm\over KM_D^{D-2}}>d(b-d)^{D-2}\ .
\ee
Maximizing the right-hand side with respect to $d$, we
find the binding condition:
\be
b< (D-1)\left[\ktd
Mm\over (D-2)^{(D-2)} KM_D^{D-2}\right]^{1/(D-1)}=\rem\ ,
\ee
which defines the {\it electromagnetic capture radius} $\rem$.  To simplify subsequent expressions, we rewrite this as
\be \label{emrad}
\rem = {1\over M_D} \left({\beta_D M\over M_D}\right)^{1/(D-1)}
\ee
where we define
\be\label{betadef}
\beta_D = {(D-1)^{D-1}\over (D-2)^{D-2}} {\ktd M_D^2 m\over K}\ .
\ee

The ratio of electromagnetic to Schwarzschild radii is given by
\be
{\rem\over R} = {\beta_D^{1/(D-1)}\over k_D^{1/(D-3)}}
\left({M_D\over M}\right)^{2/(D-3)(D-1)}\ . 
\ee
Since $K$ is governed by atomic scales, $\beta_D\gg1$, and $\rem$
exceeds $R$ for subatomic $R$. 

For low relative velocities, nuclei entering this radius can become bound to the black hole.  
Note that this will not be the case for sufficiently high relative velocity
$v$, as the free-particle capture radius $R/v$ is smaller than $\rem$ for large $v$.  However, we will consider sufficiently small
velocities that $\rem<R/v$, where competition with electromagnetic binding described by $F_E$ dominates.

While subsequent total absorption of a captured nucleus is not
guaranteed,\footnote{In particular, (chromo-)electrostatic effects apparently slow accretion early in this phase.} the most conservative assumptions for the purposes of
discussing accretion on Earth are those that lead to the fastest
accretion.  We will thus assume that  all of the mass of the nucleus is absorbed.  
In the absence of other effects, this black hole would also have the charge of the nucleus.  This
charge may discharge through the Schwinger mechanism, or be retained,
depending on assumptions.  If it is retained, the next time that the
black hole encounters a nucleus within \rem, this charge is insufficient to
prevent absorption, but with sufficient charge buildup repulsion could
become an important effect.  A positively-charged black hole will also have an
enhanced absorption rate for electrons, which works toward
neutralization.  So, while charge effects could possibly somewhat slow
the absorption rate, we will make the conservative assumption that
they don't, and that sufficient neutralization is automatic.

To determine actual capture sizes, one needs the parameter $K$.  
There are different ways of estimating a typical $K$.  One approach is
to estimate the dipole force when one separates an ion in a crystal
from the electron cloud of the bonding orbitals, as suggested earlier
by eq.~(\ref{eq:ecloud}). Assuming there that all but the outermost
electrons move coherently with the nucleus, yields the value
\be \label{eq:Kalpha}
K\sim {\alpha\over \arad^3} \sim \frac{14\, \ev}{\mathrm{\AA}^2} \ ,
\ee
where $\arad$ is the atomic radius, $\sim 1$~\AA.  Another method is to use
the relation to the Debye frequency $\omega_D$, 
\be \label{eq:Komega}
{K\over m}=\frac{\omega_D^2}{\chi}\ ,
\ee
where $\chi$ is an ${\cal O}(1)$ constant that depends on the
material.  Corresponding Debye temperatures, $T_D=\omega_D$, 
fall in the range 300-600K for typical materials forming the Earth's
interior ($T_D^{\mathrm{Fe}}=460$K, $T_D^{\mathrm{Si}}=625$K, $T_D^{\mathrm{Mg}}=320$K). 
In this case, one finds a typical $K$ of size:
\be
K =  \frac{12\, \ev}{\chi\mathrm{\AA}^2} \; \left( \frac{m}{40\gev}
\right) \; 
\left( \frac{T_D}{400\mathrm{K}} \right)^2 =1.20\times 10^{-27}  {m\over \chi}  \left( \frac{T_D}{400\mathrm{K}} \right)^2M_0^2 \; ,
\ee
consistent with (\ref{eq:Kalpha}).   Here we have introduced the \tev\ mass scale,
\be\label{tevscale}
M_0 = 1\tev\ ,
\ee
which frequently provides a useful normalization scale.
In Earth one also has semi-solid or semi-fluid layers, but these have characteristic values of $K/m$, given by pairing potentials, of similar size to those of solids.

Since $K/m$ is directly related to $T_D$, we will in fact parameterize results
in terms of this temperature.  For example,
\be\label{betafour}
 \beta_4 =  \chi {27M_4^2\over 32 \pi T_D^2} =   1.30\chi\times 10^{57}
 \left({400K\over T_D}\right)^2 \ . 
 \ee

With this prelude, we are now prepared to discuss  the subatomic
phase of  accretion.

\subsubsection{Subatomic growth laws}
\label{sec:atomic}

As we have described, in principle there can be three regimes depending on the size of the capture radius relative to nuclear and atomic scales.  The capture radius in atomic matter is given by $\rem$; let us estimate this for the minimum size black hole, with $M\sim M_D$.  For $D=11$, we find from (\ref{emrad}), (\ref{betadef}), and (\ref{eq:Komega}) that 
\be
\rem(M=M_D, D=11)\sim 3\chi^{1/10} \times 10^{-14} \cm\ ,
\ee
with larger values for smaller $D$.  Thus the subnuclear growth phase is nearly negligible, and we will (conservatively) set the corresponding time to zero.

We therefore turn directly to evolution from $\rem\sim r_N$ up to the atomic radius, $\arad\sim 1$~\AA.  Combining the general evolution equation (\ref{eq:defevol}) with the expression (\ref{emrad}) for the electromagnetic radius, the growth law for a black hole moving with velocity $v$ takes the form
\be\label{evolaw}
{dM\over dt} = {\pi \rho v} \rem^2\ .
\ee
This expression integrates to give a distance
\be \label{distom}
d = d_0 \left({M_D\over M_0}\right)^3 {D-1 \over (D-3)\beta_D} \left({\beta_D M\over
  M_D}\right)^{(D-3)/(D-1)}  
\ee 
for growth to a mass $M$, where we introduce
the characteristic distance, given via \tev\ units (\ref{tevscale})
\be\label{dodef}
d_0={M_0^3\over \pi\rho}\; .
\ee
Using the average density for Earth, $\rho_E=5.5~\gr/\cm^3$, one finds
$d_0= 3\times 10^{11}\cm=9s$,  which is much bigger than the Earth's
radius. The distance (\ref{distom}) is governed by the upper limit of the
mass, and the lower limit has thus been dropped. 
The expression
(\ref{distom}) can also be written in terms of the final $\rem$,
using (\ref{emrad}), as  
\be\label{distR}
d=d_0 \left({M_D\over M_0}\right)^{D-2}\left({D-2\over D-1}\right)^{D-2} {1\over (D-3)\ktd} {K\over m M_0^2} (\rem M_0)^{D-3}\ .
\ee

From eqs.~(\ref{radii})--(\ref{radiii}) we see that this evolution
applies to all values of $R_{EM}<\arad$ if $D\le 7$. For $D\geq8$, instead,
the radius $R_D$ of the extra dimensions is smaller than $\arad$, and
therefore as $\rem$ grows larger than $R_D$ the dimension governing
the force law changes, and one should then set $D=4$ in eq.~(\ref{distR}).
We therefore must treat these cases separately.

In performing the following estimates, using the formula
(\ref{distR}), we will assume that the black hole uniformly travels at
the escape velocity $v_E$.  Were its velocity higher, it would not be
gravitationally bound to Earth.  This is clearly a conservative
assumption, as the black hole will slow down as it accretes; if needed
one could model this slowdown by integrating the evolution of the mass
and potential energy over the lifetime of the black hole.  (We expect
that the bounds of this paper might be tightened by  a  
more complete treatment of this slowdown.)
We will also assume a uniform density $\rho_E$ for Earth.  
Of course, if the black hole spent most of its time near the center of the
Earth, one should use a higher central density. But, correspondingly,
the black hole velocity would  be lower, scaling linearly with the
distance from the Earth's center. Since the Earth's density in the
deep core is at most a factor of 2--3 higher than its average value,
the estimates using the average density and the escape
velocity should tend to (conservatively) overestimate the accretion rate.

 \subsubsection{Timescales for $r_N \lsim$\rem$\lsim\arad$
     in $D=6,7$ }
\label{sec:rntoa}
The relevant time scales for growth to atomic sizes for $D=6,7$ can be found by setting $\rem\sim 1$\AA\ in 
in eq.~(\ref{distR}), assuming constant velocity $v_E$, and substituting the values of the other parameters.
This results in the following timescales:
\ba 
t &\sim& 4.5 \times 10^{3} {1\over \chi}
\left({T_D\over 400K}\right)^2 \left({M_D\over M_0}\right)^4~s\quad ,
\quad D=6 \\ 
t &\sim& 3.0 \times 10^{11} {1\over \chi}
\left({T_D\over 400K}\right)^2 \left({M_D\over M_0}\right)^5~s \quad ,
\quad D=7 \; . 
\ea
These times are quite short, compared to geologic time scales, and
this phase for $D=6,7$ 
will therefore be regarded as negligible.
 
\subsubsection{Timescales for $r_N \lsim$\rem$\lsim R_D$ in $D\geq 8$}
\label{sec:rntord}

For $D\geq 8$, the black hole would first evolve from $\rem\sim r_N$ up to $\rem\sim R_D$ via the evolution law (\ref{evolaw}).  The relative timescales can be
obtained by equating $\rem$ in eq.~(\ref{distR}) with the expression
of $R_D$ given by eq.~(\ref{rddef}), and assuming the uniform velocity $v_E$, resulting in: 
\ba 
t &\sim& 5.4 \times 10^{6} {1\over \chi}
\left({T_D\over 400K}\right)^2\left({M_0\over M_D}\right)^{3/2} ~\yr \quad , \quad D=8 \\
t &\sim& 2.0 \times 10^{4} {1\over \chi}
\left({T_D\over 400K}\right)^2\left({M_0\over M_D}\right)^{7/5} ~\yr \quad , \quad D=9 \; \\
t &\sim& 2.2 \times 10^{2} {1\over \chi}
\left({T_D\over 400K}\right)^2 \left({M_0\over M_D}\right)^{4/3}~\yr \quad , \quad D=10 \; \\
t &\sim& 4.8 {1\over \chi}
\left({T_D\over 400K}\right)^2 \left({M_0\over M_D}\right)^{9/7}~\yr \quad , \quad D=11 \; .
\ea
The corresponding times are short compared to {\it e.g.} the solar lifetime, and become shorter
in higher dimensions, since the values of $R_D$ in these cases become
smaller and smaller, approaching $r_N$ and reducing the available 
evolution range. As we show next,
this however means that there will be more range for
the 4-dimensional evolution, which is typically slower because of the
weaker gravitational coupling in 4 dimensions. 

\subsubsection{Timescales for $R_D \lsim$\rem$\lsim \arad$ in $D\geq 8$}
\label{sec:rdtoa}
Once $\rem$ reaches ${\cal O}(R_D)$, the distance at which the black hole's
gravity competes with electromagnetic binding forces is in the region
where the black hole's field transitions to the lower dimensional
form. For concreteness, let us first neglect warping and assume that
all radii are the same so that this is a transition to the
four-dimensional regime.  One reaches this regime at the crossover
radius given by (\ref{fcross}).

The distance required to reach a given $\rem>R_C$ is then, from
the distance (\ref{distR}) and using the formula (\ref{betafour}) for
$\beta_4$  
\be\label{fddist}
d= \frac{32 \pi \, d_0}{9\chi} \, \left({M_4\over M_0}\right)^2
\left({T_D\over M_0}\right)^2(\rem M_0)\ . 
\ee
We are  interested in the corresponding time scale to reach
$\rem=1$~\AA, where evolution begins to cross over to ``macroscopic."
Introducing the numerical values for our parameters, 
along with the escape velocity $v_E$ and
 taking $\rem=1$~\AA, we then find a time  of the order of hundred
 billion years
\be\label{fdtime} t = 9.9 \times 10^{18} s {1\over \chi}
\left({T_D\over 400K}\right)^2 = 3.1 \times 10^{11} {1\over \chi}
\left({T_D\over 400K}\right)^2 \yr\ \; ,
\ee
for evolution to the ``macroscopic" crossover.

In the case of $D=8$, $R_C$ is close to 1\AA. One may therefore fear
that, should the effective $R_C$ be an underestimate by a factor of
2-3, there will be no room for this phase of evolution.  However,
notice that, aside from the factor $(D-1)/(D-3)$, the distance
(\ref{distom}) is the characteristic distance for one e-fold growth of
the mass. Since the higher-dimensional evolution law must match onto
the four-dimensional one in this region (as can be seen explicitly),
and since the growth of the radius over a single e-fold in the mass is
small ($e^{1/(D-1)}$, or $e^{1/3}$ in four dimensions), the e-fold
time for $\rem\sim1$~\AA\ sets a lower bound on the
evolution time.

\subsection{Macroscopic accretion, \rem$\gsim \arad$}

\subsubsection{Bondi accretion basics}

Once the electromagnetic capture radius of the black hole, $R_{EM}$,
grows beyond the atomic radius $\arad$ , accretion becomes a macroscopic
process, with multiple atoms falling in, and with
the gravitational range of the black hole exceeding the mean free
path.  In this regime, two effects counter the free fall of matter:
the cohesion forces that keep atoms together, and matter's finite
compressibility. We shall not attempt to provide a simple model to
describe the effect of cohesion forces on a macroscopic scale, due to
the varied composition of matter inside the Earth, with crystalline,
semisolid, and liquid phases at various depths.  Moreover, it is possible that 
once the
accretion rate reaches a certain threshold, radiation emitted from the
accreting matter can melt the surrounding material.  (This reradiation
effect is discussed in Appendix~\ref{app:rerad}, where it is found not
to be important until $\rem\gg 1$~\AA.)  Since our aim is to be
conservative and consider the fastest conceivable evolution, we shall
neglect the slow down due to cohesion forces, and treat the inside of
the Earth as a non-viscous fluid, free to fall into the black hole,
subject only to the general laws of hydrodynamics, such as the
continuity equation and energy conservation. The compressibility of
the medium, which limits the amount of matter that can be funneled
towards the black hole, is accounted for by macroscopic hydrodynamic
properties of the medium, such as its sound speed.

The description of accretion under these conditions was developed by
Bondi, Hoyle and Lyttleton\cite{BHL}.  We review the derivation of the
resulting evolution equation and extend it to incorporate the
$D$-dimensional force law in Appendix~\ref{app:bondi}; it is 
\be\label{bondires} 
{dM\over dt} = {\pi\lambda_Dc_s R_B^2 \rho} \ ,
\ee 
where we define the {\it Bondi radius} in terms of the black hole radius $R$,
\be\label{BondiR}
\rson=\left[{(D-3)\over 4 c_s^2}\right]^{1/(D-3)} R\ ,
\ee 
 $\lambda_D$ is  a  numerical constant depending on $D$ and  on the
polytropic index $\Gamma$, given in (\ref{lambdaDdef}), which can range between $3<\lambda_D<18$, and $\rho$ and $c_s$ are the density and
sound speed within the matter asymptotically far from the black hole.

\subsubsection{Matching microscopic and macroscopic regimes}

Before estimating corresponding time scales, let us first compare the
sub- and super-atomic regimes of growth at the transition point,
$R_{EM}\sim \arad$.

Using eq.~(\ref{eq:Komega}) and the relation
between Debye frequency and sound velocity $c_s\simeq\omega_D \arad=T_D \arad$,\footnote{Using $T_D=400$~K
  and $\arad=1$~\AA, we obtain $c_s\simeq T_D \, \arad \sim 5.2$~km~\isec,
  which is consistent with the sound velocities of typical Earth
  materials. E.g.  $c_s^{Fe} 
  \sim
5$km~\isec\ at atmospheric pressure; sound velocities in the liquid forms of a metal are just 20-30\%
smaller.} 
we can rewrite the
expression for \rem\ given in eq.~(\ref{emrad}) as
\be
R_{EM}\simeq  \arad\Delta \; \left({M\over M_{a,D}}\right)^{1/(D-1)}  \; ,
\ee
where
\ba
&&\Delta = \frac{D-1}{D-2} \left[
  (D-2)(D-3)\frac{\chi}{2}\right]^{1/(D-1)} \; , \\
&&\label{madef}M_{a,D} = { {c_s^2} (\arad \, M_D)^{D-3} \over k_D}\,{M_D} \; .
\ea

From equations (\ref{BondiR}) and
(\ref{schwarz}), we also find:
\be
\rson = \arad \; \left[{D-3\over 4}\right]^{1/(D-3)}\left({M\over M_{a,D}}\right)^{1/(D-3)} \; .
\ee
The evolution equations for the two regimes can therefore be equivalently rewritten as:
\ba
&&\left( \frac{dM}{dt} \right)_{EM} = 
\Delta^2 \, \pi \, \rho \, v_{EM} \; \arad^2 \, \left( \frac{M}{M_{a,D}} \right) ^{2/(D-1)}  
\\
&&\label{Bondievol}\left( \frac{dM}{dt} \right)_{B}  = \lambda_D\left[{D-3\over 4}\right]^{2/(D-3)} \, \pi
\,  \rho \, c_s \; \arad^2 \, \left( \frac{M}{M_{a,D}} \right) ^{2/(D-3)} \;.
\ea
As before, we use $v_{EM}\sim v_E$; notice that $c_s$ inside the Earth
has a comparable value.  
So, up to an overall factor of order 1, $M_{a,D}$ turns out to
be the mass value at which the two evolution rates are the
same, and the capture radii for the subatomic and the Bondi accretion
regimes coincide.  The subatomic growth is faster when $M<M_{a,D}$, while
Bondi's growth is faster when $M>M_{a,D}$.
 This means that for the purpose of being
conservative, it is justified to use the former accretion model below
$M_{a,D}$, and the latter above $M_{a,D}$.

\subsubsection{Time evolution with Bondi accretion}

We can split the numerical analysis for the Bondi 
accretion into the case of
$R_D<\arad$ ($D\geq8$), where  all the evolution for $\rem>\arad$ is
four-dimensional, and the case with $R_D>\arad$ ($D\leq7$), where we need to consider
both phases.

The $D$-dimensional Bondi evolution equation (\ref{bondires}) is
straightforward to integrate.  Since we are interested in times for evolution to given radii, it is most useful to convert it to an equation for the Bondi radius $R_B$, using (\ref{BondiR}).  This gives the following times, for evolution from an initial Bondi radius $R_{B,i}$ to a final Bondi radius $R_B$:
\ba
\label{Dtime}
t &=& d_0 \, {4 c_s\over (D-5) \, \lambda_Dk_D}
\left({M_D\over M_0}\right)^{D-2}  (M_0R_B)^{D-5} \quad ,\quad D>5  
\\
\label{Ftime}
t &=& d_0 \, {4c_s\over \lambda_5 \, k_5}
\left({M_5\over M_0}\right)^3\ln(R_B/R_{B,i})\quad , \quad D=5
\\ 
\label{FDevol}
t &=& d_0 {4c_s \over \lambda_4 k_4} 
\left({M_4\over M_0}\right)^2\left( {1\over
  M_0R_{B,i}} - {1 \over M_0R_B}\right)\quad ,\quad D=4 
\ea
Recall that  $d_0$ was defined in (\ref{dodef}).
A transition from the $D$- to 4-dimensional Bondi behaviour will
occur when $\rson$ is in the range of $R_{D,C}$. It is easy to check
that this transition is continuous, namely the values of the $D$- and
4-dimensional Bondi radii coincide, when $R_B=R_C$.
At this radius, the mass is given by 
\be \label{eq:mbd24}
M_B = {16\pi }  c_s^2 \left[4\pi\kd(D-3)\right]^{1/(D-4)}  \left(
\frac{M^2_4}{M^2_D} 
  \right)^{(D-3)/(D-4)}  M_D \; .
\ee
 The evolution time to the slightly smaller radius $R_B=R_D$ is found
 from (\ref{Dtime}), 
\be\label{Bondi2rd}
t = d_0 c_s {4\over (D-5) \lambda_Dk_D} \left({M_D\over
  M_0}\right)^{(D-2)/(D-4)} \left({M_4\over
  M_0}\right)^{2(D-5)/(D-4)}\ . 
\ee

\subsubsection{Macroscopic time scales: $D\geq8$}
\label{sec:macro}
We start with the case $R_D<\arad$, so that evolution for $R_B>a$ is
purely four-dimensional Bondi accretion.  When the black hole enters
the macroscopic regime, $R_B\approx a$, eqn.~(\ref{BondiR}) implies
that its mass is of order $10^{11}~\gr$, therefore still small
in geologic terms. Thus by this time it should
have settled deep within the Earth.  The quantity $c_s d_0$, relevant
to the Bondi evolution formulae, can be estimated from the
approximately linear relation,  known as Birch's law~\cite{Birch}, 
between sound speed and
density. This has been tested experimentally for
Fe~\cite{Brown,Fiquet} up to the  
densities of $12~\gr/\cm^3$ found in the Earth's core, giving : 
\be\label{Birch}
d_0c_s\approx 1.33\times 10^{-4} s\ ,
\ee
which is then density independent.  (One also finds values comparable
to this using, {\it e.g.}, sound speeds and densities for materials
such as iron at low pressure.) 
 Eq.~(\ref{FDevol}) then  gives the following time to double the
radius from $R_B\approx a$ to  $R_B\approx 2a$ (and the
mass from $M\approx 10^{11}$~gr to $M\approx 2\times 10^{11}$~gr):
\be \label{eq:tjm}
t = \frac{8 \pi}{\lambda_4} 
d_0 \; c_s \, \frac{1}{\arad M_0} \, \frac{M_4^2}{M_0^2}
 \sim 1.2 \; \times 10^{12} \; \frac{1}{\lambda_4}
  \; \yr \ .
\ee
As shown in Appendix~\ref{app:bondi}, the value of
 $\lambda_4$ is in the range 4-18; for $\Gamma=5/3$ (namely the adiabatic index of a
 non-relativistic electron gas), $\lambda_4=4$. 
Notice that the
parameter dependence  of (\ref{eq:tjm}) is identical to that of
eq.~(\ref{fddist}), once we use $c_s\simeq T_D a$. This is also
reflected in the similarity of the 
timescales, eqs.~(\ref{fdtime}) and (\ref{eq:tjm}). 

  As a side note, we remind the reader of discussions of the
  possibility that primordial black holes remain from the early
  universe; in the standard quantum scenario only those with masses
  $\gsim10^{15}\gr$ would have not yet evaporated.  With the present
  formalism, we can provide a bound on the lifetime of Earth, should a
  minimum-mass primordial black hole be captured within its
  gravitational field.  Using the parameters in this section, and the
  evolution law (\ref{FDevol}), we find a bound on the accretion time
  $t\gsim 47$~Myr, with shorter times for higher-mass black holes.
  These, and corresponding accretion times we will find for white
  dwarfs and neutron stars, may allow one to set limits on galactic
  densities of primordial black holes.

\subsubsection{Macroscopic time scales, $D=6,7$}
\label{sec:bondi67}
In the cases $D=6,7$, $R_D>a$ and so we have $D$-dimensional Bondi
evolution up to $R_B\sim R_D$, and then four-dimensional Bondi
evolution from $R_B=R_C$ up to infinity.  To be conservative, we model
the phase with $R_D<R_B<R_C$ by assuming $D$-dimensional evolution
with a constant Bondi radius, with
$R_B=R_C$, until the black hole mass grows to the point that the
respective $R_B$ exceeds $R_C$.  For $D=7$ the mass at $R_B\approx  a$ is of order $10^4\gr$; for $D=6$ it is much smaller.  For our approximate estimates, we again use the value of $c_sd_0$ given in (\ref{Birch}).

The times for the evolution up to $R_D$ are thus given by
(\ref{Bondi2rd}), leading to:
\ba
\label{bondisix_toRC}
t &=& 5.5\times10^4 \frac{1}{\lambda_6}    
		     \left(\frac{M_D}{M_0} \right)^{2}
		     \yr \quad D=6\\
\label{bondiseven_toRC}
t &=& 8.6\times 10^8 \frac{1}{\lambda_7}  
\left(\frac{M_D}{M_0} \right)^{5/3}
\yr \quad D=7\ .
\ea

The following phase, between $R_D$ and $R_C$, then has a time scale given by:
\be\label{Fbondiev}
t = d_0 c_s {16\pi\over [4\pi(D-3)k_D]^{1/(D-4)}\lambda_D}
\left({M_D\over M_0}\right)^{(D-2)/(D-4)} \left({M_4\over
  M_0}\right)^{2(D-5)/(D-4)}\ , 
\ee
The subsequent evolution to large sizes has a time scale determined by
the initial radius, $R_B=R_C$ in~(\ref{FDevol}), and results in an
expression identical to (\ref{Fbondiev}), with $\lambda_D$ replaced by
$\lambda_4$. The timescales for these two phases are given by: 
\ba
\label{bondisix}
t &=& 9.7\times10^4 \frac{1}{\lambda_{6,4}}
\left(\frac{M_D}{M_0} \right)^{2}
\yr \quad D=6\\
\label{bondiseven}
t &=& 1.28\times 10^{10} \frac{1}{\lambda_{7,4}}
\left(\frac{M_D}{M_0} \right)^{5/3}
\yr \quad D=7\ .
\ea
The $D=6$ time is short as compared to geologic time scales. Using
 $\lambda_7=4$ (independent of $\Gamma$) and $\lambda_4=4$
 (for $\Gamma=5/3$) we
 obtain in $D=7$ a combined time scale of approximately $(6.4, 20, 40,65,94)$ billion years for $M_D=1,\ldots,5\, \tev$.
 Since, as we discuss in Appendix~\ref{app:prod}, a conservative
 threshold for black hole formation is $M_{min}=3M_D$, these values
 correspond to minimum black hole masses
 $M_{min}=(3,6,9,12,15)$~\tev .

\subsection{Warped evolution}
In order to parameterize more general evolutions, we consider the case
of a warped scenario, as described in section \ref{sec:bhwarp}.  In
this case, we have $D$-dimensional evolution up to capture radius
$R_D$, then a warped evolution up to $R_C$, then four-dimensional
evolution from then on.  These scales are related by an expression of 
the form (\ref{capprox}).  Combining this with the maximum value of
the warping, found from (\ref{scalerat}) by taking $R_D$ to be the
minimum possible value, $M_D^{-1}$, we find that  
\be\label{rcrdrat}
R_C/R_D\lsim 10^2\ ,
\ee
so there is still not wide disparity between these scales.

As the preceding discussion has illustrated, a key question is the
location of $R_C$ with respect to the atomic scale $a$.  Macroscopic
evolution in   
the four-dimensional regime 
is dominated by the timescale from the lower endpoint  at
$R_B=R_C$, and if $R_C\lsim 200$\AA, this, 
via (\ref{FDevol}), yields a safe time scale in
 excess of  $ 3\times 10^9\yr$.  The warped growth below $R_C$ should
also yield a similar time scale, as in the preceding discussion.

On the other hand, accelerated growth is possible for $R_C$
significantly larger than $200$\AA.  We particularly saw this in the case
$D=6$.  Warped evolution presents another extreme (but
finely-tuned) scenario.  Specifically, consider the case $D=5$ with
radius just below the experimental bound, say $R_C\approx 0.2$mm.  In
this case, one finds five-dimensional evolution through subnuclear,
subatomic, and Bondi phases.  If, in line with our discussion of
scales, we take $R_D\approx 0.02$mm, the evolution time up to this
radius follows from (\ref{Ftime}).  This yields an estimate
$t_{B,5}\approx 5\times 10^{-3}s$.  At this point, the black hole
has a mass $M_{B,5}$ around $0.1\gr$.  Next, one evolves through
the warped regime.  The precise form of the evolution in this regime
is not completely understood.  The form of the linearized
potential\cite{Froiss} suggests that the radius grows as the logarithm
of the energy, but ref.~\cite{Froiss} also points out that the
corresponding solutions are possibly unstable and instead spread out
more widely on the visible brane.  Ref.~\cite{AMW} worked out details
of a possible picture of the resulting solution, which would be the
gravitational dual of a plasma ball of QCD.  Whatever the precise
evolution law is, it is quite slow, since at the end of this phase the
mass is of size $M_{B,4}\approx2\times10^{17}\gr$ while the radius
has only changed from $R_D$ to $R_C$.  Conservatively, one can take
the fastest time scale for evolution, namely that with a {\it
constant} capture radius $R_B\approx R_C$.  Finally, the time for
the four-dimensional evolution from $R_B=R_C$ to the mass of the Earth
has time scale given by (\ref{FDevol}). Both of these phases yield
time scales
\be
t \sim 3\times 10^5 \;\yr\ .
\ee
  While this time scale as well as that of the $D=6$ case only
  represent {\it lower} bounds on the accretion time, since we have
  made various conservative assumptions such as that of accretion from
  a fluid, these times are too short to
  provide comfortable constraints.  

\subsection{An Eddington limit?}

If evolution is sufficiently rapid, as is particularly exhibited by
the $D=6$ case and the extreme $D=5$ warped case, and in four
dimensions, one is naturally led to question whether there is an
Eddington limit.  This would occur if radiation from the rapidly
accreting matter produced sufficient pressure to inhibit accretion.

In particular, one can parametrize the total luminosity of the outgoing radiation in terms of the rate of mass accretion by an efficiency parameter $\eta$,
\be
L= \eta {dM\over dt}\ .
\ee
If this radiation is in photons, it will exert a force on the infalling atomic matter of Earth which can be approximated as given by the Thomson cross section $\sigma$:
\be
F_L = {\eta \dot M \sigma\over 4\pi r^2}\ .
\ee
The question is whether this can balance the force of gravity pulling the matter inwards.

We examine this question in more detail in Appendix~\ref{app:rerad}.
In short, we do not find evidence for such an Eddington limit; this is
connected to the known result (see e.g.~\cite{FrKiRa})
that spherical accretion onto a black
hole is inefficient at producing a large luminosity.\footnote{In
  certain low-collisionality astrophysical contexts, different from
  conditions dealt with here, magnetic fields
  alter this story; see {\it e.g.}~\cite{Sharma:2008pk}.} 
Nonetheless, we note that if a mechanism to produce such an Eddington
limit {\it were} found, this would lengthen the shorter accretion time
scales considerably.  In particularly, as we review in Appendix
\ref{app:rerad}, four-dimensional Eddington evolution produces
exponential growth of the mass with time, with time constant
\be
t_{Edd}= \eta {\sigma\over 4\pi m G}\ ,
\ee
where $m$ is the average atomic mass per electron, $m\approx 2m_p$.  
This produces an e-fold time scale of size $t_{Edd} \approx 2.3 \eta
\times 10^8 \yr$.

\subsection{Summary of growth on Earth}
This section has modeled growth laws on Earth, using conservative
assumptions.   

The resulting growth times for $D\geq8$ are bounded below by the
many-billion-year time scales given in eqs.~(\ref{fdtime}),
(\ref{eq:tjm}); these arise from the subatomic and superatomic
regimes, respectively, and are times that are long as compared to the
expected natural lifetime of the Sun. At these times, the black
holes masses are still in the range of $10^{11}$~\gr, and have a growth
rate of the order of 300~kW, which gives a bound on possible power output that is totally negligible in the geologic context.

The case of $D=7$ also gives times longer than the Sun's lifetime,
with an overall timescale for the evolution,
eq.~(\ref{bondiseven}), ranging from 6 to over 80 billion years. At such time scales, the growth rate
reaches approximately 10 GW, about $10^{-7}$ the solar flux on Earth, 
and much smaller than the $\sim40$ TW heat flow from the interior of
the planet; this bounds any possible thermal impact to be negligible. 
The
case $D=6$, with time scale (\ref{bondisix}), even though very long
by human standards, is much shorter than the natural lifetime of the
solar system.  As we saw, certain warped scenarios are also similarly
potentially problematic.  Therefore, in order to constrain these
scenarios we turn to their consequences for other astronomical
bodies, particularly white dwarfs and neutron stars.

\section{Stopping of cosmic ray-produced black holes}
\label{app:cr}
Collisions with center-of-mass energies comparable in energy to LHC occur frequently in the
universe. The best known and directly measured process is the
collision of high-energy cosmic rays (CRs) with the nucleons in the
Earth's atmosphere. For the collision of a CR of mass $Am_p$ to exceed
the nucleon-nucleon center of mass (CM) energy of $E_{LHC}=14$~TeV, the CR
energy should be at least $E_{min}(A)=A \, E_{LHC}^2/2m_p \sim
(10^{17}A) \ev$,  
well below the maximum value of
measured CR energies.  A simple estimate of  
the number of nucleon-nucleon interactions above
LHC energies can be obtained from an  approximate flux
relation, derived from the current   
data~\cite{Yamamoto:2007xj,Matthiae:2007zz,Abbasi:2007sv}: 
\be \label{eq:CRrates}
\frac{d\Phi}{dE} \sim 10^6 (E/\mathrm{GeV})^{-3} \; \mathrm{m}^{-2}\,
\mathrm{s}^{-1} \, \mathrm{sr}^{-1}
\,\mathrm{GeV}^{-1} \; ,
\ee
which provides a lower bound to the measured CR spectra in the
interesting region $E_{CR}>10^{17} eV$, up to the GZK
cutoff~\cite{Greisen:1966jv,Zatsepin:1966jv}  of
$E_{GZK} \sim 5\times 10^{19}$~eV. 
Confining ourselves to the part of the spectrum below the GZK cutoff,
we obtain the following integrated flux: 
\be
N(\sqrt{s}>E_{LHC}) = A\int_{E>E_{min}(A)} \; \frac{d\Phi}{dE} \, dE \sim
\frac{1.6\times 10^3}{A} \, \yr^{-1} \, \km^{-2} \, \mathrm{sr}^{-1} \; ,
\ee
where $\sqrt{s}$ is the CM energy of a nucleon-nucleon collision.
This corresponds to about $1/A \, \times 10^{22}$ collisions above the
LHC energy 
at the surface of the Earth during the course of its existence. This number
greatly exceeds the total number of collisions in the course of the LHC
operations at its highest intensity (about $10^9$~\isec\ over a $10^8$s
period), even assuming a cosmic ray flux dominated by Fe nuclei.

If black holes can be produced at the LHC, they will therefore be
copiously produced in such CR-induced collisions with astronomical
bodies.  Thus, stability of such bodies on astronomically long time
scales offers the prospect of ruling out rapid accretion scenarios.
In order to provide such bounds, one needs to check that a CR-produced
black hole will slow sufficiently to be trapped in such an object, so
that it begins accreting.  After that, one  needs to check the relevant
accretion time scale.  This section will focus on the former question.
We will find that while, as briefly described in section
\ref{sec:crEarth}, we cannot guarantee that Earth is an efficient
target for trapping hypothetical CR-produced black holes in all
scenarios, white dwarfs and neutron stars do provide very useful
targets.

\subsection{Production kinematics}
Let $E$ be the energy of a cosmic ray nucleon hitting an astronomical target. 
Black hole production would arise from collisions of two partons, with 
center-of-mass momentum fractions
$x_1$ and $x_2$ for incident parton and target parton, respectively.
The mass $M$  of the resulting black hole is given by
\be\label{prodM} M^2 \approx 2 \, x_1 \, x_2 y^2\, E \, m_p \ee
where $y\leq1$ is an efficiency factor, parametrizing inelasticity (energy loss) due to radiation in the collision process, and 
$m_p$ is the proton mass. The bulk of the production is at $x_1
\sim x_2 \sim x= M/(y\sqrt{2Em_p})$. After being produced, the black hole in the
fixed-target frame will carry energy $x_1y E \sim M \sqrt{E/2m_p}$, and thus has a Lorentz $\gamma$ factor 
\be \gamma \simeq \sqrt{x_1\over x_2}\sqrt{E\over  2m_p}.
\ee
Since $x_1,x_2,y\leq1$, eq.~(\ref{prodM}) gives
\be \label{eq:gamma-prod}
\gamma ={M\over 2x_2  m_py} > M/2m_p\ .
\ee
The black hole will therefore be highly relativistic.  These boosts
range up to typical values of size $\gamma\sim3M/m_p$; for example,
in the extreme case of interest for the LHC, $M=14\tev$,
fig.~\ref{fig:gamma} shows significant production at
$\gamma\lsim4.5\times 10^4$.

If the black hole is  very weakly interacting, it could travel
across the object unimpeded, thus preventing limits from being set. We
therefore concentrate first on this issue.

\subsection{Stopping: neutral black holes}
\label{app:bhtrapping}

In line with our discussion of section \ref{sec:instab}, we will
explore the assumption that black holes interact only via their
gravitational field.  Collisions can slow such a black hole via two
mechanisms.  First, in a typical collision the black hole will
gravitationally scatter a particle in its asymptotic field, thus
losing some of its momentum.  We 
henceforth refer to this gravitational mechanism
as Coulomb
slow-down.
Second, for smaller impact parameters, a black hole can absorb a
particle, in the process possibly emitting some radiation, and
typically increasing its mass.  This also slows the black hole, and
will be referred to as accretion slow-down. 
 In what follows we denote
by $\Eprod$ and $\pprod$ the energy and momentum of the black holes at
the time of production by the cosmic ray.

\subsubsection{Accretion slow-down}

For the smallest black holes, such as those in the early stage of
the accretion, the target particles should be thought
of as partons, moving relativistically inside the nucleon. 
In the black hole rest frame,
denoted by a prime,  incident partons have energy and momentum
\be
E_p'\approx p_p'\approx {E\over M} (E_p-p_p)\ ,
\ee
where $E_p$ and $p_p$ are the parton energy and parallel momentum component
in the star rest
frame. Note that for the first few collisions $E_p'\sim M$, since
$\Eprod/M\sim M/m_p\sim M/E_p$. But as the black
hole  slows,  $E_p'$ becomes much smaller than $M$.  

If black holes can capture matter via their gravitational fields and accrete within Earth, then their gravitational fields will similarly capture and scatter matter while moving through any astronomical object, and we will consistently work within such a framework.
Some general features of gravitational scattering and/or capture of
relativistic particles in the field of a $D$-dimensional black hole
are reviewed in Appendix~\ref{app:scattering}. There we find the
minimum impact parameter, $\bhm R$, below which relativistic
particles  enter the capture regime.   This can be calculated classically, or defined quantum-mechanically in terms of the capture cross section,
\be
\sigma_c = \pi (\bhm R)^2\ ,
\ee
with closely corresponding results in the cases of interest.  A capture collision should result in the parton transferring its momentum, and
is expected to result in much of the parton energy also being absorbed, although it is also likely that some of the parton energy will be radiated in the process.
The change in the black hole mass and momentum are thus parameterized as
\be\label{deltapm}
\Delta p' = \cac p_p'\quad ;\quad \Delta M \simeq\Delta E'=  \cacm E_p'\ .
\ee
We expect $\cac\simeq1$, and have $0<\cacm<{\rm Min}(1,\cac)$
parameterizing the fraction of energy absorbed.\footnote{Indeed, we
  expect that (chromo-)electrostatic effects significantly reduce
  $\cacm$, below an energy threshold $E_p'\sim \alpha_s/R$. Notice
  that any energy that is not absorbed must be re-rediated, thus
 contributing to black hole momentum loss. } 
Back in the star frame, the momentum  
change will then be
\be
\Delta p = \cac p_p - (\cac-\cacm){E^2\over M^2} (E_p-p_p)
\ee
 where  the terms proportional to $p_p$ contribute zero when averaged over different parton momenta.
One can combine these equations with the capture cross section to determine
the accretion contribution to the momentum and mass variation.  For
a black hole of velocity $v\sim 1$ in a matter distribution with
parton density $n$, one finds: 
\ba\label{pacevol}
\left({dp\over dt}\right)_{ac} &=& n \pi[\bhm R({\sqrt s})]^2 \Delta p
\;, \\
\label{Macevol}
\left({dM\over dt}\right)_{ac} &=& n \pi [\bhm R({\sqrt s})]^2 \Delta M
\;,
\ea
where
\be\label{sdef}
s=M^2+2(EE_p - pp_p) + m_p^2\ .
\ee
This leads to the following evolution as a function of path length $\ell$:
\be\label{acloss}
\left({dp\over d\ell}\right)_{ac} 
= - (\cac-\cacm)\bhm^2 \pi\rho{E^2\over M^2}   R^2(\sqrt s)\ ,
\ee
\be\label{mgain}
\left({dM\over d\ell}\right)_{ac} 
= \cacm \bhm^2 \pi\rho {E\over M} R^2(\sqrt s)\ ,
\ee
where we introduced  the energy density $\rho \simeq n\langle
E_p\rangle$.  We see that in the limit $\cacm=\cac$ (``perfect accretion"), there is no average momentum transfer.

\subsubsection{Coulomb slow-down}
Consider now the case of gravitational elastic collisions.  For the
 earliest collisions, $R\gg1/p_p'$, which is the regime of classical
 particle scattering.  Once $\gamma$ has decreased by a factor of
 approximately ten, the wavelength becomes longer than $R$ and we
 enter the quantum regime.

Thus, we consider incident partons outside the capture regime, classically described by
impact parameter $b>\bhm R$.
The momentum loss of
the black hole is
\be\label{elastd}
\Delta p =-2{E^2\over s}(E_p-p_p)\sin^2{\theta\over 2}
\ee
where 
$\theta$ is the CM scattering angle.  (For all but the first collisions, the CM frame is well-approximated by the black hole frame.)
One can then sum over collisions, as in the accretion case, to obtain
the differential momentum loss.  
In terms of the differential cross section, this takes the form 
\be
\left({dp\over d\ell}\right)_{sc} = -{E^2\over s} \rho
\int_{\cos\theta_c}^1 \, d\cos\theta {d\sigma\over d\cos\theta} \,
2\sin^2 {\theta\over 2}\ , 
\ee
where $\theta_c\sim 1$ represents the maximum angle avoiding capture.  
Here the parton momenta have averaged to zero.
In parallel with (\ref{acloss}), we write this as
\be\label{elloss}
\left({dp\over d\ell}\right)_{sc} = -\cel\bhm^2 \pi \rho {E^2\over s} R^2(\sqrt s)\ .
\ee
where
\be
\cel= {1\over \sigma_c} \int_{\cos\theta_c}^1 \,d\cos\theta
      {d\sigma\over d\cos\theta} \, 2\sin^2 {\theta\over 2}\ ; 
\ee
the corresponding classical expression is
\be\label{classcsc}
\cel={1\over\bhm^2} \int_{\bhm}^\infty d{\hat b}^2 \, 2\sin^2 {\theta\over 2}\ .
\ee

The parameter $\cel$ is estimated in Appendix~\ref{app:scattering}, yielding for the quantum case
the values (0.5, 0.25, 0.17) for $D$=5-7. 
Note that as a result of $\bhm>1$, gravitational Coulomb scattering plays a subdominant role.

\subsubsection{Slow-down to $\gamma\sim 1$.}
We now combine the Coulomb and accretion stopping rates to determine
the length required to slow-down the black hole to the
non-relativistic regime.  
As noted, initially $EE_p\sim M^2$, but as the
energy falls, the mass term in (\ref{sdef}) dominates.  Then the
Coulomb and accretion stopping formulas, (\ref{elloss}) and
(\ref{acloss}), have the same form.  Moreover, dividing (\ref{mgain})
by their sum, and defining 
\be
c'={\cacm\over \cac-\cacm + \cel}\ ,
\ee
we find in this
regime
\be
{dM\over dp} = -c' {M\over E}\ ,
\ee
with solution
\be\label{MGrowth}
{M\over \Mprod} = \left({p\over \pprod}\right)^{-c'}\ ,
\ee
This, together with the dependence $R(M)=\Rprod(M/\Mprod)^{1/(D-3)}$, then
allows us to integrate the sum of (\ref{elloss}) and (\ref{acloss})
and deduce the distance $d$ to evolve to a given final momentum $p$ and mass $M$ related by (\ref{MGrowth}):
\be
 \int^d \rho d\ell \simeq {1\over  \cel+\cac+\cacm(D-5)/(D-3)} {1\over \bhm^2}
    {M\over\pi R^2(M)}{M\over p} \ . 
\ee
The left hand side defines the column density, $\delta(d)$, 
as a function of  $d$.

The momentum where the black hole becomes
near-relativistic, $p\approx M= \Mnr$, is obtained from
eqn.~(\ref{MGrowth}) as:
\be\label{Mnreq}
\Mnr=\Mprod \left({\pprod\over \Mprod}\right)^{\cacm/(\cac+\cel)}\ .
\ee
Since, as we have seen, the initial boost $\gamma_i=p_i/M_i$ is large, growth of the mass in this phase can be significant.
The corresponding column density is
\be\label{stopdist}
\delta_R(d) \simeq
 {1\over  \cel+\cac+\cacm(D-5)/(D-3)}{1\over \bhm^2}
{\Mnr\over \pi  R^2(\Mnr)} \ .
\ee
For rough benchmarks we can replace the right hand side by
$M_0^3/\pi$.  If we work with constant density, we see that the
stopping distance is then approximated by $d\sim d_0= {M_0^3/\pi \rho}$, as
defined in (\ref{dodef}). 

In view of the value for Earth $d_0(E)\approx 3\times 10^{11}\,\cm$,
these mechanisms cannot efficiently slow down neutral CR-produced black holes in 
Earth, or in other bodies such as planets and ordinary
stars\footnote{As a consequence of this, neutral black holes produced during
  head-on 
  collisions of cosmic rays within the galaxy will freely escape the
  galaxy, not being trapped by either collisions with the interstellar
  medium and stars, or by the galactic
  magnetic field. Therefore arguments such as those used in
  Ref.~\cite{Dar:1999ac} to rule out the production of strangelets
  do not seem to easily apply in this context.}.  For
the same reason, typical black holes produced at the LHC are expected
not to be captured by the Earth (see Appendix~\ref{app:boundBH}), posing no risk;
however, there is small but finite probability for them to be produced
with velocities small enough to become gravitationally bound to the
Earth and, in the hypothetical case of stability, to begin accreting.

On the other hand, for a neutron star with densities surpassing $10^{14} \gr/\cm^3$, one has $d_0(NS)\lsim 0.01\, \cm$.
Thus neutron stars can promptly slow down such black holes, and then
quickly bring them to below the escape velocity, which for a neutron
star is close to $v\sim 1$. Finally, for
white dwarfs, whose central density can exceed $\rho= 10^7
\gr/\cm^3$, one finds $d_0(WD)\sim 1.5\, \km$, compared to radii in the
$10^3-10^4\, \km$ range.  Thus, in order to establish stopping in
white dwarfs, we need to make a complete numerical analysis,
considering  also the nonrelativistic phase of the slow down. 

\subsection{Stopping in white dwarfs}
\label{sec:WDstop}
Stopping scales in white dwarfs are not enormously far from their radii,
motivating a more complete treatment. We begin by noting that once a
black hole reaches the near-relativistic regime, $p\sim M$, it must be
further slowed to below the escape velocity in order to be trapped.
Thus we must understand nonrelativistic slowing.

We consider white dwarfs with masses $M\approx \Msun$, which have
radii $R_{WD}\approx 5500\km$.  Such a white dwarf has an escape
velocity $v_{WD}\sim 2 \times 10^{-2}$.  
The material in such white dwarfs is
described as a fluid of degenerate electrons, in which are embedded
atomic nuclei, for example carbon and oxygen.  (Later, at time scales
$\sim$ Gyr, this material can crystallize.)

\subsubsection{Nonrelativistic stopping}
\label{sec:NRstop}
At the end of the relativistic regime, the black hole has a mass
$\Mnr\approx p$ given by (\ref{Mnreq}).  
In addition to $R$, capture dynamics can in principle be governed by
the capture radius for free particles, $R/v$, or $\rem$, defined using
white dwarf parameters, $a_{WD}\sim 10^{-10}\cm$.  However,  
in the regime $v\gsim.02$, these radii are all less than or of order $r_N$ (see
Appendix~\ref{app:accretion}),  so the capture is dominantly subnuclear.  

As the black hole moves through the stellar material, it collides with nucleons at a rate
\be\label{nuccolrat}
{dn\over dl} = \pi {\rho\over m_p} r_N^2\ .
\ee
When a collision occurs, it moves non-relativistically through the nucleon.  As it does so, it is bombarded by the relativistic partons within the nucleon.  The average mass and momentum collected during such a transit is given using (\ref{pacevol}),  (\ref{Macevol}), times the average time $\Delta t = 4r_n/3v$ of the transit; combining with (\ref{nuccolrat}) gives
\ba
{dp\over d\ell} &=& -(\cac-\cacm) \bhm^2 \pi\rho R^2\\
\label{Mnrevol}
{dM\over d\ell}&=& \cacm \bhm^2 \pi\rho {R^2\over v}\ .
\ea
Note two possible caveats to these formulas.  First, in a collision with
a single nucleon, $\Delta M$ cannot be bigger than $m_p$.  We show
that this is true in Appendix~\ref{app:accretion}. 
Second, there could be enhancements of $\Delta M$ due to the
fact
that if the black hole captures a parton, the QCD string can pull in
more energy, whether or not it breaks.  (Moreover, since most of the
nucleons are in nuclei, it may even be that it pulls in more
of the nucleus.)  Indeed, one might expect a minimum energy captured
of $\sim 100$~MeV if the BH captures one parton.  Let us, however,
stick with this simple and conservative estimate. 

Coulomb stopping may be suppressed in this regime, and we conservatively neglect it.  As before, one can find an equation relating $M$ and $p$:
\be
\left({M\over M_{NR}}\right)^{1-\cacm/\cac} = \left({p\over M_{NR}}\right)^{-\cacm/\cac}\ .
\ee
Thus, for ``perfect accretion," $p$ remains a constant.
We also find
\be
v=\left({M\over M_{NR}}\right)^{-\cac/\cacm}\ .
\ee

One can then integrate (\ref{Mnrevol}) to find the column density to a given final mass $M_f$:
\be
\delta_{NR}(d) = {1\over \cacm \bhm^2} \int_{\Mnr}^{M_f} {dM\over \pi R^2} \left({\Mnr\over M}\right)^{\cac/\cacm}\ .
\ee
This can be evaluated using the scaling (\ref{schwarz}) of $R$ with $M$, to find
the distance travelled to reach mass $M_f$: 
\be  \label{eq:NRstop}
\delta_{NR} (d)={1\over \cac-\cacm(D-5)/(D-3)}{1\over \bhm^2}
 \left\{1-\left({\Mnr\over M_f
 }\right)^{[{\cac}/{\cacm}-(D-5)/(D-3)]}   \right\}
	       {\Mnr\over 
  \pi R^2(\Mnr)}  \; ,
\ee
an evolution law governed by larger velocities.  Note that this gives
the same scale as the evolution to $\Mnr$, as in (\ref{stopdist}).  This
is of course a {\it conservative} scale, since we have completely
neglected any momentum loss due to scattering, and also have neglected
possible enhancements due to binding effects with  nuclear
fragments.

\subsubsection{Stopping bounds}

From these equations we can compute
the column densities necessary for 
stopping. We first note that, by virtue of the fact that
$\gamma_i=p_i/M_i$ is large, the stopping distance grows with
increasing $\cacm/\cac$, due to the exponential dependence in
(\ref{Mnreq}).  Therefore, we set it to its maximum value,
$\cacm=\cac$.\footnote{We note that this appears quite conservative,
  in that we expect $\cacm$ to become small below the
  (chromo-)electrostatic threshold mentioned previously.  This in turn
  would significantly reduce the effective vale of $M/R^2$ that enters
  the expressions (\ref{stopdist}) and (\ref{eq:NRstop}) for the
  needed column densities.} 

Notice also that stopping distance increases with decreasing $\cel$.  Thus, one is tempted to set this to zero.  However, even small $\cel$ plays an important role.  Specifically,  consider the bound on the nonrelativistic stopping,
\be\label{NRbound}
\delta_{NR} (d)< {D-3\over 2\cac\bhm^2}{\Mnr\over
  \pi R^2(\Mnr)} \ .
  \ee
To test its sensitivity to our physical expectation $\cac\simeq1$, let
us see how much the column density for given $\cel$ changes if we take
$\cac=1/4$, as compared to its value for $\cac=1$ and $\cel=0$.  Note
from $\cacm\leq\cac$ that this would correspond to a reduction in the
mass accretion rate of $1/4$. 
From (\ref{NRbound}) we easily find less than 25\% variation in   the
resulting bound on $\delta_{NR}$ as long as 
\be
\cel>{1\over 4}{1\over (D-5)\ln \gamma_i/[(D-3)\ln(16/5)] -1}\ .
\ee
The tightest constraints on stopping parameters arise for largest $M_i$ and thus, as we see from (\ref{eq:gamma-prod}), large boost.  Using that equation, or alternately from fig.~\ref{fig:gamma}, we see that there is a large production efficiency for $\gamma_i \lsim 3 M_i/m_p$, or with $M_i=14\,\tev$,  around $\gamma_i\sim 4.5\times 10^4$.  Thus, we find the variation of the stopping distance bounded in this fashion so long as $\cel>(.12,.07)$ for $D=6,7$.  Moreover, this bound neglects the fact that any reduction of $\cac$, corresponding to a reduction of accretion, should lead to an {\it increase} of scattering, parametrized by $\cel$, thus improving the bound.  These features arise from the exponential dependence in  (\ref{Mnreq}); a similar statement is slightly stronger for $\delta_R$, as a consequence of its additional dependence on $\cel$ through its denominator.    Rough values of $\cel$ for $D=6,7$ are
given (see Appendix~\ref{app:scattering}) by (.25,.17).  Thus, in addition to the physical expectation
$\cac\simeq1$, we find the statement that even for small 
$\cel$, one does not increase the stopping distance by varying $\cac$ over a wide range.  
We thus take $\cel=0$ and $\cac=1$, and will rely on the resulting column densities to not be more than $25\%$ higher, although we expect that they could be significantly lower.

Taking these values, and combining our bounds from (\ref{stopdist})
and (\ref{NRbound}), we find 
\be\label{colestT}
\delta_T(d)= \delta_R(d)+\delta_{NR}(d) < {(D-3)^2\over 2(D-4)} {1\over  k_D^{2/(D-3)} \pi \bhm^2} \left({M_D\over
  M_0}\right)^3 \left({\gamma_i M_i\over M_D}\right)^{(D-5)/(D-3)} M_0^3\ .
\ee
The column density $M_0^3$ converts to $4.6\times 10^{12}\gr/\cm^2$.
For our numerical estimates
here we shall confine ourselves to black holes produced with 
$\gamma_i \lsim 3 M_i/m_p$, as above.
Since the evolution is
dominated by the phase where the wavelength of the incident particle
is large compared to the black hole radius, we use the values of the
parameters $\bhm$ corresponding to the quantum absorption, as given in
Appendix~\ref{app:quantum}. To maximize the needed column density, we also
use the maximum value of $M_D$ corresponding to a given mass $M_i$,
namely $M_D=M_i/3$.  Equation~(\ref{colestT}) then leads to the
maximum column densities shown in table~\ref{tab:WDstop} for various
black hole masses.

{\renewcommand{\arraystretch}{1.1}
\begin{table}
\begin{center}
\ccaption{*}{\label{tab:WDstop}
\it Column densities $\delta_{T}$, in units of $10^{15}$gr/cm$^2$,
 required to stop a black hole of given masses.}
\vskip 2mm
\begin{tabular}{l|llll}
\hline \hline
$\delta_{T}/10^{15}$gr/cm$^2$  &  $D=5$ & $D=6$ & $D=7$ & $D=8$  \\ \hline
$M=7$~TeV
 &     0.09 &     0.65 &     1.8 &     3.3\\
$M=8$~TeV
 &     0.13 &     1.0 &     2.9 &     5.3\\
$M=9$~TeV
 &     0.19 &     1.5 &     4.3 &     8.2\\
$M=10$~TeV
 &     0.25 &     2.1 &     6.2 &    11.9\\
$M=11$~TeV
 &     0.34 &     2.9 &     8.7 &    16.8\\
$M=12$~TeV
 &     0.44 &     3.9 &    11.8 &    23.0\\
$M=13$~TeV
 &     0.56 &     5.1 &    15.6 &    30.7\\
$M=14$~TeV
 &     0.70 &     6.5 &    20.2 &    40.0\\
\hline \hline

\end{tabular}
\end{center}
\end{table} }
 Integration of the column density of a
$M_{WD}=\Msun$ white dwarf along a diameter, using the density
 profiles shown in fig.~\ref{fig:WDprofile}~\cite{mesa},
yields a column density
$\delta_{WD}=2\int_0^{R_{WD}} \rho d\ell = 13\times
 10^{15}\gr/\cm^2$.  This number increases to  $\delta_{WD}=21\times
 10^{15}\gr/\cm^2$ for $M_{WD}=1.1\,M_\sun$, and to $\delta_{WD}=38\times
 10^{15}\gr/\cm^2$ for $M_{WD}=1.2\,M_\sun$.
The systematic uncertainty on these values,
 determined by varying the parameters of the white dwarf such as
 temperature and composition, is of the order of
 $10\%$~\cite{Lars}.

\begin{figure}
\begin{center}
\includegraphics[width=0.48\textwidth,clip]{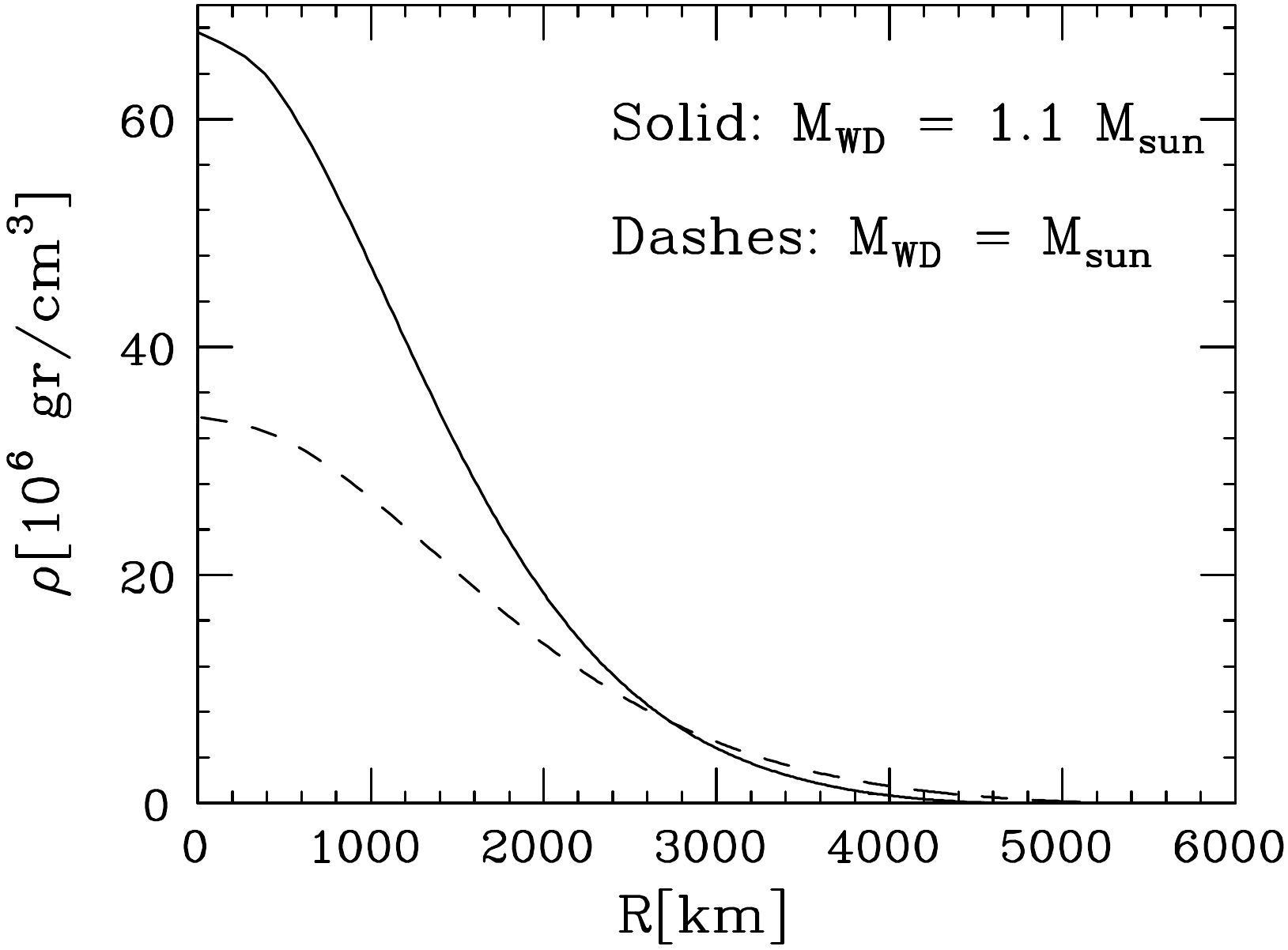}
\hfil
\includegraphics[width=0.47\textwidth,clip]{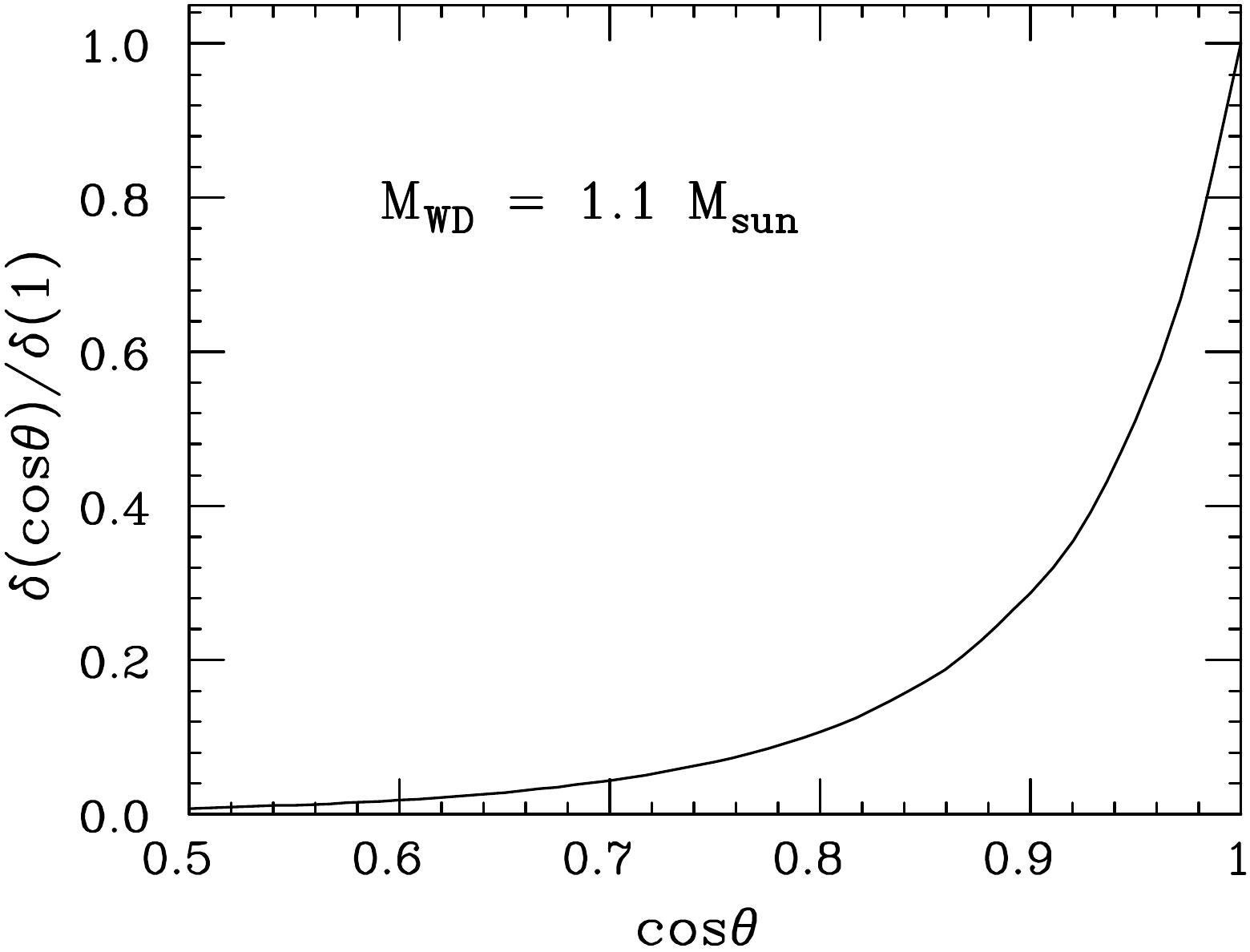}
\end{center}
\vskip -1cm
\ccaption{*}{\label{fig:WDprofile} \it Left: density profile for a solar-mass
  white dwarf (courtesy K. Shen). Right: column density, as a function
  of the penetration angle $\theta$ with respect to the zenith,
  normalized to the column density at $\theta=0$.}
\end{figure}

A comparison between the required stopping column densities, and the
available stopping power of white dwarfs, is shown in
fig.~\ref{fig:wdstop}.
\begin{figure}
\begin{center}
\includegraphics[width=0.68\textwidth,clip]{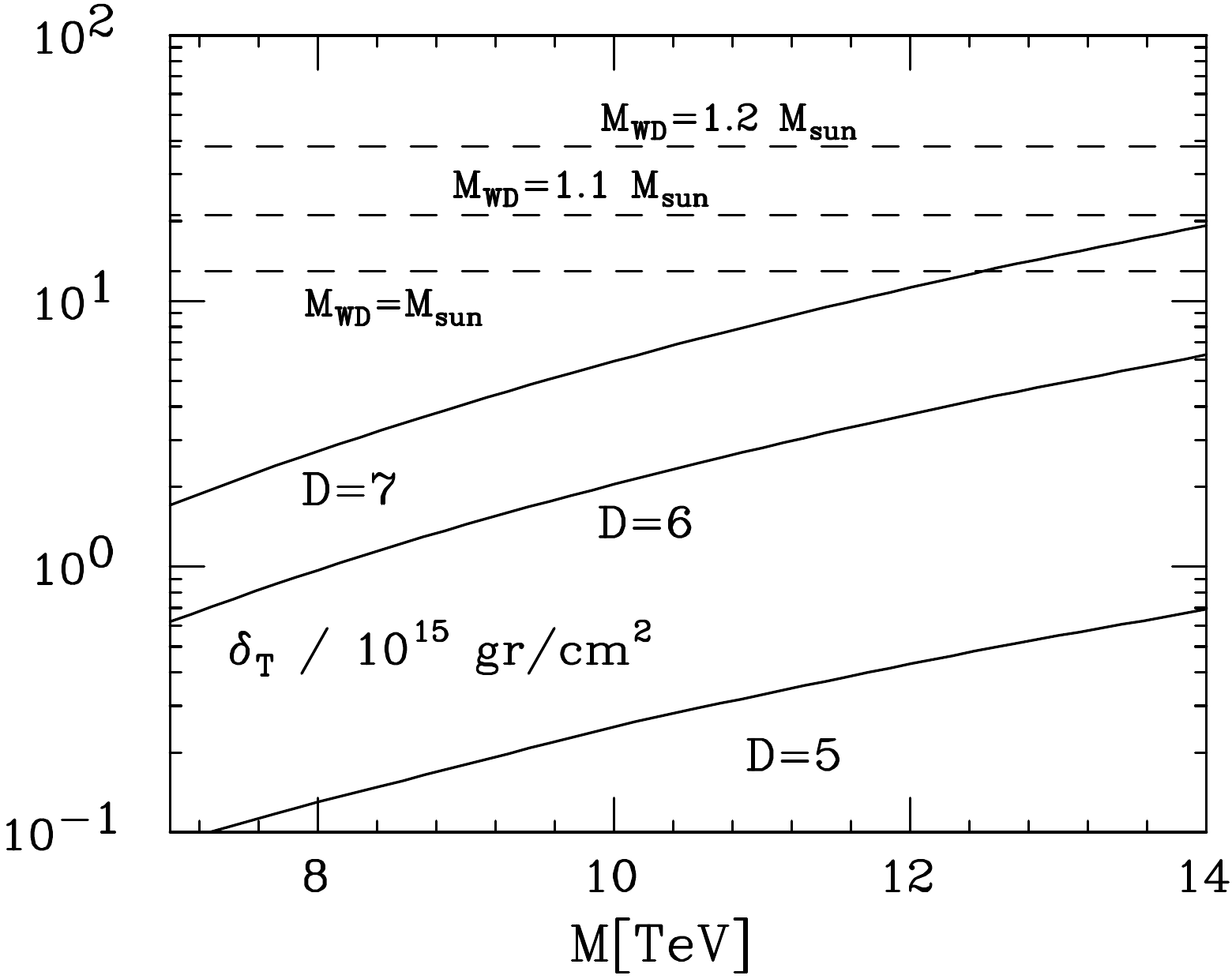}
\end{center}
\vskip -1cm
\ccaption{*}{\label{fig:wdstop} \it Column densities required to stop
  black holes of different masses in different space-time dimensions
  $D$ (solid lines) and integral column densities for white dwarfs of
  mass $M_{WD}=(1, \; 1.1, \; 1.2)$ solar masses.}
\end{figure}

 We thus conclude that a solar-mass white dwarf can efficiently stop
black holes. In the case of $D=5$ and 6, black holes corresponding to
the most conservative LHC scenario, with masses up to 14~TeV and $M_D$
accordingly large, will stop within a fraction of the maximum column
depth for white dwarfs at or above $M_\sun$. In the case of $D=7$,
one needs stars heavier than approximately 1.1 solar masses in order
to achieve stopping up to 14~TeV. However, as indicated in
Section~\ref{sec:bondi67}, $D=7$ black holes above 6~TeV give rise to
accretion lifetimes on Earth in excess of 20 billion years.  Our
calculated stopping column densities for masses below this are well
below the column density of a solar mass white dwarf.

As shown in table~\ref{tab:WDstop}, the column densities required
   to stop the heaviest black holes for $D\ge 8$ exceed the stopping
   power of even the most massive white dwarfs, and therefore we
   shall only state empirical constraints on such scenarios when discussing
   the neutron stars case.

\section{Black hole production on white dwarfs}

In this section we briefly describe the expected rates of black hole
production on white dwarfs; for more details see Appendix
\ref{app:prod}.  Before discussing such production rates, however, we
must discuss the effects of magnetic fields.

\subsection{Magnetic screening}

White dwarfs and neutron stars are known to have significant magnetic
fields, which can have important effects on the charged cosmic-ray
primaries.  (For more details see Appendix~\ref{app:synch}.)  For
example, in the case of a dipole field of polar strength $B_p$, 
an incident ray of charge $Ze$
and momentum $p$ perpendicular to the magnetic axis will have a Larmor
radius that depends on the distance $r$ from the center of the
object as follows:
\be
r_L(r) = {2 p\over ZeB_p}\left({r\over R_0}\right)^3\ ,
\ee
where $R_0$ is the radius at the surface.
While even for neutron stars $r_L$ can be greater than the radius of
the star,  as shown in Appendix~\ref{app:synch},
with the very high energy cosmic rays required one must
consider the effects of synchrotron radiation.  This is studied in
more detail in Appendix~\ref{app:synch}, with the result that a ray at
incident angle $\theta$ and with mass number $A$ will have an energy
at $R_0$ determined in terms of its incident energy $E_\infty$ by
\be
E_{R_0} = E_{max}{E_\infty \over E_\infty+E_{max}}\ ,
\ee
where
\be
E_{max} = \frac{60 A^4 m_p^4}{2(Ze)^4(\sin\theta B_p)^2 R_0}\ .
\ee
Thus, $E_{max}$ sets an effective maximum energy for cosmic rays that
penetrate to the surface of the star; for protons, and normalizing to
typical white-dwarf parameters, we find that
\be\label{WDEmax}
E_{max}(\theta=\pi/2)
= 3.6\times 10^{18} \ev {5000\km\over R_0} \left({10^6 G\over B_p}\right)^2\ .
\ee
Since, as we will find, optimal bounds come from considering cosmic
ray energies up to $\sim 10^{20}\ev$, we see that in order to avoid
significant magnetic screening, we must consider white dwarfs with
magnetic fields $B_p\lsim \mathrm{few}\times 10^5$~G.  Bounds from
stars with larger fields are still achievable, since cosmic rays
incident at angles closer to
the magnetic poles will experience reduced energy loss,
but this leads to a reduction of rates for acceptable cosmic rays (see
Appendix~\ref{app:synch}).  We will return to the question of magnetic
fields for neutron stars, where we will find them to be somewhat more
problematic.
 
\subsection{Production rates}
\label{sec:bhprodrates}
Production rates can be computed using the known fluxes of cosmic
rays, together with cross sections obtained by convolving parton-level
cross sections with parton distribution functions.  We will briefly
summarize these calculations here; more details are supplied in
Appendix~\ref{app:prod}.

The basic parton level production cross section is of the form
$\sigma\approx \pi R^2(\sqrt {\hat s})$, where $\sqrt{\hat s}$ is the
CM energy of the pair of partons forming the black hole.  This is only
an estimate; one must take into account that not all the collision
energy is captured by the black hole, and even less is captured as impact parameters
grow comparable to
$R(\sqrt {\hat s})$.  Thus, to be more
precise, we account for this via an inelasticity factor,
$y$; we conservatively summarize trapped-surface
calculations of the
inelasticity~\cite{Eardley:2002re,D'Eath:1992qu,Yoshino:2002tx,Yoshino:2005hi}\footnote{We  
  also note that the refined  estimates in~\cite{D'Eath:1992qu}, which
  are based on perturbatively-calculated radiation, have in the case
  of zero impact parameter recently been checked via numerical
  relativity techniques~\cite{Sperhake:2008ga}, with good agreement. 
}
by a 
simple dependence like that in~\cite{Giddings:2007nr}
\be\label{exinel}
E_{BH}= y\sqrt{\hat s}\ ,\ b<0.5~R\quad;\quad
E_{BH}=0\ ,\ b>0.5~R\ .
\ee
To be conservative for cosmic ray production rates, we use a lower than expected value,
$y=0.5$~\cite{Giddings:2007nr}, for  the inelasticity.  This corresponds to a limit of  7~\tev\ for
the maximum value of black hole mass that can be produced at the LHC.
The actual value of $y$ may be higher, and its reach could be slightly extended due to quantum fluctuations.
For this reason we also consider values in the range $0.5<y<1$ (the
upper limit being an unrealistic
extreme~\cite{Yoshino:2005hi,Meade:2007sz,Giddings:2007nr}) to allow for
black hole production at the LHC all the way up to the largest
available energy, namely 14~TeV. In making the estimates of cosmic-ray
production rates we shall conservatively choose the value of $y$
corresponding to the smallest possible inelasticity compatible with
production of a given mass value at the LHC, namely
$y=M_{min}/14~\tev$.  Furthermore, again to be conservative in our rate
estimates, we take the minimum black hole mass to be $M_{min}=3M_D$;
for example, \cite{Giddings:2001bu} used as a benchmark the
less-conservative value $M=5M_D$.

The resulting cross sections are then convoluted with the
CTEQ6M~\cite{Pumplin:2002vw}
parton distribution functions.  The resulting nucleon cross sections
are, in turn, convolved with the measured ultrahigh-energy cosmic ray
flux, extracted from  the latest Auger
spectra~\cite{Yamamoto:2007xj}.  In the case of cosmic ray
primaries that are nuclei of mass number $A$, one must also use a nucleon energy reduced by this factor.  The rate calculations
of Appendix~\ref{app:prod} are carried out for two test cases: that of
incident protons, and that of incident iron nuclei (A=56).  Partial
results are summarized in table \ref{tab:wdrate}, for production on a
white dwarf of radius 5400 \km, and more results are presented in
Appendix~\ref{app:prod}.  (In that Appendix we also present figures
resulting from a   20\%  hypothesized overestimate of cosmic ray energies, to model
the quoted Auger energy resolution of $\pm
20\%$~\cite{Yamamoto:2007xj,Ridky:2007zz}. )

{\renewcommand{\arraystretch}{1.1}
\begin{table}
\begin{center}
\ccaption{*}{\label{tab:wdrate}
\it Black hole production rates, per million years, induced by cosmic rays
  impinging on a $R=5400$~km white
  dwarf. $N_p$ refers to the case of 100\% proton composition,
  $N_{Fe}$ refers to 100\% Fe.  $M_D=M_{min}/3$ and $y=M_{min}/14$~\tev.}
\vskip 2mm
\begin{tabular}{l|lll}
\hline \hline
$D=$   &  5 & 6 & 7  \\ \hline
$N_p$/Myr, $M_{min}=7$~\tev
&   $2.1\times 10^7$ & $4.3\times 10^7$  & $6.7\times 10^7$\\
$N_{Fe}$/Myr, $M_{min}=7$~\tev
 &  $7.2\times 10^4$  & $1.6\times 10^5$ & $2.6\times 10^5$  \\
\hline
$N_p$/Myr, $M_{min}=14$~\tev
&   $2.3 \times 10^6$ & $5.9 \times 10^6$  & $1.0 \times 10^7$ \\ 
$N_{Fe}$/Myr, $M_{min}=14$~\tev
&   $7.3 \times 10^3$ & $2.1 \times 10^4$  & $3.8 \times 10^4$ \\ 
\hline \hline
\end{tabular}
\end{center}
\end{table} }

 According to the stopping calculations of section~\ref{app:cr}, not
all cosmic-ray produced black holes are stopped by a white
dwarf. In order to allow for sufficient column density, we must
require the cosmic rays to reach the white dwarf at an angle
sufficiently close to the azimuth as to force the black hole to
traverse a sufficient fraction of the full column density of the
star. The angular dependence of the column density is shown in the
right plot of fig.~\ref{fig:WDprofile}. In the case of $D=5$, our calculated column densities required for stopping are a mere few percent of those available and we can therefore easily accept a reduction to 10\% of the available column density. 
In the cases of $D=6$ and $7$, to be conservative we can
allow for at least 30\% and 80\% of the full available column density,
leading to a reduction of the useful cosmic ray flux to a level of 10\%
and 2\%, respectively.  Multiplying these efficiencies by the number
of events given in table~\ref{tab:wdrate}, and integrating over a
period of 10 million years, leads to total numbers of accumulated
black holes larger than  $5000$, even in the totally extreme case of 100\%
Fe composition of the cosmic rays. 
 Notice that even in the most conservative case of rates obtained for
$M=14$~\tev\ with $y=0.5$ (see table~\ref{tab:wdrate05} in
Appendix~\ref{app:bhrates}),  the number of events in 50 million years
still exceeds 100 for each value of $D$.

\section{Black hole catalysis of white dwarf decay}

The conclusion of the preceding sections is that white dwarfs with solar-size masses, as well as neutron stars, have sufficient ability to stop a cosmic-ray produced black hole, and that significant production rates for such black holes can in particular be achieved on white dwarfs.  
 We will seek  bounds from the statement that 
any stable black holes that could be produced on Earth will be produced and trapped in cosmic ray collisions with these astronomical objects.  The remaining problem is to examine the subsequent evolution.

We have in particular argued in section \ref{sec:WDstop} that a black
hole will be brought to zero velocity, largely through accretion and
scattering, deep within a white dwarf.  By this time the mass of the
black hole may grow significantly from its initial value.  The
surrounding medium consists of nuclei, for example, carbon and oxygen,
embedded in a sea of relativistic 
degenerate electrons.  A lower bound on the
relative velocities between the black hole and these nuclei is given
by the thermal velocity; for temperatures $\gsim10^7K$, these are at
least of size $v_T\gsim 3\times 10^{-4}$. At times ${\cal O}(0.6~\mathrm{Gyr})$
white dwarf cores begin to crystallize, 
but we will find accretion to be relevant well before this time.

\subsection{Subatomic accretion}

A small black hole will only exert an influence when within the
nucleus, as described by our ``bag-model" discussion in section
\ref{sec:NRstop}.  However, once larger it will have a longer-scale
influence.  In general its gravitational force must compete with
electrostatic forces between the nuclei and their surroundings.  We
can estimate these forces as in the discussion of atomic matter on
Earth --- a displacement of a nucleus from its equilibrium position is
expected to produce a dipole force resulting from interactions with
the ambient electron cloud, and this is estimated to be of a size
given by (\ref{eq:ecloud}), where one uses the typical internuclear
separation $a_{WD}\approx 10^{-10}\cm$ corresponding to $\rho\approx
10^7 \gr/\cm^3$.  This leads to a characteristic electromagnetic
capture radius $\rem$ as given in (\ref{emrad}), with $\beta_D$ given
by (\ref{betadef}), together with the approximate value 
\be
{K\over m M_0^2} \approx 1.2\times 10^{-18}\ .
\ee

For a black hole evolving from $M\sim M_D$, the radius $\rem$ will initially be smaller than both the nuclear scale and the velocity-dominated capture radius $R/v$.  One can readily check when $\rem$ exceeds the latter; this happens at 
\be
R_{EM}=R/v_T=\sqrt{ {(D-3)(D-1)^{D-1}\over 2(D-2)^{D-2}}} \sqrt{m\over K} v_T^{(D-3)/2}\ .
\ee
Thus, for $D\geq 6$, by the time $\rem$ reaches $\sim 1\, \fm$ it governs capture, though for $D=5$ there is a brief phase where $R/v_T>1\,\fm$ governs the capture rate.  Both the subnuclear phases and this phase are governed by very short time scales.

We will focus on cases where $R_D>a_{WD}$, as they include scenarios
with short evolution times on Earth.  These in particular include the
non-warped cases of $D=6,7$.  The evolution time up to the scale
$\rem=a_{WD}$ is then given by $d/v_T$, with $d$ the electromagnetic
evolution distance given in (\ref{distR}).  This yields time scales
for this phase 
\ba
t_{EM}&=& 1.5 \times 10^{-7} s \left({M_D\over M_0}\right)^3\ ,\quad D=5\ ,\\
t_{EM}&=& 0.09  s \left({M_D\over M_0}\right)^4\ ,\quad D=6\ ,\\
t_{EM}&=& 6 \times 10^4 s \left({M_D\over M_0}\right)^5\ ,\quad D=7\ ,
\ea
which are quite short.

\subsection{Bondi accretion}
Once $\rem>a_{WD}$, one enters the Bondi accretion phase.  Since, by
assumption, $R_D>a_{WD}$, this is initially $D$-dimensional up to
$R_D$, and then four-dimensional beyond $R_C$.  In cases with warping,
there can also be a warped phase intermediate between $R_D$ and $R_C$.   

Consider first the unwarped cases, with $D=6,7$.  The relevant time
scales are given by eqns.~(\ref{Bondi2rd}), (\ref{Fbondiev}) for
evolution up to $R_D$, and from $R_C$ onward.
To be conservative, we now model the transition phase
between $R_D$ and $R_C$ by assuming the {\it slowest} evolution,
namely a 4-dimensional growth, with a
constant Bondi radius equals to $R_D$, until the mass reaches the
value corresponding to $R_B=R_C$. This gives an evolution time scale:
\be\label{FbondievWD}
t = d_0 c_s {16\pi [4\pi(D-3)k_D]^{1/(D-4)} \over \lambda_D}
\left({M_D\over M_0}\right)^{(D-2)/(D-4)} \left({M_4\over
  M_0}\right)^{2(D-5)/(D-4)}\ , 
\ee

The evolution time scales are determined by the 
white dwarf parameters (for a $M=M_\odot$ white dwarf~\cite{Shen}) $d_0\approx 1.4\times 10^{-6} s$ and $c_s\approx
1.4\times 10^{-2}$. The combination $d_0 \, c_s$ has a value that is
approximately $1.5\times 10^{-4}$  times the Earth's value (\ref{Birch}). The resulting time
scales will therefore be  comparatively  shorter:
\ba
t(R_B<R_D) &=& 8\,\frac{1}{\lambda_6} \left({M_D\over M_0}\right)^2  \yr\ ,\ \ \quad\quad\quad D=6\\
t(R_D<R_B<R_C) &=& 4\times 10^2\,\frac{1}{\lambda_4} \left({M_D\over M_0}\right)^2 \yr\ ,\ \ \quad\quad\quad D=6\\
t(R_B>R_C) &=& 15\,\frac{1}{\lambda_4} \left({M_D\over M_0}\right)^2
\yr\ ,\ \ \quad\quad\quad D=6 
\ea
\ba
t(R_B<R_D) &=& 1.3\times 10^5 \,\frac{1}{\lambda_7} \left({M_D\over M_0}\right)^{5/3} \yr\ ,\ \ \quad\quad\quad D=7\\
t(R_D<R_B<R_C) &=& 6 \times 10^7\,\frac{1}{\lambda_4} \left({M_D\over M_0}\right)^{5/3}\yr\ ,\ \ \quad\quad\quad D=7\\
t(R_B>R_C) &=& 1.9\times 10^6\,\frac{1}{\lambda_4}\left({M_D\over M_0}\right)^{5/3} \yr\ ,\ \ \quad\quad\quad D=7\ .
\ea
 Taking the largest value of relevance for
the LHC, $M_D\sim 4.7$~TeV, and using the value of $\lambda_{D}$
appropriate for a relativistic electron gas, the longest possible
phase of the evolution in $D=7$ does not exceed  80 million
years. We recall that, in this largest-$M$ condition, the Earth's
lifetime for $D=7$   exceeds 10 billion years by a large margin.
We also notice that if we consider the case of more massive and thus
denser white 
dwarfs, the timescales are reduced. For example, the
central densities of white dwarfs of mass $M=1.1~\Msun$ and
$M=1.2~\Msun$ are 2 and 4 times, respectively, larger than for
$M=\Msun$~\cite{Shen}, leading to accordingly shorter evolution
times. 

From this analysis we conclude that Bondi accretion time scales for
the unwarped $D=6,7$ scenarios are quite short, especially as compared
to known white dwarf lifetimes that exceed 1 Gyr.  We have argued in
Appendix~\ref{app:rerad} that these are not modified by an
Eddington limit, at least until accretion macroscopically disrupts the
star.  If there were an Eddington limit, Appendix~\ref{app:rerad}
argues that one is even more likely to find one for Earth.  Moreover, as discussed in Appendix~\ref{app:rerad},
radiation of an ensemble of black holes at Eddington fluxes would
interfere with white dwarf cooling, providing an independent argument
against this possibility.   
We note, parenthetically, that in the true macroscopic regime, when
the black hole starts to exert large-scale effects on its host body,
the evolution may well not be Bondi, but in any case would disrupt the
object in question. 

\subsection{Generalized scenarios}

We next consider a more general warped scenario, with $R_C\gsim 15$~\AA.  In this regime, we find from (\ref{capprox}) that $R_C/R_D\lsim 20$, so $R_D>a_{WD}$.  
The growth will then be governed by Bondi evolution that is $D$-dimensional up to $R_D$, warped between $R_D$ and $R_C$, and four-dimensional between $R_C$ and large mass.

For $R_D\ll R_C$, the
bound on the first phase is significantly smaller than the other two.
As discussed in the case of accretion in Earth, the evolution in the
warped regime in general involves a large change in mass in a
relatively small change of radius.  For such a slow-growth law, the
growth time is governed by the upper limit of the radius range,
$R_B=R_C$.  This implies
 that the timescale for growth through this
phase is given by
\be\label{Bwarpt}
t_w\approx {M(R_C)\over \pi \rho c_s R_C^2} = {16\pi c_sd_0\over
  \lambda_4} \left({M_4\over M_0}\right)^2 {1\over M_0 R_C}\ . 
\ee
The time for growth from $R_C$ to large mass is the same, from
(\ref{FDevol}).   
For $R_C\gsim 15$~\AA, this yields timescales $t_{WD}\lsim 5\times10^6\,yr$. In
fact, even in the unrealistic case of evolution via a {\it constant} radius
relationship $R_B=R_D$ over the entire warped range up to just below
$R_C$, this would only enhance this time scale by $(R_C/R_D)^2
\sim400$, and thus the slightly larger value $R_C\gsim 30$\AA\ would 
still be in the range constrained by experimental bounds ($\lsim$~1~Gyr).   Indeed,
one can directly compare the white dwarf evolution in the range near
$R_B=R_C$ to that on Earth; from (\ref{FDevol}) we find 
\be
{t_{\rm Earth}\over t_{\rm WD}} \approx {d_0({\rm Earth})c_s({\rm
    Earth}) \over d_0({\rm WD})c_s({\rm WD})}{\lambda_4({\rm WD})\over
  \lambda_4({\rm Earth})}\ . 
\ee
Taking for example $\Gamma=5/3$ for Earth and $\Gamma=4/3$ for a white
dwarf, this yields a ratio of accretion times 
\be\label{accrat}
{t_{\rm Earth}\over t_{\rm WD}} \approx1.9\times 10^4\ . 
\ee

\subsection{Summary of white dwarf constraints}

This and the preceding sections have argued that in the hypothetical
TeV-scale gravity scenarios possibly relevant to LHC, 1) cosmic rays
will produce significant numbers of black holes on white dwarfs of
low ($\lsim \mathrm{few}\times 10^5G$) magnetic fields, on time
scales short as compared 
to known white dwarf lifetimes; 2) such  black holes, even if neutral
and with the highest masses accessible at the LHC,
will be stopped on white dwarfs with masses $M\gsim \Msun$,
by
accreting and scattering the dense matter of the star during their
transit; and 3) the white dwarf will then be accreted.  Accretion of a
white dwarf has been argued to be more rapid than that of Earth.
Different considerations reinforce this statement.  First, before the
$\sim$Gyr
 time scale, white dwarf matter is in a liquid form, in contrast to
significant solidity of matter in Earth.  Second, white dwarfs pack
the mass of the Sun into a region the size of the Earth, so are much
more dense, and have much higher internal pressures, assisting
accretion.

Several surveys of low-magnetic-field white dwarfs exist in the
literature. The use of Zeeman spectropolarimetry, in particular, has
allowed detection of fields down to the level of few kG.  
White dwarf masses are determined by spectral measurements of surface
gravities, with parallax and gravitational redshifts serving as
cross-checks; see e.g. \cite{KepBrad,FBB}.  Ages are determined
through white dwarf cooling; a textbook account appears in  
\cite{ShapTeuk}, with further discussion in
\cite{FBB,Liebert:2004bv,Althaus:2007zv}. 
Several
known white
dwarfs satisfying our criteria of mass $M\gsim \Msun$, $B_p\lsim
\mathrm{few} \times 10^5G$, and age $T\gsim 100$~Myr can be found,
for example, in~\cite{Schmidt:1995eh,Cuadrado,Kawka,Madej}. When not
given explicitly, the ages can be inferred from the mass-temperature
relations, as discussed e.g. in~\cite{Liebert:2004bv,Althaus:2007zv}. 
Some examples of
relevant stars are:\footnote{Where $B_\ell$ appears, it
  represents the measurement of the average
  longitudinal field, $B_\ell$, whose definition and relation to $B_p$
  can be found, e.g. in~\cite{Schmidt:1995eh}}
\begin{itemize}
\item WD0346-011~\cite{Schmidt:1995eh,Kawka}, with  parameters
 $M=1.25\Msun$, $B_{p}< 1.2
  \times 10^5$~G, and $T \sim 100$~Myr;
\item WD1022-301~\cite{Kawka}, with $M= 1.2 \Msun$, $B_{p}< 1.2 \times
  10^5$~G, 
and $T \gsim 100$~Myr;
\item WD1724-359~\cite{Kawka}, with $M=1.2\Msun$, $B_{p}< 1.2 \times
  10^5$~G, 
and $T \sim 150$~Myr;
\item WD2159-754~\cite{Kawka}, with $M=1.17\Msun$,~\footnote{See
  however~\cite{Subasavage:2008if} for a photometric determination of
  the surface gravity, leading to a lower mass value. Parallax
  determinations of the absolute distance are underway to confirm the
  mass assignment.} $B_{p}< 3 \times 10^4$~G, and $T \sim 2.5$~Gyr;
\item WD0652-563~\cite{Kawka}, with $M=1.16\Msun$, $B_{p}< 2.7 \times
  10^5$~G, and $T\sim100$~Myr
\item WD1236-495~\cite{Kawka}, with $M=1.1\Msun$, $B_{p}< 3 \times
  10^4$~G, 
and $T \gsim 1$~Gyr;
\item WD2246+223~\cite{Schmidt:1995eh}, with $M=0.97\Msun$~\cite{Madej},
  $B_{\ell}=1.5\pm 13.8 \times 10^3$~G,
and $T \sim 1.5$~Gyr.
\item WD2359-434~\cite{Cuadrado}, 
with $M=0.98\Msun$~\cite{Kawka}, $B_{\ell}=3\times 10^3$~G,
and $T \sim 1.5$~Gyr;
\end{itemize}

The above arguments thus state that comparison of respective accretion
rates, together with survival of white dwarfs to observed time scales
$\gsim 10^8 $~years, implies survival of Earth for a significantly
longer time, and in particular longer than the natural solar lifetime.

Finally, as in section~\ref{sec:macro}, we can estimate the lifetime
of a white dwarf, should it capture a minimum-mass
 primordial black hole.  With
Bondi evolution from the corresponding initial radius, one finds a
lifetime $\approx 1.8$~Gyr.

\section{Bounds from neutron stars}

\subsection{Production on neutron stars}

Neutron stars are very common in the Universe, 
and in fact provide robust examples of long-lived objects in other galaxies.
They also represent the
highest known densities of matter that have not undergone
gravitational collapse to a black hole.  Since they are 
particularly close to densities beyond which black holes are expected
to form, one might expect that introduction of a microscopic stable
black hole into a neutron star would rapidly catalyze its decay into a
macroscopic black hole. The known stability of NSs, with lifetimes
significantly exceeding $10^9$ years, therefore offers the prospect of
limits on microscopic black hole stability and accretion power.

However, known neutron stars have strong magnetic fields, which are
observed to range upwards from $\sim 10^8$~G.  In the case of a field
of $10^8$~G and a radius $R_0=10~\km$, (\ref{WDEmax}) yields a maximum
energy $1.8\times 10^{17}\ev$ for protons impinging perpendicular
to the field axis,  giving collisions
just above  
the LHC CM energy; the maximum energy is only about sixteen
times higher for heavy elements.  One can avoid this limit for cosmic
rays incident near the magnetic poles, but the acceptance for
protons of energies in the optimal range of $\sim 5\times 10^{18}$~eV
 is estimated in
Appendix~\ref{app:synch} to lead to a reduction of acceptable flux by
a factor of approximately $10^{-3}$, considerably weakening the
resulting bounds.

\subsubsection{Production in binary systems}
This suppression suggests we consider a more reliable way to
inject CR produced black holes into a neutron star.  Many NS binaries
are known, and in particular parameters and evolution of low-mass
binaries are well-understood.  Moreover, in such binaries, the
companion to the NS can subtend a significant solid angle in the sky
of the NS, as described in Appendix~\ref{app:binaries}.  Cosmic rays which would hit the NS but whose direction
intersects the companion will therefore scatter on the companion.  In
our TeV-scale gravity scenarios, part of the time they will therefore
convert to black holes, which then impact the neutron star.  Since we
only need bounds if stable black holes are neutral, the magnetic field
of the NS is irrelevant for these.  Such a production mechanism produces a
``full-coverage equivalent" (FCE) given by
\be\label{FCEdef}
FCE = \int dt {\Delta \Omega(t)\over 4\pi}\ ,
\ee
where $\Delta \Omega$ is the solid angle subtended by the companion,
and where we allow for time dependence due to evolution of the binary
system.  In order to compute the actual production rate on the neutron
star, we use the uncorrected rates of Appendix~\ref{app:prod}, times
the number of years of $FCE$.  A survey of known classes of binary
systems (see Appendix~\ref{app:binaries}) reliably yields FCE's in the
2~Myr range, resulting from systems with a 1~Gyr lifespan.  The neutron
star production rates  are exhibited in  table~\ref{tab:nsrate}
and  in fig.~\ref{fig:nsrate} of Appendix~\ref{app:bhrates}. A
summary of that table, focusing on the most interesting cases of $D\ge
8$, is shown here in table~\ref{tab:nsrate8}. 
{\renewcommand{\arraystretch}{1.1}
\begin{table}
\begin{center}
\ccaption{*}{\label{tab:nsrate8} \it 
Summary of 
black hole production rates, per million years, induced by proton cosmic rays
impinging on a $R=10$~km neutron star. 
$M_D=M_{min}/3$ and $y=M_{min}/14~\tev$.  }
\vskip 2mm
\begin{tabular}{l|llll}
\hline \hline
$D=$   &  8 &    9   &   10    & 11 \\ \hline 
$M_{min}=7$~\tev
  & 323 & 422 & 526 & 633 \\   
$M_{min}=10$~\tev
  & 129 & 172 & 218 & 265 \\
$M_{min}=12$~\tev
  & 80  & 109 & 139 & 171 \\
$M_{min}=14$~\tev
  & 54  & 74  & 95  & 118 \\
\hline \hline
\end{tabular}
\end{center}
\end{table}}
We find that in the example of a flux of even only 10\% protons, we have a
rate for the extreme case of 14 TeV black holes that is $\approx
5/$Myr, and so 2~Myr of FCE would indicate that typical such systems
have experienced sufficient black hole production to initiate the
accretion.  Less-extreme (and still
quite robust) binary scenarios provide
significantly higher
rates. However, a greater dominance of heavy elements reduces the range of such bounds.

At energies below the GZK cutoff, there are indications of a
significant component of heavy elements.  There are both theoretical
and experimental indications that one transitions to a significant
proton component at the GZK cutoff.  On the theoretical side one can
cite both the match of the observed spectrum to that from models of
proton acceleration, and expectations that gamma-ray bursters (GRBs) and
active galactic nuclei (AGN) primarily accelerate protons (see
\cite{Kashti:2008bw,Anchordoqui:2007fi,Hooper:2008pm} for more
discussion.)  On the 
experimental side, 
indications of a predominantly light composition include both measured
penetration depths of showers~\cite{Unger:2007mc,Hires} 
and,  most recently, correlations of
arrival directions with known AGN~\cite{Abraham:2007si,Cronin:2007zz}
(see however~\cite{Gorbunov:2008ef}).   
Thus, while not all scenarios are definitively eliminated by such a bound, 
it appears likely that these bounds will be strengthened with future data on composition.

\subsubsection{Production via cosmic neutrinos}
Primary cosmic rays propagating through the 3K cosmic microwave
background photons will experience significant interactions above the
GZK energy $\approx 5\times 10^{19}\ev$.  These interactions produce a
``guaranteed" flux of neutrinos (see
e.g.~\cite{Engel:2001hd,Fodor:2003ph,Anchordoqui:2007fi}), which
avoid the synchrotron losses of charged cosmic rays.  Using these
fluxes, Appendix~\ref{app:prod} calculates production rates on neutron
stars.  For example, in the very conservative scenario of requiring
black holes to have 14 TeV mass, and using both $D=5$ and our
most conservative inelasticity assumptions, $y=0.5$, one finds
production rates  $\gsim$5000/Myr, as shown in table~\ref{tab:nsrate-nu}.

These thus suggest a very robust bound for production on neutron
stars.  While we believe it is quite good, we will not take this bound
with absolute certainty, for two reasons.  First, while the physics of
the GZK effect is quite robust, and moreover appears to be in the
process of being experimentally confirmed via correlations of the
highest energy cosmic rays with 
AGN~\cite{Abraham:2007si,Cronin:2007zz}, experiments have not
reached sensitivity sufficient to measure the cosmic neutrino
flux.\footnote{Suggestions that models with extra dimensions suppress
neutrino fluxes~\cite{Lykken:2007kp} do not appear relevant to the scenarios
for which we require bounds.}  Second, there exist proposals that
baryon number conservation is enforced in higher-dimensional brane
world models through reduced interactions between neutrinos and
quarks by virtue of these living on different branes~\cite{Stojkovic:2005fx}.
While these models are not compelling, they would seem to raise a small
possibility that neutrino cosmic rays would not produce black holes
the same way that nucleons do.

\subsection{Catalysis of neutron star decay}

Due to the immense pressures inside a neutron star, one expects
introduction of even a microscopic black hole to rapidly catalyze its
decay.  To understand this process, we note that neutron stars have
different layers, a crust extending to a depth of $\sim1~\km$, and
under this, matter at nuclear densities.  Since treatment of accretion
is simplest in this inner region, we would like to understand whether
a BH can penetrate to this distance.

\subsubsection{Penetration to core}

The slowing distance (\ref{stopdist}), together with a subsequent
phase of slowdown to sub-escape velocities, $v\lsim 0.1c$, may or may
not permit penetration to depths $\gsim 1\km$, depending on details.
Note that the characteristic distance $d_0$ can be rewritten in
appropriate nuclear units as  
\be
d_0(\rho) = 9 \times 10^{-4} \cm/\rho[m_p/\fm^3]\ ,
\ee
and that crustal densities range from $10^{-6}$--$10^{-1} m_p/\fm^3$.

Even if penetration does not occur during slowdown, there is a different argument that it takes place on rapid time scales.  To see this, instead assume that a sufficiently slow black hole could become temporarily bound in the crust, by absorbing a parton and thus binding to the medium by strong forces.  This binding should, however, be temporary, as the scenario we wish to constrain is that where black holes don't remain charged, but instead discharge through Schwinger production.  Even ignoring this, if a black hole is bound to a nucleon via the color force, it will absorb the remaining partons of the nucleon, and thus become color neutral, on a relatively short time scale.  One can readily estimate this time scale.  A nucleon has a parton density of order $1/\fm^3$, and partons within the nucleon travel at speeds $\approx c$.  With the smallest possible absorption cross section, of order $\sigma\sim \pi/\tev^2$, we find a characteristic absorption time $t_{\rm abs}\sim 10^7\fm$.  

The neutralized black hole will then continue to fall in the net gravitational field of the neutron star until another such binding/neutralization event.  The characteristic distance between such events is
\be
d_{\rm free} \approx {8x\times 10^{-7} \cm\over  \rho[m_p/\fm^3] r_c^2[1/TeV]}\ ,
\ee
where $x$ parameterizes the typical nucleon energy fraction per parton, and we have expressed the capture radius $r_c$ in $\tev^{-1}$ units.  The corresponding time between such collisions is given by
\be
t_{free} \approx \sqrt{2d_{\rm free}/g_{NS}}\ ,
\ee
where $g_{NS}$ is the gravitational acceleration near the surface of the neutron star,
\be
g_{NS}\approx {GM_{NS}\over r_{NS}^2}\ .
\ee
This characteristic time scale is much longer than $t_{\rm abs}$, and thus sets the speed with which the black hole can penetrate the crust.  The corresponding average velocity is
\be
v_{\rm av}\approx {d_{\rm free}\over t_{\rm free}} \approx \sqrt{g_{NS} d_{\rm free}/2}\ .
\ee
The mean free path $d_{\rm free}$ reaches a minimum near the bottom of the crust, and consequently the average velocity is slowest there.  We can therefore bound the crust penetration time by using $\rho=m_p/\fm^3$ to derive a minimum gravitational drift velocity and time:
\be
t_{\rm crust} \lsim d_{\rm crust}/v_{\rm av}[\rho=m_p/\fm^3] \approx 10 s\, r_c[1/\tev]\ .
\ee
Thus the black hole should rapidly penetrate the crust and enter the neutron-fluid region of the core.

\subsubsection{Accretion from within a neutron star}
\label{sec:NSaccret}
Like in atomic matter, one might expect different phases for black
hole accretion within a NS, depending on the relative range of the
black hole's influence as compared to the radii $R_D$, $R_C$
representing crossover to lower-dimensional behavior, and as compared
to $1~\fm$, the characteristic separation between nucleons, which
delineates the crossover from microscopic  to macroscopic absorption in
the NS context.

To better understand these points, let us begin by comparing the force due
to a black hole to typical forces between the neutrons in the NS,
which are of size $\gev/\fm$.  Specifically, the force on a nucleon of
mass $m_p$ at distance $r$ is of order
\be
F_G \sim -\frac{m_p}{r} \left( \frac{R}{r} \right)^{D-3}
\ .
\ee
Equating this to the typical nuclear force, we find that gravity beats
such a force at scales 
\be
R_N = R \left({r_N\over R}\right)^{1/(D-2)}\ ,
\ee
which are only moderately larger than the Schwarzschild radius in the
regime $R<r_N$.  Thus a conservative ({\it i.e.}, for the purposes of
NS evolution, {\it slow}) estimate of the evolution is given by simply
taking the capture radius in (\ref{genevol}) to be $r_c=R$ in the
subnuclear regime $R<r_N$.  To estimate the accretion rate, one needs
the flux $F$. This receives contributions both from the velocity of
the black hole, and from the Fermi motion of the partons in the
nuclei.  The latter produces a flux $F\sim \rho$, in units where
$c=1$, and thus a geometric rate law
\be\label{geolaw}
{dM\over dt} = \pi \rho R^2\ .
\ee
The evolution equations are of the same form as eq.~\ref{bondires}, with
the replacements $c_s=1$,
$\lambda_D=1$, and $R_B\rightarrow R$.   The  resulting time is analogous to eq.~(\ref{Dtime}), and evaluated at  
$R=r_N\sim 1$~fm gives:
\be
t = {d_0\over k_D}\left({M_D\over M_0}\right)^{D-2} {D-3\over D-5} \left(
r_N \, M_0\right)^{(D-5)} \;  ,
\ee
leading to timescales ranging from a fraction of a second to at most few
weeks for $6\le D \le 11$. 

When the black hole enters the regime $R\gsim r_N $, the absorption
becomes macroscopic -- the black hole is capable of absorbing multiple
nuclei, and its gravitational range exceeds mean free paths.  As
Appendix~\ref{app:rerad} describes, an Eddington limit would be even
more problematic for a neutron star, given its high density and
opacity, and so evolution is described as Bondi until the black hole
reaches a scale where it disrupts the star.

The corresponding growth laws are those given in section
\ref{sec:Earthacc}.  In unwarped scenarios, we find for evolution from
$R_B=r_N$ to $R_D$ the growth time (\ref{Bondi2rd}), and for evolution
from $R_C$ up to large scales the comparable time scale
(\ref{Fbondiev}), with $\lambda_D\rightarrow\lambda_4$. For the stage
in between, we use, as in the case of the white dwarfs, the
conservative time in eq.~(\ref{FbondievWD}).  Taking a typical value
$\rho=2\times 10^{14}\gr/\cm^2$ gives $d_0=7\times 10^{-3}s$, and the
speed of sound is of magnitude $c_s\sim 0.1$.  The time scales are
therefore about $10^{-6}$ times smaller than those of the white
dwarf. The longest evolution corresponds to the 4-dimensional phase
from $R_D$ to $R_C$ in $D=11$ dimensions, with a time scale of $\sim10$
million years; for $D\leq7$, times are $\lsim 50 \yr$.

In the more general warped case, the growth times are, as in the white
dwarf case, dominated by the upper end of the warped phase and its
crossover to four-dimensional accretion, with time scale
(\ref{Bwarpt}).  For neutron star parameters and a value $R_C\geq
5$~\AA, this yields times of size $t_{NS,w}\sim 20\yr$.

Combining the results of this and the preceding subsections, we find
that the rate-limiting step to destroy a neutron star is the time
required for a black hole to be produced and reach the surface of the
neutron star.  Once it reaches the core, the accretion times are very
rapid compared to the neutron star's lifetime, ${\cal O}$(Gyr).  These
bounds appear quite challenging to avoid.  In order to do so, one
would need a significant deficit of light cosmic ray primaries,
together with a heavy ($\gsim 7\,\tev$) minimum black hole mass and
only systems with low $FCE$, {\it and} one would have to have a
neutrino flux that is either suppressed by unknown mechanisms or is
unusually non-reactive.

As a final note, in our framework we can estimate the lifetime of a
  neutron star that captures a primordial black hole of mass ${\cal
  O}(10^{15}\gr)$.  Our parameters yield a time of order $3\times
  10^5yr$.  We note that such processes have been proposed as the
  origin of some gamma ray bursts~\cite{Derishev:1999vj}, with roughly
  comparable accretion times.  The present analysis, in
  addition to   giving the analogous accretion time scales for Earth and
  white dwarfs, lends further detail to such a possibility through
  our description via Bondi accretion, and through our arguments
  against an Eddington limit.

\section{Summary and conclusions}

In this paper we have studied accretion of hypothetical stable
TeV-scale  black holes in two primary contexts: the Earth, and
compact stars -- white dwarfs and neutron stars.

For Earth, we identified two main evolution domains: that where the
black hole's gravitational range of influence is less than the atomic
scale, and that where it is greater.  An important distinction occurs
depending on where the crossover radius $R_C$ to four-dimensional
behavior lies relative to the atomic scale.  In particular, if $R_C$
is in the subatomic regime, evolution is four-dimensional at both
subatomic and macroscopic scales; this case includes unwarped
scenarios with $D\geq8$.  In this case we have argued that this
evolution occurs on times longer than the expected natural solar
lifetime, in two different ways: via a microscopic argument, and via a
macroscopic, hydrodynamic argument.  In both approaches, we used
conservative assumptions, leading to the largest accretion rates and
to fastest evolution.  At the end of the first phase, at times ${\cal
O}(10^{11})$ years, the mass of the black hole is still small, with a
mass of less than a megaton.
Such statements extend to more general warped scenarios, and crossover
scales up to $\simeq200$\AA\ lead to accretion times longer than the
Earth's natural lifetime.

On the other hand, those cases where $R_C\gsim 15$~\AA\ have been treated by deriving
bounds from white dwarfs, and also from neutron stars.  In
particular, we have argued that in such scenarios cosmic rays will
produce black holes on such astronomical objects, and that these
objects will stop even these very high-momentum black holes.  We then
studied accretion, showing that accreting black holes will disrupt
such objects on time scales short as compared to their observed
lifetimes.  In particular, we found a general relationship
(\ref{accrat}) between accretion times for Earth and for white dwarfs,
which, when combined with white dwarf ages exceeding $10^9$ years,
provides a very strong constraint.  Thus, the implication of these
arguments is that such scenarios, where Earth could be disrupted on
time scales short as compared to its natural lifetime, are ruled out.

We summarize here the origin of our constraints, as a function of
dimensionality $D$ and of black hole mass $M$.
\begin{description}
\item{$D=5$:} The bounds on
evolution scales on Earth for $D=5$ in the case of maximum allowed crossover radius $R_C$ are quite short.
This is a result of the higher-dimensional force law extending well into the macroscopic regime.
On the other hand,  the greater interactivity of
  $D=5$ black holes makes it possible for those produced by cosmic
  rays to get promptly trapped in both white dwarfs and neutron
  stars. In white dwarfs they are produced abundantly, with build-up
  time scales of the order of few thousand years even assuming a
  cosmic ray composition of 100\% Fe, and at the largest  Planck
  mass of interest. After being produced and trapped, in these extreme scenarios they quickly accrete to masses comparable to that of  the star on time scales that can be short when measured in years.
This would make
  it impossible for any white dwarf with a mass of the order of one
  solar masses to have survived longer than  a few thousand years,
  contrary to  observations.  Scenarios with increased warping have correspondingly lower $R_C$ and longer accretion times.  In particular, once $R_C\lsim 200$\AA, accretion times on Earth exceed its future lifespan.
\item{$D=6$:} The bounds on evolution times on Earth for $D=6$ are of the order
  tens of thousand years, thus short on geological time
  scales. Once again the main reason is the large extra-dimensional radius, and the
  high capture rates. As in the $D=5$ cases, such black holes produced
  by cosmic rays can be
  stopped inside dense stars. The production rates are even larger
  than in $D=5$, and the star accretion time scales for unwarped $D=6$ are comparable to the maximum-$R_C$ version of that scenario.  With increased warping, again $R_C$ decreases and accretion times increase.  
\item{$D=7$:} The bound on time scale for their macroscopic evolution on Earth
  is in the range of 6--80 billion years, depending on the black hole
  mass. Furthermore,
  $D=7$ black holes would be produced plentifully by cosmic rays
  on white dwarfs, and be stopped inside their surface, if the white dwarf mass
  is larger than 1.1 solar masses. The evolution
  times would be longer than in $D=5$ and 6, reflecting the lower
  growth rate that keeps them microscopic for billion
  years inside the Earth. But within the very conservative estimate of
  80 million years (for a $M=\Msun$ star, and shorter in the more
  massive cases) their
  accretion process of the white dwarf would be completed. Massive white
  dwarfs older than a few hundred million years would therefore be
  ruled out in these scenarios, once again contrary to observation.  Again, warping decreases $R_C$, thus increasing accretion times.
\item{$D\ge 8$:} For these black holes the bound on evolution time on Earth is
  extremely long, with times of size 100 billion years. This is due to the
  radius of the extra dimensions being smaller than 1~\AA, thus forcing
  most of the evolution to take place in 4 dimensions, where gravity
  is a totally negligible force. Warping only magnifies this effect.  In spite of this slow growth, these
  black holes would still grow fast enough inside a neutron star to
  consume it within about ten million years. The significant production
  rates on neutron stars when $D\ge 8$, and the existence of
  billion-year old X-ray binary systems, provide therefore additional evidence that such black holes either do not exist, or
  decay promptly. 
\end{description}

We also note that these bounds likely extend in case other objects are imagined
that could result from high-energy collisions in the relevant
energy ranges, that have weak-scale cross sections, and that could
threaten the long-term stability of matter.

We conclude by first summarizing the conditions needed for our bounds
to be necessary to rule out a possible risk.  In order for our bounds
to have relevance, a sequence of unlikely things would have to be
true.  First, TeV scale gravity, with a Planck scale no higher than a
few TeV, would have to be correct, so that black holes can be produced
at LHC.  Most workers consider this to be a fascinating possibility,
but also a somewhat unlikely possibility.  Second, black hole
radiance, which has been deeply studied from a number of theoretical
perspectives, would have to be wrong, {\it and} more general quantum
mechanical arguments for black hole instability would have to be
wrong.  Most workers consider this to be an exceedingly improbable, if
not impossible, scenario. Finally, one would need a mechanism to shut
off the quantum effects responsible for Hawking radiation, but still
leave intact either the quantum effects responsible for Schwinger
discharge, or some other neutralization mechanism that acts to
discharge the resulting stable black holes.  It is very difficult to
conceive of a consistent physical framework that provides such a
mechanism.

In the event that all these conditions are  satisfied, 
one can turn to the considerations of this
paper to assess the possible impact on Earth.  This paper has argued
that in order for such a scenario to have an impact on Earth at time
scales short as compared to the natural lifetime of the solar system,
in the five billion year range, the configuration of extra dimensions
would have to be such that gravity doesn't transition to
four-dimensional behavior until around the $200$\AA\
scale.  This apparently requires additional 
fine-tuning, reducing the likelihood even further.
 But beyond that, this paper has argued
that such scenarios are ruled out by the longevity of known white dwarfs,
on billion-year time scales.  In such a scenario, cosmic ray-produced
black holes should have catalyzed white dwarf destruction on
significantly shorter time scales.  

Moreover, decay of observed
neutron stars would also have been catalyzed, unless both of two
unlikely possibilities are
 realized, namely that the composition of
ultrahigh-energy cosmic ray primaries is 
dominantly heavy elements, and
ultrahigh energy cosmic ray neutrinos either are not produced, or have
suppressed gravitational interactions with partons.
To summarize, the present study argues for the following {\it additional}
layers of safety, beyond those that would have to fail to make this
study relevant:

\begin{description}
\item[{\bf 1.}] {\bf Only in scenarios such that the crossover scale
 to four-dimensional gravity is larger than about $ 200$~\AA\ does one have
 significant accretion at times short as compared to the natural
 lifetime of Earth.  This is a-priori unlikely, due to the additional
 fine-tuning required to realize such a TeV-scale gravity scenario.}
 
 \item[{\bf 2.}]{\bf In these scenarios where the bound on black hole accretion
 time on Earth is short as compared to natural time scales, white
 dwarfs would likewise be accreted, on much shorter time scales, in
 contradiction to observation.}
 
\item[{\bf 3.}]{\bf Unless cosmic rays have dominantly
a very heavy composition,
and moreover either the expected neutrino flux doesn't exist or has
unusual gravitational couplings to hadronic matter, neutron star decay would
likewise be catalyzed on time scales contradicting observation.}
\end{description}

In short, this study finds no basis for concerns that TeV-scale black
holes from the LHC could pose a risk to Earth on time scales shorter
than  the Earth's  natural lifetime.  Indeed, conservative arguments based on detailed calculations and the best-available scientific knowledge, including solid astronomical data, conclude, from multiple perspectives, that there is no risk of any significance whatsoever from such black holes.

\vskip.2in

\noindent{\bf Acknowledgements} We are grateful to many colleagues who
helped us in the course of the nine months of this project, providing
valuable guidance and advice on the many facets of our study, helping
us identify the key issues and pointing us to the relevant literature
sources.  Among these: J. Arons, J. Ellis, M. Fairbairn, G. Giudice,
G. Horowitz, D. Ida, Y. Kanti, G. Landsberg, D. Marolf,
J. March-Russell, K.-Y. Oda, S. Park, 
J. Polchinski, T. Rizzo, S. Rychkov, 
M. Salaris, M. Srednicki, N. Toro, I. Tkachev, M. Vietri,
U. Wiedemann and T. Wiseman.  The input of L. Anchordoqui, A. Dar,
A. De Rujula, S. Dodelson, D. Hooper and G. Gelmini helped us clarify
several issues related to the cosmic ray composition, and that of
L. Balents and D. Scalapino, questions of atomic structure relevant to
subatomic accretion.  We would particularly like to thank O. Blaes for
a number of important discussions on Bondi and Eddington accretion,
K. Shen and G. Schmidt for guidance on white dwarf structure and
populations, and L. Bildsten both for many crucial discussions and for
supplying information about neutron-star binary systems.  We would
also like to thank J. Hartle for earlier collaboration on this work,
and for many useful discussions.  The work of SBG was supported in
part by the U.S. Department of Energy under Contract
DE-FG02-91ER40618, and by grant RFPI-06-18 from the Foundational
Questions Institute (fqxi.org).

\appendix

\section{Bondi accretion}\label{app:bondi}

Bondi accretion~\cite{BHL,ShapTeuk}
 describes the steady flow of continuous matter into a black hole,
assuming spherical symmetry and hydrodynamical conditions. Its basic
ingredients are the continuity equation, guaranteeing a conserved matter flow, 
energy conservation, which gives the
accretion velocity, and accretion boundary conditions at the surface
of the black hole, which provides the sink for the continuous matter
infall.  As is well known, many features of accretion depend on matter properties in the non-relativistic region, far from the black hole horizon, and so a non-relativistic treatment is warranted.  In particular, one may describe the gravitational dynamics in terms of a general potential $\phi(r)$ that becomes strong in the TeV regime, as discussed in section \ref{genpers}.
Accretion from within a neutron star approaches the relativistic regime; relativistic corrections, which are typically small, are described for example in \cite{ShapTeuk}.  

The continuity equation is easily expressed as
\be \label{eq:cont}
\frac{dM}{dt} = 4\pi \, \rho(r) \, r^2 \, v(r) = \mathrm{constant}
\ee
where $\rho(r)$ and $v(r)$ represent the matter medium density and velocity,
with positive $v(r)$ corresponding to radial flow towards the black hole at $r=0$. The Euler
equation,
\be \label{eq:euler}
v\frac{dv}{dr} \; +\frac{1}{\rho}\,\frac{dp}{dr} = -\partial_r\phi
\ee
describes energy conservation. The first term is clearly the
variation of kinetic energy for an element of the medium as it moves. The second term represents the work done by the pressure gradients. As matter falls towards the black hole, its pressure will increase in response to increase in density. This energy,
as well as the kinetic energy, are drawn from the gravitational
potential of the black hole.  Indeed, this equation can be integrated to give the Bernoulli equation,
\be\label{Berneq}
0={v^2\over 2} -\int_r^\infty {dp\over \rho} + \phi(r)\ ,
\ee
where we neglect the kinetic and gravitational energy densities far
away from the black hole.

More highly compressible matter results in a higher accretion rate.  We parameterize matter properties via a general polytropic equation
of state, assuming adiabatic evolution, as
\be
p=K \, \rho^\Gamma
\ee
where $K$ and $\Gamma$ are constants. The density change as a function
of a pressure variation is related to the sound speed in the medium,
according to
\be
c_s^2 = \frac{dp}{d\rho} = \frac{\Gamma p}{\rho}\ .
\ee
This equation of state also provides the value of the integral in the Bernoulli equation:
\ba
&&\int_r^\infty {dp\over \rho} = {c_s^2\over \Gamma-1}\Bigg\vert^\infty_r\quad,\quad \Gamma\neq1\\
&& \int_r^\infty {dp\over \rho} = K\ln\left[{\rho(\infty)\over \rho(r)}\right]\quad,\quad \Gamma=1\ .
\ea

Using these definitions, $\rho$ can be eliminated by combining the continuity equation and Euler's equation, with the result
\be
\hf \partial_r v^2 \left( 1-{c_s^2\over v^2}\right) = -\partial_r\phi + {2c_s^2\over r}\ .
\ee
As one approaches the black hole, $v$ increases until it reaches the local sound speed $c_s$; this equation gives a relation defining the resulting {\it sonic horizon}, at $r=r_s$:
\be
{1\over r_s} = {\partial_r \phi(r_s)\over 2 c_s^2(r_s)}\ .
\ee
For example, in the case where the potential transitions from higher-dimensional form
\be
\phi(r) = -{k_DM\over 2M_D^{D-2} r^{D-3}}\quad ,\quad r<R_D
\ee
to four-dimensional form
\be
\phi(r)= -{G_4M\over r}\quad,\quad r>R_C\ ,
\ee
the sonic horizon is given by 
\be\label{Dson}
r_s^{D-3}={D-3\over 4} {k_DM\over M_D^{D-2} c_s^2(r_s)}
\ee
for $r_s<R_D$, and by
\be
r_s = {G_4M\over 2c_s^2(r_s)}
\ee
for $r_s>R_C$.

The continuity equation (\ref{eq:cont}) implies that the accretion rate is determined by quantities at the sonic horizon as
\be
{dM\over dt} = 4\pi \, \rho(r_s) \, r_s^2 \, c(r_s)\ .
\ee
This can then be computed by relating these quantities to those at large distances.

For example, evaluation of Bernoulli's equation (\ref{Berneq}) at $r=r_s$ yields 
\be
c_s^2(r_s) \left(\hf + {1\over \Gamma-1}\right) +\phi(r_s) = {c_s^2(\infty)\over \Gamma-1}\ .
\ee
In a D-dimensional regime, the formula (\ref{Dson}) for the sonic radius then gives
\be
c(r_s) = \sqrt{2(D-3)\over D+1-\Gamma(7-D)} c_s(\infty)\ .
\ee
From the equation of state relation $c_s^2\propto \rho^{\Gamma-1}$ we then find the $D$-dimensional relation between asymptotic fluid parameters and those at the sonic horizon:
\be\label{sonval}
\rho(r_s) = \rho(\infty) \left[{2(D-3)\over D+1-\Gamma(7-D)}\right]^{1/(\Gamma-1)}\quad,\quad p(r_s) = p(\infty) \left[{2(D-3)\over D+1-\Gamma(7-D)}\right]^{\Gamma/(\Gamma-1)}
\ee
Also, define the {\it Bondi radius} for a given mass by the equation
\be\label{BondiRn}
R_B(M) =  \left[{(D-3)\over 4c_s^2(\infty)} {k_D M\over M_D^{D-2}}\right]^{1/(D-3)}\ .
\ee
Note that the sonic horizon radius and Bondi radius defined in (\ref{BondiRn}) also differ by an ${\cal O}(1)$ proportionality constant.
Combining these quantities gives the Bondi accretion rate,
\be\label{Dbondin}
{dM\over dt} = \pi R_B^2(M) \rho(\infty) c_s(\infty) \lambda_D
\ee
where $\lambda_D$ is a $D$-dimensional constant,
\be\label{lambdaDdef}
\lambda_D \; = \;
4\left[ \frac{2(D-3)}{D+1-\Gamma (7-D)} \right]^{{[D+1-\Gamma(7-D)]/}[{2(\Gamma-1)(D-3)]}}\ .
\ee
In the range $1\le \Gamma \le 5/3$, numerical values fall between $4\le \lambda_4 < 18$ for $D=4$, and $3<\lambda_D<6.6$ when $D=5,\dots,11$.

Proper understanding of the accretion process also requires knowing the radial dependence of the physical parameters.  Eq.~(\ref{sonval}) shows that their values at the sonic horizon are close to their asymptotic values.  Within the sonic horizon, where the $v$ becomes supersonic, the relation $v^2\approx 2\phi$ and the continuity equation give the radial dependence of the density,
\be\label{raddens}
\rho(r) \simeq \rho(r_s)\, {r_s^2\over r^2} \sqrt{\phi(r_s)\over \phi(r)}\ ,
\ee
from which other quantities follow via the equation of state.  In particular, note that in a $D$-dimensional non-warped regime, 
\be
\rho(r)\simeq \rho(r_s) \left({r_s\over r}\right)^{(7-D)/2}\ .
\ee
From this we see that matter is compressed more strongly at lower $D$, remains constant density in $D=7$, and is rarified for $D>7$.  Also, from (\ref{wpert}), note that in a warped regime, the density scales as
\be
{\rho(r)\over \rho(r_0)}\simeq \left({r_0\over r}\right)^{(8-D)/2} e^{j_D(r-r_0)/2R_D}\ ,
\ee
which implies rarification in the region $R_D<r<R_C$.

\section{ Effects of radiative transport }
\label{app:rerad}
Acceleration and compression during accretion can cause the infalling
medium to radiate.  This raises the possibility of new effects.  For
example, in the case of accretion on Earth, such radiant energy could
melt the material surrounding the black hole, and thus has potential
to increase the accretion rate.  (In our rate bounds for accretion
within Earth, we guarantee that we have accounted for melting by
treating the problem as accretion from a fluid.)  However, pressure
from the outgoing radiation also has the potential to decrease the
accretion rate, and in particular one should check for the possibility
of an Eddington-limited rate.

The actual process of
radiation and its reabsorption is somewhat complicated, but can be  modeled based on simple considerations.  In the microscopic regime, one expects radiation resulting from accelerations of infalling particles.  In the macroscopic regime, another reradiation mechanism is heating of the material through its compression, resulting in thermal bremsstrahlung.
In general, one can parametrize the reradiation luminosity as 
\be\label{lumin}
L= \eta {dM\over dt}
\ee
where $\eta<1$ is the fraction of absorbed energy that is reradiated.

\subsection{Subatomic regime}

In the subatomic context, where accretion is treated as absorption of
individual particles, such particles can in general radiate some of
their energy via bremsstrahlung during absorption.  The precise
spectrum depends on details of the capture process, but a
characteristic scale is the horizon radius.  Thus, for horizon radii
below $1$\fm, one expects nucleons to be broken up, and to emit gluon
radiation that hadronizes into pions.  For radii above the Fermi
scale, one instead expects primarily emission of photons, with typical
energies approaching $\sim 1/R$.  We expect such radiation effects to
be small, due to small couplings of gauge boson amplitudes in the
vicinity of the black hole due to spatial wavefunction/gray body
factors.

In this limit where individual particles are absorbed one at a time, one certainly doesn't expect an Eddington limit, but can ask whether emitted energy is sufficient to melt atomic matter.  
Note that if the
accretion is driven by black hole motion, then one naturally converts
the luminosity (\ref{lumin}) into an energy deposition per unit length
of travel, 
\be
{dE\over dl} = \eta \sigma \rho\ ,
\ee
with $\sigma$ the capture cross section of the black hole.

Photons have characteristic absorption lengths in the $\cm$ range at
sufficiently high energies.  Let $\ell_a$ be the average absorption
length for the reradiated photons.  In this case, the energy density
in the vicinity of the black hole's track has size 
\be
{\cal E} = {\eta \sigma \rho\over \pi \ell_a^2}\ .
\ee

This can be estimated in the region $\rem\sim 1$~\AA.  In this case, one has
\be
{\cal E} = 10^{-16}\rho {\eta\over \ell_a^2(\cm)}\ .
\ee 
For this value of $\rem$, the Schwarzschild radius for $D=8-10$ is in the range
$10^{-11}-10^{-12}\cm$, so a characteristic energy is expected to be in the
\mev\ range.  In this range, $\ell_a$ indeed is of order $\gsim1 \cm$.   
For comparison, the latent heat of fusion for materials such as iron,
rock, {\it etc.} are of size
\be
{300 {\mathrm{kJ}}\over \mathrm{kg}} = 3\times 10^{-12}  \ .
\ee
To reach a threshold where reradiation could contribute significantly to
melting, one should therefore wait until the black hole grows to increase $\sigma/(\pi \ell_a^2)$ by a
factor of at least $\sim 10^4$, beyond the atomic scale.

In addition to photons, as  noted  for sufficiently small $R$, one can also have some fraction of the energy
emitted as strongly interacting particles (mostly pions). These have a
longer mean free path than photons, and should therefore contribute
even less to melting. 

\subsection{Macroscopic accretion}
\label{sec:macacc}

When the Bondi radius exceeds the internuclear separation $a$ in atomic matter, or $r_N$ in nuclear matter, accretion transitions to that of a continuous medium.

In this context, for spherically-symmetric accretion, microscopic motion deviating from the overall radial macroscopic flow is parametrized by the temperature of the medium.  As the medium is compressed, it heats up, and can radiate through thermal bremsstrahlung; elementary accounts of this mechanism appear in \cite{ShapTeuk,FrKiRa}.  In particular, the resulting luminosity can be estimated as resulting from free-free scattering.  Let $\Lambda_{ff}$ be the corresponding emissivity; for example, \cite{ShapTeuk} gives the relativistic form of this, 
\be\label{emff}
\Lambda_{ff} \sim {\alpha^3\over m_e^2m_p^2}\rho^2 T\ ,
\ee
up to factors of order one.  The total luminosity from a range of $r$ is then computed from the integral 
\be\label{lumint}
L_{ff}\approx \int \Lambda_{ff} 4\pi r^2 dr
\ee
over that region.  If one integrates this quantity from sonic horizon to event  horizon in a four-dimensional accreting medium, using the radial density dependence (\ref{raddens}), and the temperature dependence of a perfect fluid,
\be
T\propto{p\over \rho}\propto \rho^{\Gamma-1}\ ,
\ee
one finds\cite{ShapTeuk} that the integral is dominated near the event horizon,
\be
L_{ff}\approx {4\pi R^2} \Lambda_{ff}(R){\rm Min}[l(R), R/3]\quad ,\quad (D=4)\ ,
\ee
where $l(r)$ is the radius-dependent photon mean free path.
On the other hand, one can readily check that for a more rapidly increasing potential, {\it e.g.} $D\geq5$, and for $\Gamma\leq 5/3$, the integrated luminosity is dominated by the integrand at the {\it maximum} value of $r$ for the region.

The radiative transfer of the resulting radiation depends on properties of the medium such as its opacity.  Particularly relevant is  the photon mean free path $l(r)$, as compared to radius $r$ -- this determines whether radiation thermalizes, or can instead escape to regions with weaker gravitational potential.  (In the regimes we consider, free-free absorption is a typical opacity mechanism.) This ratio is  given by
\be
{l(r)\over r} = {1\over r\kappa(r) \rho(r)}\ ,
\ee
where $\kappa(r)$ is the opacity.    Using the equation (\ref{raddens}) for the density, we find that this parameter varies like
\be
{l(r)\over r} \simeq {1\over r_s \kappa(r) \rho(r_s)}  \sqrt{r^2\phi(r)\over r_s^2 \phi(r_s)}
\ee
in the supersonic region.  For constant opacity, we thus find different behaviors at decreasing $r$, depending on how the potential grows.  For four-dimensional growth, $l(r)/r$ decreases as $\sqrt r$, signifying that the medium becomes increasingly optically thick.  For $D=5$, this quantity stays constant at decreasing $r$, and for faster growth of the potential characteristic of $D>5$ or warped regimes, $l(r)/r$ increases for decreasing $r$ -- the medium becomes more transparent.

The size of the opacity is often approximately given by the Thomson cross section, $\kappa\approx \sigma_T/m$, where $m$ is the medium mass per electron ($\approx Am_p/Z\approx 2m_p$ for non-hydrogenic atomic matter).  However, this can vary depending on density and temperature.  It also depends on the frequency of the radiation, which will typically be given by the local temperature of the medium.

We will first illustrate these considerations by describing the case of white dwarfs, and then briefly summarize the corresponding story for Earth and for neutron stars.

\subsubsection{Radiative effects in white dwarfs}
\label{radwd}
Let us first compare the relevant scales.  The ambient mean free path, estimated via the Thomson value for the opacity, at $\rho=10^7 \gr/\cm^3$, is approximately $l_0\approx 5\times 10^{-7}\cm$.  More careful calculations of the opacity\cite{KiMa} yield a value $l_0\simeq 5\times 10^{-8}\cm$ at the temperature $10^8K$ characteristic of a young white dwarf, and values smaller by about $1/100$ for the temperature $10^7K$ characteristic of the 1~Gyr age.  

For the moment we restrict attention to the case $R_D\gsim l_0$.  From the above values, we find that this covers most situations with $R_D$ larger than the atomic scale, and specifically the cases we particularly would like to bound via white dwarfs, those of $D\leq7$ and the warped cases such as $D=5$ with large radius.

This means that there are three regimes for the evolution: first as $r_s$ evolves from the internuclear separation $a_{WD}\sim 10^{-10}\cm$ to $l_0$, then on to the scale $R_D\sim R_C$, then the four-dimensional regime $r_s>R_C$.  We consider them in turn.

In the first regime, the medium is effectively transparent, and is
governed by the higher-dimensional force law.  For constant opacity,
the preceding arguments indicate that the medium becomes more
transparent closer to $R$.  Initially, since the temperature is
nonrelativistic, the opacity evolves via a Kramers law, $\kappa\propto
\rho/T^{7/2}$, and thus drops slightly.  Then, at the relativistic
temperature $\sim 6\times 10^9$K, the free-free emissivity
(\ref{emff}) indicates $\kappa\propto \rho/T^3$, giving an essentially
constant value in the degenerate regime $\Gamma=4/3$.  Since the
medium is degenerate, one also expects a correction factor due to
Fermi blocking of the available energy levels ; this is included in
the results of \cite{KiMa}, and evolves as  $\sim T/ \rho^{1/3}$, thus
also remaining essentially constant in the degenerate regime. 

Accretion is determined by the balance of forces in the vicinity of the sonic horizon.  There will be a luminosity pressure from the ambient luminosity of the gas, 
\be\label{lumpress}
p_\gamma\sim T^4 \ ,
\ee
 in the downward direction towards $r=0$.  Below $r_s$, the medium is being evacuated by the accretion and does not have time to thermalize.  It will produce a luminosity, estimated for example by $L_{ff}$ as in the preceding discussion.
Since this is higher-dimensional accretion the integral will be dominated at the upper end; however, the result will not produce a value that competes with the downward pressure, since the integral contributes less upward pressure than would be present at the surface $r=r_s$ in the absence of a black hole.  Moreover, even if this pressure were competitive with the downward pressure, we can estimate the importance of them both via (\ref{lumpress}).  For the early temperature $T_0\sim 10^8K\sim 9\times10^3 eV$, and a typical pressure, we find 
 \be
 {T_0^4\over p_0}\approx {(9\times10^3 eV)^4\over 5\times 10^{-3}\mev^4}\approx 10^{-6}\ .
 \ee
 Thus these photon pressures cannot compete with the pressure of the degenerate fluid.

The next regime is evolution from $r_s=l_0$ up to $R_D$.  In this regime, the medium is optically thick when crossing the horizon, and thus approximately thermal.  With falling $l(r)/r$, it may become optically thin before reaching the horizon.  There, as above, we would find the inward thermal luminosity  overcomes the outward luminosity from the thin region at lower $r$.  Any resulting radiation cannot reach $r=r_s$.  Moreover, even its effect at $r<r_s$ is small.  For example, the ratio of photon pressure to medium pressure scales as
\be\label{pressrat}
{p_\gamma(r)\over p(r)} = {p_\gamma(r_0)\over p(r_0)} \left[{\rho(r)\over \rho(r_0)}\right]^{3\Gamma-4}\ 
\ee
and thus remains relatively small.  Furthermore, in the regime $r<r_s$, gravitational attraction dominates over the medium pressure.

As $r_s$ passes $R_D,R_C$, one reaches the four-dimensional regime.   The medium is opaque at the sonic horizon, but, if the true horizon lies below $R_C$, may or may not become transparent below that depending on parameters.  Once the true horizon reaches $R_C$, the fluid remains optically thick to the horizon, and thus any outward photon flux thermalizes.  Again, by estimating $p_\gamma/p$ down to the horizon, we see that its effect is small even on the local fluid evolution at $r<r_s$.

As one final check, one can compare the mean outward photon diffusion velocity to the inward flow velocity in the four-dimensional accretion regime.  This tells us whether the photons are {\it trapped}.  (For more discussion, see \cite{Begelman}.)  We find the resulting trapping condition $r^2<RR_B^3/l_0^2$.  This is satisfied for all $r$ out to $r_s$ if $R_B\gsim  l_0/c_s$.  Thus even very small four-dimensional black holes exhibit photon trapping, preventing their escape to the region near the sonic horizon.  

From these considerations we conclude that there are good reasons to rule out important radiation effects that could produce an Eddington limit for accretion within a white dwarf, although one cannot state for certain that some  form of dissipation would not play such a role.  However, we note that even in media that are less optically thick, efforts to produce luminosity that reaches the Eddington limit
for spherical black hole accretion in astrophysical contexts have failed, as described in \cite{ShapTeuk,FrKiRa}.
Radiation from spherical accretion onto a black hole seems to be quite inefficient, even in optically thinner
 situations, which we can find in accretion from Earth.\footnote{In
   certain low-collisionality contexts, magnetic fields can change
   this situation; see {\it e.g.} \cite{Sharma:2008pk}.  However, one
   can check that inside the low-magnetic field white dwarfs we
   consider, the dynamics is collisional, with typical Larmor radii
   greatly exceeding mean free paths. }

\subsubsection{Radiative effects in Earth}
In the case of Earth, the mean free path is much longer; for a rough estimate, using the Thomson cross section, and a density $\sim 10\gr/\cm$ characteristic of the interior of the Earth, one finds a photon mean free path of size $l_0 \sim 0.5cm$.  Thus, in cases relevant to Earth, we must explore another regime, $R_C<l_0$.  This changes the discussion as follows.

First, the phase from $r_s=a$ to $r_s=R_D$ is similar to the first phase for white-dwarf accretion:  the medium begins in the optically thin regime, and can get thinner as it nears the horizon.  Thus the upward luminosity pressure should not be competitive with the downward pressure.  Estimating relative sizes of photon vs. medium pressures for material near the center of the Earth, we find
\be
{T_0^4\over p_0}\sim {(.5 \ev)^4\over 2\times 10^{10} \ev^4}\sim 4\times 10^{-12}\ .
\ee
Thus, radiation pressure is negligible.

The next possible regime is from $r_s=R_C$ to $r_s=l_0$, and is new: it involves four-dimensional accretion from an optically thin medium.  In this case, the contribution to the luminosity from the integral (\ref{lumint}) is dominated by its lower limit, as long as that is in the four-dimensional regime.  Thus, initially it will be dominated by $r\approx R_C$.  However, as the black hole grows, the opacity at $r=R_C$ typically grows.  Before $R$ reaches $R_C$, the optical thickness $\tau(R_C)$ from $R_C$ to $r_s$,  defined via
\be
\tau(r) = \int_{r}^{r_s}{dr\over l(r)}=\int_r^{r_s} dr \kappa(r)\rho(r)\ ,
\ee
reaches unity. Subsequently the radius $r_T$ defined by $\tau(r_T)=1$ increases past $R_C$.  This means that the medium is optically thick and thermalizes inside this radius.  Thus, the integral (\ref{lumint}) should be cut off at the greater of $R_D$, $r_T$.  The net luminosity can be estimated as due to the contribution of that integral, dominated by the resulting radius, plus the luminosity of the thermal region inside $r_T$, 
\be
L_T\approx 4\pi r_T^2 T^4(r_T) \ .
\ee

This case is similar to the case of accretion of an optically thin gas onto a black hole, treated for example in \cite{ShapTeuk,FrKiRa}.  This situation, in which radiation can readily propagate from regions of high compression and temperature, seems to have the highest prospect of producing an Eddington limit among the scenarios we consider.  However, 
while a significant amount of luminosity can be generated, it appears difficult to attain an Eddington limit, due to the cutoff on free-propagation of photons at the radius $r_T$.  In particular, if one assumes an Eddington limit with a small efficiency $\eta$, this implies that $r_T \approx R/\eta^2$.  This thus requires $r_T$ to be large, 
which in turn limits  the temperature at the last-scattering surface.  This basic inconsistency is related to those explained for example in \cite{FrKiRa}.
Thus, presence of an Eddington limit would  require satisfying a non-trivial set of consistency conditions, in the absence of other dissipative mechanisms in the medium.  (One can incidentally check that there is not sufficient infall time for neutronization to provide a {\it cooling} mechanism.)

Finally, ultimately $r_s$ reaches $l_0$.  Above this value, the accreting medium is optically thick at the sonic horizon, and typically gets thicker as $r$ decreases.   If the true horizon is still inside $R_D$, there is a possibility that the  medium might become optically thin in a region before reaching $R$.  Whether or not this happens, the significant optical depth in the regime just inside the sonic horizon apparently precludes an Eddington limit.

\subsubsection{Radiative effects in neutron stars}
\label{radns}
A rough estimate of the photon mean free path in the neutron star case, using the Thomson cross section, yields $l_0\lsim 1\fm$.  Thus in all cases where we seek a bound (namely, if $R_D\gsim1$~\AA), we are in a situation analogous to that of the second and third regimes for a white dwarf, but with an even higher opacity.  Moreover, the range of $r$ between the sonic radius and the horizon is quite limited; in general, we have
\be
{R_B\over R}\lsim {1\over c_s^2}\ ,
\ee
with a typical sound speed $c_s\gsim 0.17$.  Thus we typically don't expect the medium to ever become optically thin.  
Finally, asymptotic interior temperatures are expected to be in the range $10^4-10^5\ev$, and pressures are in the range of $\mev/fm^3$.  Thus, asymptotically, $p_\gamma/p \lsim 10^{-11}$, and from (\ref{raddens}) and (\ref{pressrat}) we find that the radiation pressure remains negligible down to the horizon.  

The relative unimportance of radiative pressure, and its inability to stream outward, thus indicate that the neutron star evolution should also be governed by Bondi accretion, without an Eddington limit.

\vskip .3in

We also note here that our arguments against an Eddington limit in subsections~\ref{radwd}-\ref{radns}, and therefore for a Bondi description of accretion, likewise apply to the case of primordial four-dimensional black holes, which are expected to have masses $\gsim 10^{15}\gr.$

\subsection{Eddington evolution}

For completeness, we will give a rudimentary account of accretion in the presence of an Eddington limit.
We model this effect as follows.\footnote{For a textbook treatment of
some aspects of the Eddington limit, see \cite{ShapTeuk}.}  The
reradiation luminosity leads to a flux of energy
\be
S= {\eta \dot M\over 4\pi r^2}
\ee
through the spherical surface at radius $r$ from the BH.  As
described, the reradiation consists of light particles such as
photons, pions, etc.  These outgoing particles scatter on the
accreting matter, producing an effective outward force.  If the
scattering cross section on a given infalling particle of mass $m$ is $\sigma$,
the average force on this incident particle takes the form
\be
F_L= {\eta \dot M\sigma\over 4\pi r^2}\ .
\ee
 For Earth, the incident matter is atoms.  In the case of a white dwarf or neutron star interior, generally the incident matter is a degenerate electron liquid or the n-p-e liquid of neutron star interiors.  For horizon sizes above the scale $1\fm$, the radiation is expected to be primarily photons, and thus its force can be estimated using the Thomson cross section for photon-electron scattering.  

This reradiation force has the effect of modifying the Euler equation (\ref{eq:euler}) to
\be \label{eq:eulern}
v\frac{dv}{dr} \; +\frac{1}{\rho}\,\frac{dP}{dr} = -\partial_r \phi + {\eta \dot M\sigma\over 4\pi mr^2}\ .
\ee
This yields a modified Bernoulli equation; for a D-dimensional potential, 
\be \label{eq:nbern}
\frac{1}{2} v^2 +\frac{1}{\Gamma-1} c_s^2  = \left[{1\over 2}\left( \frac{R}{r}\right)^{D-3} - {\eta \dot M\sigma\over 4\pi mr}\right]
+ \frac{1}{\Gamma-1} \, c_s^2(\infty) \; .
\ee

This equation provides an important constraint on accretion flows.  In
particular, notice that its left hand side is positive semidefinite.
Therefore, if the right hand side vanishes for some $r$, the density
and velocity of the accreting fluid must go to zero, effectively
shutting off accretion.  There are two cases, depending on whether the
Bondi radius $R_B$ defined in equation (\ref{BondiRn}) is greater or
less than the crossover radius to four-dimensions, $R_C$.

Notice that the two positive terms on the RHS of  (\ref{eq:nbern}) are of the same size when
\be\label{poseq}
r=\left[{2(\Gamma-1)\over D-3}\right]^{1/(D-3)} R_B\ ,
\ee
and consider first the case $R_B\gg R_C$.  
Then the gravitational term dominates in the range $R_B\gsim r>R_C$.  The condition that the negative Eddington term not negate the four-dimensional gravitational force is then the usual Eddington limit,  
\be\label{eddlim}
\dot M \leq {4\pi mG\over \eta \sigma} M\ .
\ee

In the case $R_B<R_C$, the positive terms become of comparable size in the higher-dimensional regime.  If the negative term is smaller than the other two terms at the radius given by (\ref{poseq}), then it will also clearly be subdominant for both larger and smaller $r$.  This yields the relevant higher-dimensional Eddington limit, 
\be\label{HDedd}
\dot M \leq {f(\Gamma,D)}{8\pi  mR_B\over \eta\sigma} {c_s^2(\infty)}\ ,
\ee
where for $D\geq5$
\be\label{fdef}
f(\Gamma,D) = 2\left({2\over \Gamma-1}\right)^{(D-4)/(D-3)} \left({1\over D-3}\right)^{1/(D-3)}
\ee
is an ${\cal O}(1)$ coefficient.   In the case of $D=4$, (\ref{HDedd}) subsumes (\ref{eddlim}) with $f(\Gamma,4)=1$.

The Eddington limit becomes relevant when the Bondi accretion rate (\ref{Dbondin}) exceeds the Eddington 
rate (\ref{HDedd}).   This occurs for
\be\label{Eddcr}
R_B\gsim R_{Edd}= {8  mc_s(\infty) \over \lambda_D \eta\sigma\rho}\ ,
\ee
defining the {\it Eddington radius} $R_{Edd}$.
Growth of the mass at the four-dimensional Eddington limit
(\ref{eddlim}) is exponential, with a time constant
\be\label{Tedd}
t_{Edd} = \eta{\sigma\over 4\pi m G}\ .
\ee

We have given arguments about the difficulty of achieving such an Eddington limit in Earth, and even more so in white dwarfs, at least until one reaches large black hole sizes which disrupt the  large-scale structure of the body in question.  Moreover, if such a mechanism were to become operative in white dwarfs, then each black hole within the dwarf would be emitting at the characteristic Eddington luminosity $L_{Edd}\simeq 8\pi mR_Bc_s^2/\sigma$.  This would also be evident through interference with white dwarf cooling.  Typical cooling rates are in the range $10^{-1}-10^{-3} L_\odot$, where the solar luminosity is $L_\odot=4\times 10^{33} erg/s$.  As an example, we find that the Eddington output of $N$ black holes of Bondi radii $R_B$ would exceed $10^{-2}L_\odot$ for
\be
NR_B/\cm \gsim 60\ .
\ee
Given the large numbers of black holes that would be produced,  on relatively short time scales one would find a buildup of black holes that have a major impact on cooling, even for a relatively large value like $\eta=.01$.

\section{Gravitational scattering of relativistic 
 particles in a Schwarzshild field}
\label{app:scattering}
This Appendix focuses on the dynamics of a test particle in the background of a $D$-dimensional black hole. 
In the classical context, we consider  the trajectory of a test particle in motion with
positive energy with respect to a $D$-dimensional black hole. The goal
is to establish the features of the particle's dynamics as a function of
the initial energy and impact parameter, and to define the
conditions under which the test particle, in the encounter
with the black hole, is scattered or is absorbed. We shall be
interested in applying this discussion to the study of energy loss,
slow-down and stopping of
black holes produced by cosmic rays. In this phase the black hole is
still sufficiently small that it only interacts, microscopically, with
the partons inside the nucleon. Since these are in relativistic
motion, we need to consider the case of relativistic test particles. 
Two kinematical regimes are then potentially
relevant: the classical one, in which the de Broglie wavelength of the
test particle is small in comparison to the black hole radius, and the
quantum regime, where the black hole itself is small compared to the
probe's wavelength. In the context of black holes produced by cosmic
rays, the classical regime is relevant early on, when the Lorentz
$\gamma$ factor of the black hole is very large, and later on, when its
size has grown significantly. If we consider first the problem in the rest frame
of the black hole, the momentum of the infalling parton is of order
$\gamma m_p$. At production, $\gamma > M/m_p$ (see
eq.~(\ref{eq:gamma-prod})), and therefore the wavelength of the
projectile is indeed smaller than $1/M$, and thus smaller than the
$D$-dimensional black hole radius, which is larger than
$1/M$. As the black hole slows down, $\gamma$ drops, and we 
enter  the quantum regime, where we stay until the black hole grows
to a size of $>1~\fm$.  
In the quantum case, we can use the known total quantum scattering cross section.  Where we need the differential cross section as a function of angle, outside the capture regime, we will use the classical result; the two are known to agree for $D=4$ Rutherford scattering.

\subsection{Classical trajectories and capture}
The equation of the particle's trajectory, in the rest frame
of a $D$-dim Schwarzschild potential, is given by:
\be
\varphi = \int \; \frac{L}{r^2} 
\left\{ E^2
 -(m^2+\frac{L^2}{r^2})\left[1-\left(\frac{R}{r}\right)^n\right]
 \right\}^{-1/2} \; dr  
\ee
where $n=D-3$, $E$ is the projectile 
energy at infinity, and $L=p\, b$ is the angular
 momentum at infinity ($b$ being the impact parameter). Setting
 $E=m\gamma$, $p=mv\gamma$ and defining $\rhat = r/R$ and $\bhat=b/R$ gives:
\be 
\varphi = \int \; \frac{v\bhat}{\rhat^2} 
\left[ 1
  -(\frac{1}{\gamma^2}+v^2
 \frac{\bhat^2}{\rhat^2})\left(1-\frac{1}{\rhat^n}\right)\right]^{-1/2}\; d\rhat  \; .
\ee
In the relativistic limit, $\gamma\to \infty$, this becomes:
\be \label{eq:relscat}
\varphi = \int \; \frac{\bhat}{\rhat^2} 
\left[ 1 -
 \frac{\bhat^2}{\rhat^2}+\frac{\bhat^2}{\rhat^{n+2}}\right]^{-1/2} \; d\rhat \; .
\ee
The orbit is thus entirely defined by the initial impact parameter $\bhat$.
Scattering states only exist if the impact parameter is large enough
that the term in square brackets 
admits a zero for $\rhat>0$, defining 
the point of closest approach, $\rhat_{min}$, of the trajectory. At this point,
\be
\bhat^2 = \frac{\rhat_{min}^{2+n}}{\rhat_{min}^n-1} \quad , \quad \rhat>1 \;.
\ee
It is straightforward to prove that this relation admits a 
real-valued  solution
for $\rhat_{min}$ only if 
\be \label{eq:bmin}
\bhat>\bhm=\frac{(2+n)^{(2+n)/2n}}{\sqrt{n} \, 2^{1/n}} \;.
\ee
Thus $b_{min}=\bhm R$ represents the minimum
impact parameter for scattering, 
below which the projectile falls inside the event
horizon. This therefore defines the classical {\it capture
radius}\footnote{For the $D=4$ case, we recover the usual results,
$b_{min}=3\sqrt{3}R/2 $, and $r_{min}=3R$, leading to a capture cross
section of $27\pi R^2/4$.
}.
Approximate values  for $D=5,\dots,11$ are
given by $\bhm=(2, 1.8, 1.6, 1.5, 1.5, 1.4, 1.4)$.

\subsection{Quantum capture}
\label{app:quantum}
The absorption cross section of 4-dimensional 
spin-1/2 and spin-1 fields in the field of a
 $D$-dimensional black hole has been calculated
in~\cite{Kanti:2002ge,Kanti:2002nr,Ida:2002ez,Ida:2006tf,IOPPC},  
generalizing the $D=4$ results
of~\cite{Page:1976df,Unruh:1976fm,Sanchez:1977si}.
Considering spin-1/2 fields, 
the largest contribution is given by the $s$-wave states, 
resulting in capture cross sections $\sigma_c$ given by the following equation:
\be
\sigma_c= 2^{(3D-13)/(D-3)} \; \pi R^2 \; .
\ee
These can be thought of as determining the capture impact parameter
$\bhmq$, like in the  classical  case, via the equation
\be
\sigma_c = \pi \bhmq^2 R^2\ . 
\ee
The quantum capture radius $\bhmq$ grows  from $\sqrt{2} $ for $D=5$ to $\sim 2\sqrt{2} $
for large $D$. The approximate values in the range of interest are
given by $\bhmq=(1.4, 1.8, 2.0, 2.1, 2.2, 2.3, 2.4)$. Notice that, with
the exception of $D=5$, these are slightly larger than the
classical capture radii given above.

\subsection{Coulomb scattering}

Outside the capture region, the projectile is deflected by the gravitational field.
In the large-$\bhat$ approximation, the approximate 
 expression for the classical scattering angle $\theta\equiv \pi-2\varphi$ 
is given by:
\be \label{eq:smallangle}
\theta_{app}=-\frac{1}{\bhat^n} \sqrt{\pi}
 \frac{\Gamma[(n+3)/2]}{\Gamma[(n+2)/2]} 
 \equiv -\frac{1}{\bhat^n} \; \alpha_{n+2} \; ,
\ee
which correctly reproduces the classical deflection of light in $D=4$, 
$\theta=-2R/b$. Notice that this approximate result
underestimates the exact deflection angle in the region around $b\sim
b_{min}$, as shown fig.~\ref{fig:appscat}. Notice also in the figure
that, as $D$ grows, large-angle scattering only takes place for
projectiles with impact parameter very close to the minimum value
$b_{min}$, as a result of the rapidly falling gravitational field. 

As
discussed in the main text, for the momentum loss due to elastic scattering
we need the quantity
\be
\cel= {1\over \sigma_c} 
\int_{\cos\theta_c}^1 \,d\cos\theta {d\sigma\over d\cos\theta} 
\, 2\sin^2 {\theta\over 2}\ .
\ee
Without an explicit formula for the quantum differential cross section, we will estimate it using the classical expression (\ref{classcsc}), combined with the small-angle formula (\ref{eq:smallangle}).  This yields the approximate value
\be
\cel\approx {\theta^2_c\over 2 (D-4)}\ .
\ee
  If $\theta_c\sim1$, we find the approximate values $\cel=(.5, .25,.17)$ for $D=5-7$.
\begin{figure}
\begin{center}
\includegraphics[width=0.68\textwidth,clip]{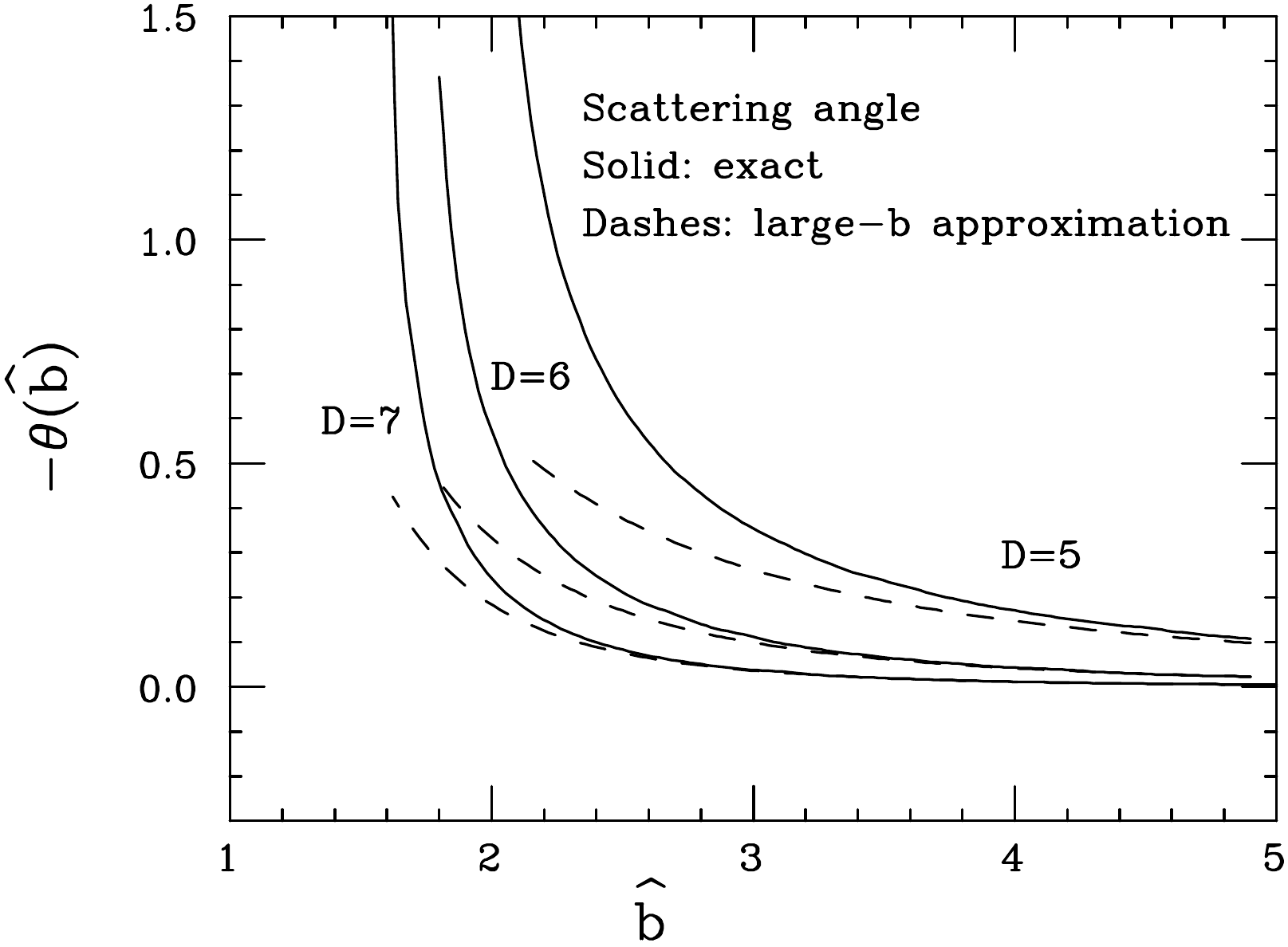}
\end{center}
\vskip -1cm
\ccaption{*}{\label{fig:appscat} \it Comparison of exact (solid) and
  approximate (dashes) relations between the scattering angle (which for an attractive force is negative) and the
  normalized impact
  parameter $\hat{b}=b/R$, for $D=5,6,7$.}
\end{figure}

\section{Nonrelativistic accretion checks}
\label{app:accretion}
The discussion in section~\ref{sec:WDstop} assumes the validity of two
facts, which we establish here:
\begin{enumerate}
\item  the mass accreted per
nucleon-crossing should not exceed the proton mass. 
\item the capture radius remains smaller than 1~fm throughtout this
  phase.
\end{enumerate}
The first statement is stronger than the second one. If the first is
true, the second one is also true. Here is the proof:

The first statement requires the following:
\be
\Delta M = \pi  \rho_n \frac{r_n }{v} R_c^2 < m_p
\ee
where $R_c$ is the capture radius, $R_c= \hat{b}_{min}R$. Writing $m_p=(4\pi /3)\rho_n r_n^3
$, we obtain $R_c<\sqrt{4v/3} r_n \lsim r_n$. Thus, we just need
to show that  the accretion rate during a single crossing
does not saturate $m_p$.

The maximal mass relevant to nonrelativistic stopping in a white dwarf is the mass at the gravitational trapping velocity, which is bounded by $M<\Mnr/v$, where we recall from (\ref{Mnreq}) that $\Mnr<\gamma_i M_i$, and where $v\simeq .02$.  Thus, our condition becomes
\be
{\hat b}_{min} R(\gamma_i M_i/v)<r_n\sqrt{4v\over 3}\ .
\ee
Recall also that a typical boost for a large initial black hole, $M=14 \tev$, is $\gamma_i\approx 4.5\times 10^4$.  For $D\geq6$, this inequality can then easily be checked to hold for all $v>.02$ and $M_D>1\tev$.  For $D=5$, even for the extreme case $M_D=1\tev$, the condition holds for black holes well above $M_i= 5 M_D$ (with corresponding reduction in initial boost), down to the low end of the velocity range, $v\simeq.05$ and of course is easier to satisfy at larger $M_D$.

\section{Black hole production rates}
\label{app:prod}
\subsection{Elementary cross sections}
In the scenarios considered here, black holes are  formed
in the collisions of partons with center-of-mass energies $\sqrt{\hat
s}\gg M_D$, and with impact parameters comparable to the
Schwarzschild radius $R=R(\sqrt{\hat s})$~\cite{Banks:1999gd,Giddings:2001bu,Dimopoulos:2001hw,Eardley:2002re}.  The
resulting cross section estimate, $\sigma \sim \pi R(\sqrt{\hat s})^2$, 
should be
improved to account for two effects.  On one side, the fraction $y$ of
partonic energy $\sqrt{\hat s}$ that is absorbed by the black hole is
expected to be less than unity.  The rest is radiated
off~\cite{Eardley:2002re,D'Eath:1992qu,Yoshino:2002tx,Yoshino:2005hi}.
Furthermore, the maximum impact parameter $b$ of the partonic
collision that can lead to a black hole formation above  the threshold mass $M_{min}$ for black holes to exist 
 is typically smaller than the radius $R(\sqrt{\hat s})$.  Recent estimates~\cite{Giddings:2007nr}, based
on~\cite{Yoshino:2005hi}, give inelasticity $y$ of
the order of 0.6--0.7, depending on $D$, dying off beyond impact
parameters of about half the radius $R(\sqrt{\hat s})$.  We
implement these constraints by allowing black hole production only for
partonic collisions with impact parameter $b<0.5~R$, and considering
inelasticities as small as $0.5$. 

For a given inelasticity $y$, the LHC would only be able to make black holes at masses
$M_{min}\leq y E_{LHC}$.  When one calculates rates for cosmic rays to produce black holes, those rates increase with $y$.  Thus, for the purposes of setting lower bounds on those rates, for a given $M_{min}$, a conservative choice is to take the inelasticity to be the smallest value that would be compatible with LHC black hole production, 
\be\label{yLHC}
y=M_{min}/E_{LHC}\ .
\ee
 If indeed the inelasticity took this value, that would correspond to {\it zero} production at LHC, due to lack of kinematical range.  For example, if the actual inelasticity were $0.5$, LHC would produce no black holes above $7~TeV$.  For purposes of exploring the possible range of $y$, we will where appropriate let $y$ range from $0.5$ up to unity.

A related issue is what is the minimum black hole mass that can arise
for a given value of the extra-dimensional Plank mass $M_D$. Several
criteria are discussed in~\cite{Giddings:2001bu}; there it was
advocated that one particularly useful criterion is that the entropy
of the black hole be large, so that a thermal approximation begins to
make sense. A non-rotating hole of mass $M$ has entropy:
\be
S_{BH}= \frac{R^{D-2} \, \Omega_{D-2}}{4G_D} =
\left[ \frac{2M}{(D-2)M_D} \right]^{(D-2)/(D-3)} \;
\left[ \frac{(2\pi)^{(2D-7)}}{\Omega_{D-2}} \right]^{1/(D-3)} \; .
\ee
For example, for the representative cases of
$D=6$ and $D=10$, a black hole with mass $M=5 M_D$ has entropy
$S_{BH}\sim 24$, a plausible threshold to assume a semiclassical
behaviour. Since for a fixed value of $M$ the black hole radius, and
thus the production cross sections, decrease with increasing $M_D$, to
be conservative in our estimates of production rates by cosmic rays we
shall loosen this constraint, and allow for $M_{min}/M_D$ to be as small as
3.  In the primary cases of interest for cosmic ray bounds, $D\leq7$, this lowest value corresponds to a Schwarzschild radius that is less than twice the Planck radius, $1/M_D$.

The production cross sections at the LHC are then obtained from the
simple formula:
\be \label{eq:LHC-BHrate}
\sigma_{BH}(M>M_{min}) = \sum_{ij} 
\int_{\tau_{min}}^1 \,
d\tau \int_{\tau}^{1} \, \frac{dx}{x} 
f_i(x) f_j(\tau/x) \hat\sigma(\rshat) \; ,
\ee
where ${\hat s} =x_1 x_2 s$, and, as discussed above, $\hat\sigma(\rshat)=\pi 
R^2(\rshat)/4 $,  
$M_{min}=3 M_D$, and
\be
\tau=x_1 x_2 > \tau_{min}=M_{min}^2/(y^2 s) \; .
\ee
$x_{1,2}$ are the
momentum fractions of the colliding nucleons carried by the partons 
$i$ and $j$,  which are taken to span the full
set of quarks, antiquarks and gluons.
For the numerical evaluations throughout this work
we shall use the CTEQ6M set of parton
distribution functions~\cite{Pumplin:2002vw}, calculated at the
factorization scale $Q=1/R$~\cite{Giddings:2001bu}.

The event rates at the LHC, integrated over the 1000~fb$^{-1}$
luminosity expected to be collected during the experiments lifetime,
are given in fig.~\ref{fig:lhc}. We show here both the cases of $y=1$
and of $y=0.5$.  We use $y=1$ only to overcompensate for uncertainty in the precise value of this parameter; the value $y=1$ is an extreme case, and recent analyses\cite{Giddings:2007nr} have argued for the more realistic value $y=0.6-0.7$, or even a lower value\cite{Meade:2007sz}.

\begin{figure}
\begin{center}
\includegraphics[width=0.68\textwidth,clip]{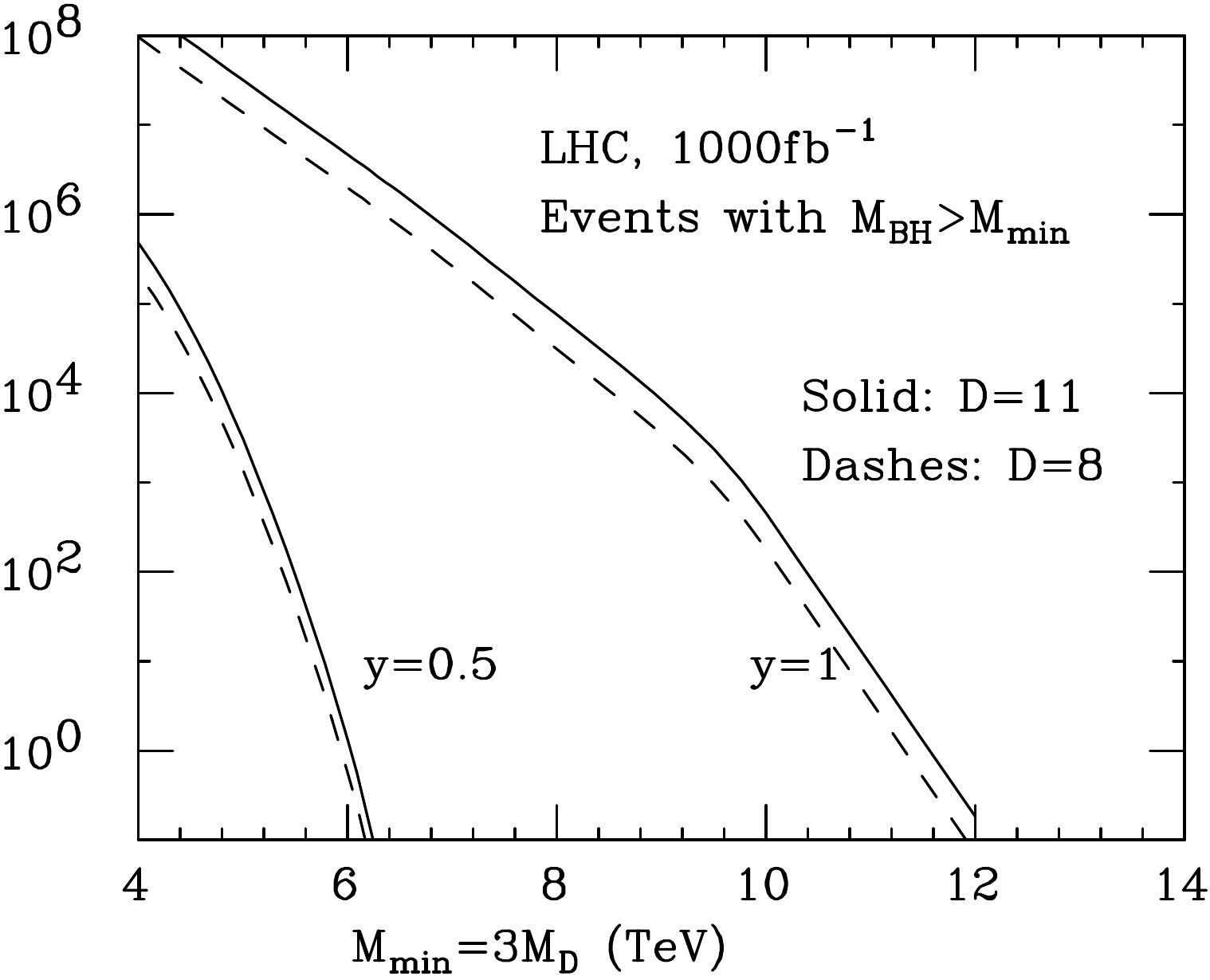}
\end{center}
\vskip -1cm
\ccaption{*}{\label{fig:lhc} \it Black hole production rates at
  the LHC, with inelasticity $y=1$ (upper curves) and $y=0.5$ (lower
  curves), and for $D=8,11$. }
\end{figure}

\subsection{Black hole production by cosmic rays}
\label{app:bhrates}
We present here the estimates of production rates and properties of
black holes produced by cosmic rays hitting the surface of
astronomical bodies. The rates presented in this section refer
to the exposure of the full area of the star to cosmic
rays coming from zenith angles between 0$^o$ and 90$^o$. 
Possible sources of reduction of acceptance must be considered:
the stellar magnetic fields, and the need for the
black hole to traverse a sufficient amount of material in order to
stop inside the star. These effects will be discussed in appropriate parts of the main text.

We assume cosmic rays to be composed of nuclei
with atomic number $A$, and will consider the two extreme cases of
$A=1$ (protons) and $A=56$ (Fe). When $A>1$, only a fraction $1/A$ of
the primary energy is available for the nucleon entering the hard
collision that will produce the black hole, thus leading to a
significant reduction in rate. 
The kinematics of the production is
therefore as follows: 
\be 
A(E)+N(m_p) \to i(x_1 E/A) + j(x_2 m_p) \to BH(M^2=2y^2 x_1 x_2 m_p E/A) \;
\ee
where $E$ is the energy of the cosmic ray primary; $N(m_p)$ is a
nucleon in the target, with rest energy $m_p$.  For each value of the primary
energy $E$, the kinematics for the production of a black hole with
minimum mass $M_{min}$ is 
defined by the constraints $E>E_{min} \equiv M^2_{min} \times A /
(2m_p \, y^2)$ and $\tau=x_1 x_2 > \tau_{min} \equiv 
E_{min}/E$. The resulting number of produced black holes (per area, per time) is then 
expressed in terms
of the cosmic ray flux $d\Phi/dE$ and by the total nucleon-nucleon 
inelastic cross section, $\sigma_{NN}=100$~mb, as follows:
\be \label{eq:CR-BHrate}
N_{BH}(M>M_{min}) = A \frac{1}{\sigma_{NN}}\int_{E_{min}}^{E_{max}} 
\frac{d\Phi}{dE} \; dE \; \sum_{ij} 
\int_{\tau_{min}}^1 \,
d\tau \int_{\tau}^{1} \, \frac{dx}{x} 
f_i(x) f_j(\tau/x) \hat\sigma(\rshat) \; .
\ee

We describe the cosmic ray energy spectrum $d\Phi/dE$ using the latest
results by Auger~\cite{Yamamoto:2007xj},  using the reported values of
the flux, and linearly interpolating in energy. We 
allow $E_{max}$ to  extend  only  up to the largest value
for which data exist, namely $E_{max}=2\times 10^{20}$~eV.

We are interested in excluding the existence of stable black holes
with masses within the reach of the LHC. The presence of an
inelasticity $y$ limits the mass reach to the range $M\lsim y\times
14$~TeV. We allow $y$ to take values in the range $(0.5,1)$ in order
to cover the full kinematic range up to 14~TeV, and 
calculate the
cosmic ray rates corresponding to the smallest possible inelasticity
compatible with production at a given mass value at the LHC, given by  (\ref{yLHC}), and
with the largest possible value of $M_D$, namely $M_D=M_{min}/3$. By
so doing we obtain the lowest possible cosmic-ray-induced rates
for black holes of any given mass that can be produced at the LHC.

\begin{figure}
\begin{center}
\includegraphics[width=0.48\textwidth,clip]{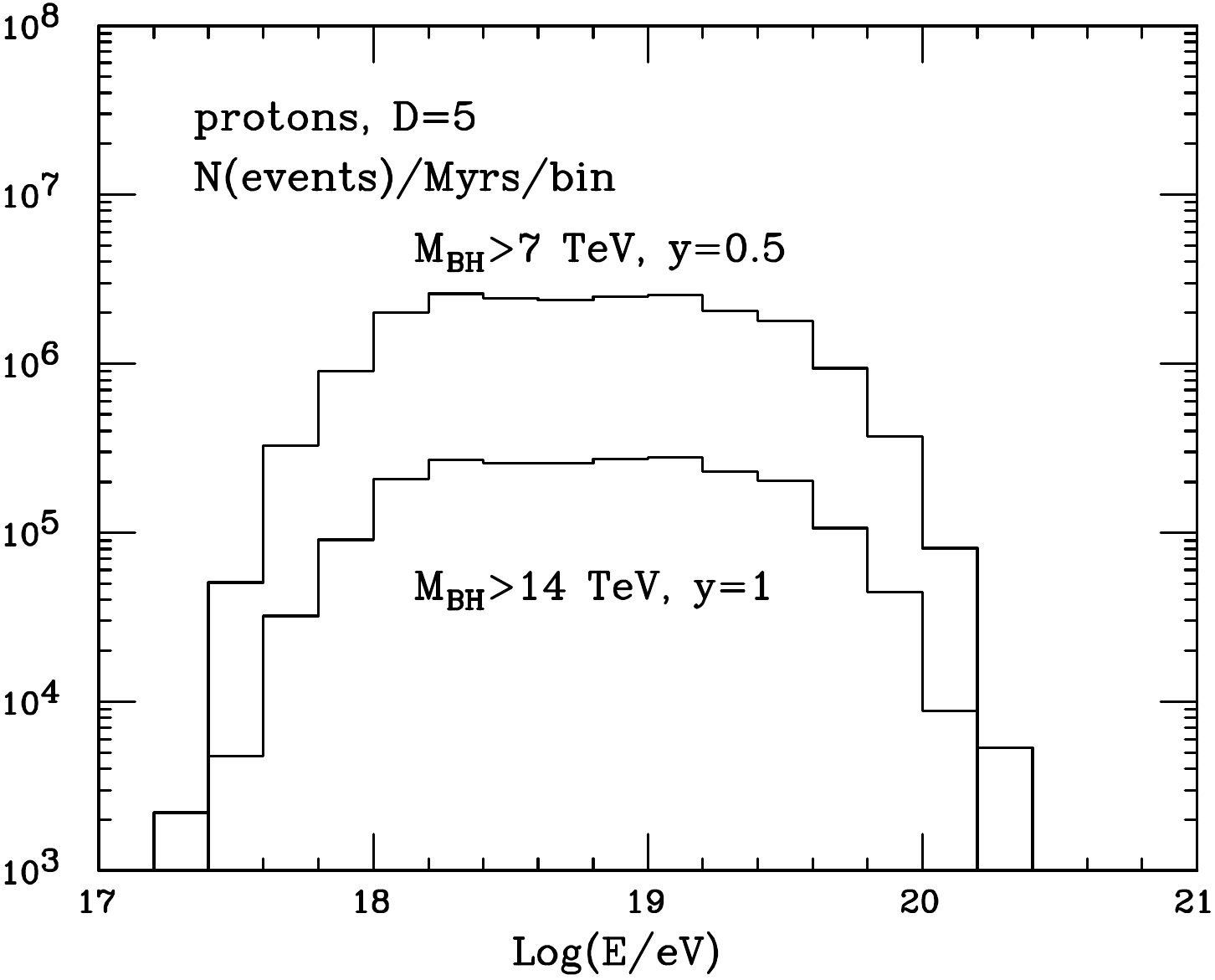}
\hfill
\includegraphics[width=0.48\textwidth,clip]{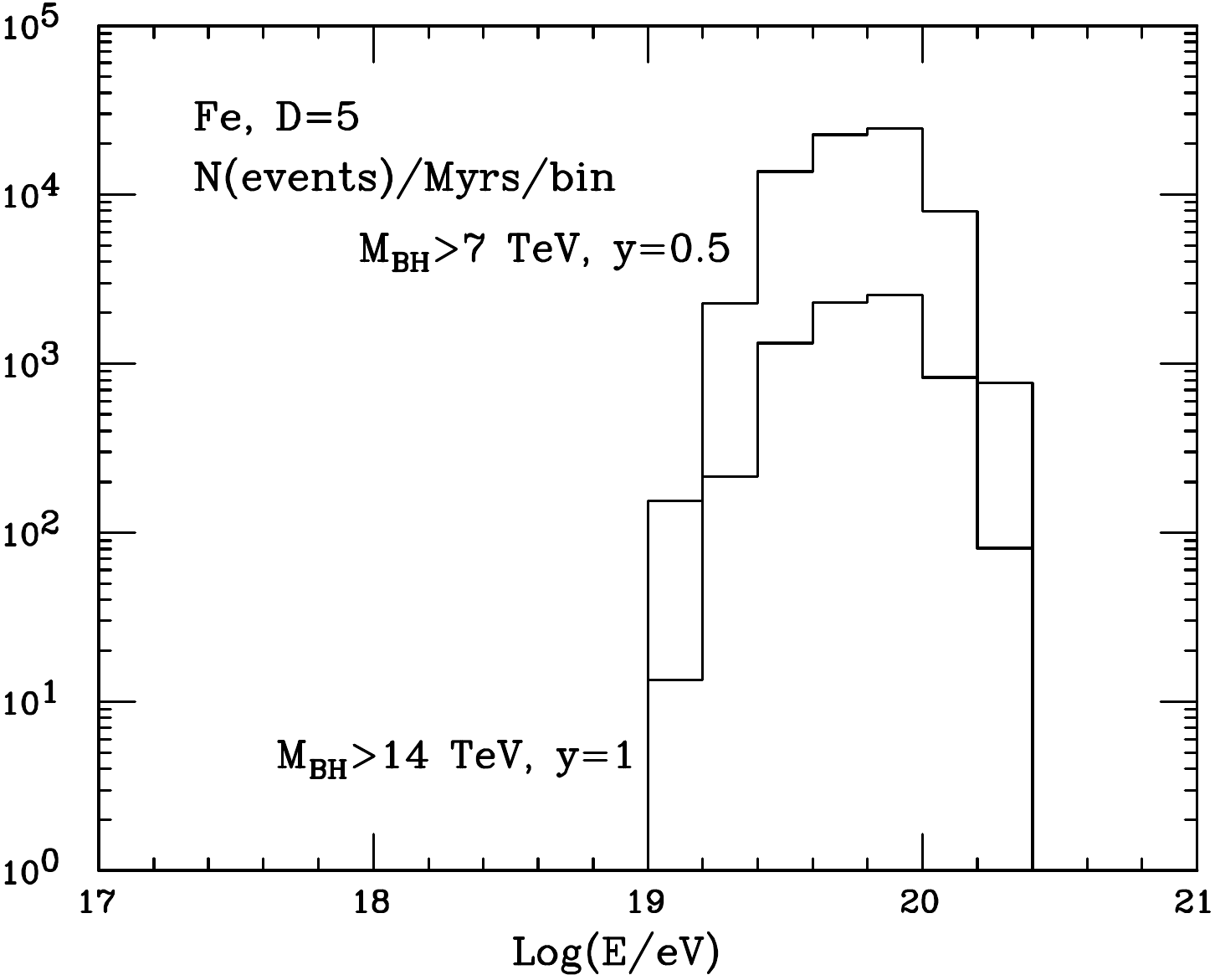}
\end{center}
\vskip -1cm
\ccaption{*}{\label{fig:wdrate} \it Black hole production rates
  by cosmic rays impinging on a 5400~km radius white dwarf. The rates
  correspond to the number of events, in one million years, in each energy
  bin. On the left we have a pure-proton cosmic ray composition, on
  the right pure Fe. The two curves correspond to minimum mass values of
  7 and 14~TeV. In all cases,  $M_D=M_{min}/3$ and
  $y=M_{min}/14$~\tev.}
\end{figure}

As a reference, we shall consider the case of a white dwarf, with
radius of 5400~km. The production rates, as a function of the energy
of the primary cosmic ray, and integrated over an exposure of the full
star surface of 1 million years, are shown in
fig.~\ref{fig:wdrate}. They are given for the case where the
production rate is the lowest, namely $D=5$, and for the values of
$M_{min}=7$~TeV, corresponding to inelasticity $y=0.5$, and
$M_{min}=14$~TeV, with $y=1$.  The integrated rates for $D=5-7$
 are shown in Table~\ref{tab:wdrate-p} (for a pure
proton composition) and in Table~\ref{tab:wdrate-Fe} (for a pure
Fe composition).  The rates
increase  due to the increasing black hole radius (for a given $M$, $M_D$) with larger $D$.  
{\renewcommand{\arraystretch}{1.1}
\begin{table}
\begin{center}
\ccaption{*}{\label{tab:wdrate-p} \it 
Black hole production rates, per million years, induced by proton cosmic rays
impinging on a $R=5400$~km white
dwarf. $M_D=M_{min}/3$ and $y=M_{min}/14~\tev$.}
\vskip 2mm
\begin{tabular}{l|lll}
\hline \hline
$D=$   &  5 & 6 & 7 \\ \hline 
$M_{min}=7$~\tev
&   $2.1 \times 10^7$ & $4.3 \times 10^7$  & $6.7 \times 10^7$ \\ 

$M_{min}=8$~\tev
&   $1.4 \times 10^7$ & $2.9 \times 10^7$  & $4.7 \times 10^7$ \\ 

$M_{min}=10$~\tev
&   $6.7 \times 10^6$ & $1.5 \times 10^7$  & $2.6 \times 10^7$ \\ 

$M_{min}=12$~\tev
&   $3.7 \times 10^6$ & $9.1 \times 10^6$  & $1.6 \times 10^7$ \\ 

$M_{min}=14$~\tev
&   $2.3 \times 10^6$ & $5.9 \times 10^6$  & $1.0 \times 10^7$ \\ 
\hline \hline

\end{tabular}
\end{center}
\end{table}}

{\renewcommand{\arraystretch}{1.1}
\begin{table}
\begin{center}
\ccaption{*}{\label{tab:wdrate-Fe} \it 
Black hole production rates, per million years, induced by Fe cosmic rays
impinging on a $R=5400$~km white
dwarf. $M_D=M_{min}/3$ and $y=M_{min}/14~\tev$.}
\vskip 2mm
\begin{tabular}{l|lll}
\hline \hline
$D=$   &  5 & 6 & 7 \\ \hline 
$M_{min}=7$~\tev
&   $7.2 \times 10^4$ & $1.6 \times 10^5$  & $2.6 \times 10^5$ \\ 

$M_{min}=8$~\tev
&   $4.6 \times 10^4$ & $1.1 \times 10^5$  & $1.8 \times 10^5$ \\ 

$M_{min}=10$~\tev
&   $2.2 \times 10^4$ & $5.5 \times 10^4$  & $9.7 \times 10^4$ \\ 

$M_{min}=12$~\tev
&   $1.2 \times 10^4$ & $3.2 \times 10^4$  & $5.9 \times 10^4$ \\ 

$M_{min}=14$~\tev
&   $7.3 \times 10^3$ & $2.1 \times 10^4$  & $3.8 \times 10^4$ \\ 
\hline \hline

\end{tabular}
\end{center}
\end{table}}

As expected
the rates for pure protons are much larger, since in the case of Fe,
to achieve sufficient energy for the nucleon-nucleon collisions, one
is forced to use cosmic rays in the tail of the data distribution.
Compositions intermediate between protons and Fe will lead to
distributions contained within these two extremes. In particular, it
is straightforward to evaluate the production rates resulting from
some specified fraction of cosmic-ray protons, by convoluting the
rates we show in fig.~\ref{fig:wdrate}, 
bin-by-bin, with the experimental determination of the
proton fraction as a function of energy. Current experimental data on
the penetration and development of the shower (see
e.g.~\cite{Unger:2007mc,Hires}) provide evidence for a
mixed composition, at least in the region where such data are
statistically significant, namely below $4 \times 10^{19}$eV. Data are
inconsistent with being fully protons, or fully Fe, and provide an
estimate of $\langle A \rangle \sim 5$~\cite{Matthiae:2007zz}. 
On the other hand, the
uncertainty of these analyses is such that one cannot separate the
individual components that contribute to the average of $\langle A
\rangle$. Phenomenological descriptions or theoretical models 
of the  highest energy
cosmic ray sources,
fitted~\cite{Anchordoqui:2007fi,Hooper:2008pm,Arisaka:2007iz,Dar:2006dy} 
using the
latest Auger spectra, 
as well as the Auger data\cite{Abraham:2007si,Cronin:2007zz} 
on the correlation
between the origin of cosmic rays around the GZK cutoff and remote
AGN,
point to a significative proton
fraction, of the order of at least 10\%, and higher at super-GZK
energies.

Proton fractions  as low as only  10\% lead to huge black hole
production rates, with sufficient accumulation inside a white dwarf
within a few years.

We point out that even assuming the most pessimistic scenario in which
100\% of the cosmic rays are made of Fe, a scenario that is
inconsistent with both data and with the modeling of cosmic ray
sources, there would still be a large number of black holes that can
be accumulated on the timescale of a few thousand years, very short
compared to the natural lifetime of a white dwarf. As a further
robustness check, we provide in Table~\ref{tab:wdrate-resc} the rates
obtained if one assumes a  20\%  systematic overestimate in the extraction of the
primary energy, to simulate the  possible  impact of the Auger
$\pm 20\%$ energy resolution~\cite{Ridky:2007zz}.
{\renewcommand{\arraystretch}{1.1}
\begin{table}
\begin{center}
\ccaption{*}{\label{tab:wdrate-resc} \it Black hole production rates, per
  million years, induced by cosmic rays impinging on a $R=5400$~km
  white dwarf, with the cosmic ray energies rescaled such that
  $E_{exp}=1.2 \times E_{true}$. $y=M_{min}/14~\tev$.}
\vskip 2mm
\begin{tabular}{l|lll}
\hline \hline
$D=$   &  5 & 6 & 7 \\ \hline 
$N_p$/Myr, $M_{min}=7$~\tev
&   $1.2\times 10^7$ & $2.5\times 10^7$  & $3.9\times 10^7$\\
$N_{Fe}$/Myr, $M_{min}=7$~\tev
 &  $3.2\times 10^4$  & $7.0\times 10^4$ & $1.2\times 10^5$  \\
\hline
$N_p$/Myr, $M_{min}=14$~\tev
&   $1.3\times 10^6$ & $3.4\times 10^6$ & $6.0 \times 10^6$ \\
$N_{Fe}$/Myr, $M_{min}=14$~\tev
 &  $3.2\times 10^3$  & $9.0\times 10^3$ & $1.7\times 10^4$  \\
\hline \hline
\end{tabular}
\end{center}
\end{table}}

Finally, we note that the production rates are still significant even
if we set the inelasticity $y=0.5$ in the calculation of the cosmic
ray rates for black holes of 14~TeV. Notice that with $y=0.5$ the
black hole mass reach at the LHC drops to zero at around
$M=7$~TeV. The number of produced black holes on our reference white
dwarf is given in table~\ref{tab:wdrate05}, 
and the distribution as a function of the
cosmic ray energy is shown in fig.~\ref{fig:wdrate05}.
{\renewcommand{\arraystretch}{1.1}
\begin{table}
\begin{center}
\ccaption{*}{\label{tab:wdrate05}
\it Black hole production rates, per million years, induced by cosmic rays
  impinging on a $R=5400$~km white
  dwarf. $N_p$ refers to the case of 100\% proton composition,
  $N_{Fe}$ refers to 100\% Fe.  $M_D=M_{min}/3$ and inelasticity
  $y=0.5$.} 
\vskip 2mm
\begin{tabular}{l|lll}
\hline \hline
$D=$   &  5 & 6 & 7  \\ \hline
$N_p$/Myr, $M_{min}=7$~\tev
&   $2.1\times 10^7$ & $4.3\times 10^7$  & $6.7\times 10^7$\\
$N_{Fe}$/Myr, $M_{min}=7$~\tev
 &  $7.2\times 10^4$  & $1.6\times 10^5$ & $2.6\times 10^5$  \\
\hline
$N_p$/Myr, $M_{min}=14$~\tev
&   $2.8\times 10^5$ & $5.7\times 10^5$  & $9.1\times 10^5$\\
$N_{Fe}$/Myr, $M_{min}=14$~\tev
 &   35 &  80 & 135  
\\
\hline \hline
\end{tabular}
\end{center}
\end{table}}

\begin{figure}
\begin{center}
\includegraphics[width=0.68\textwidth,clip]{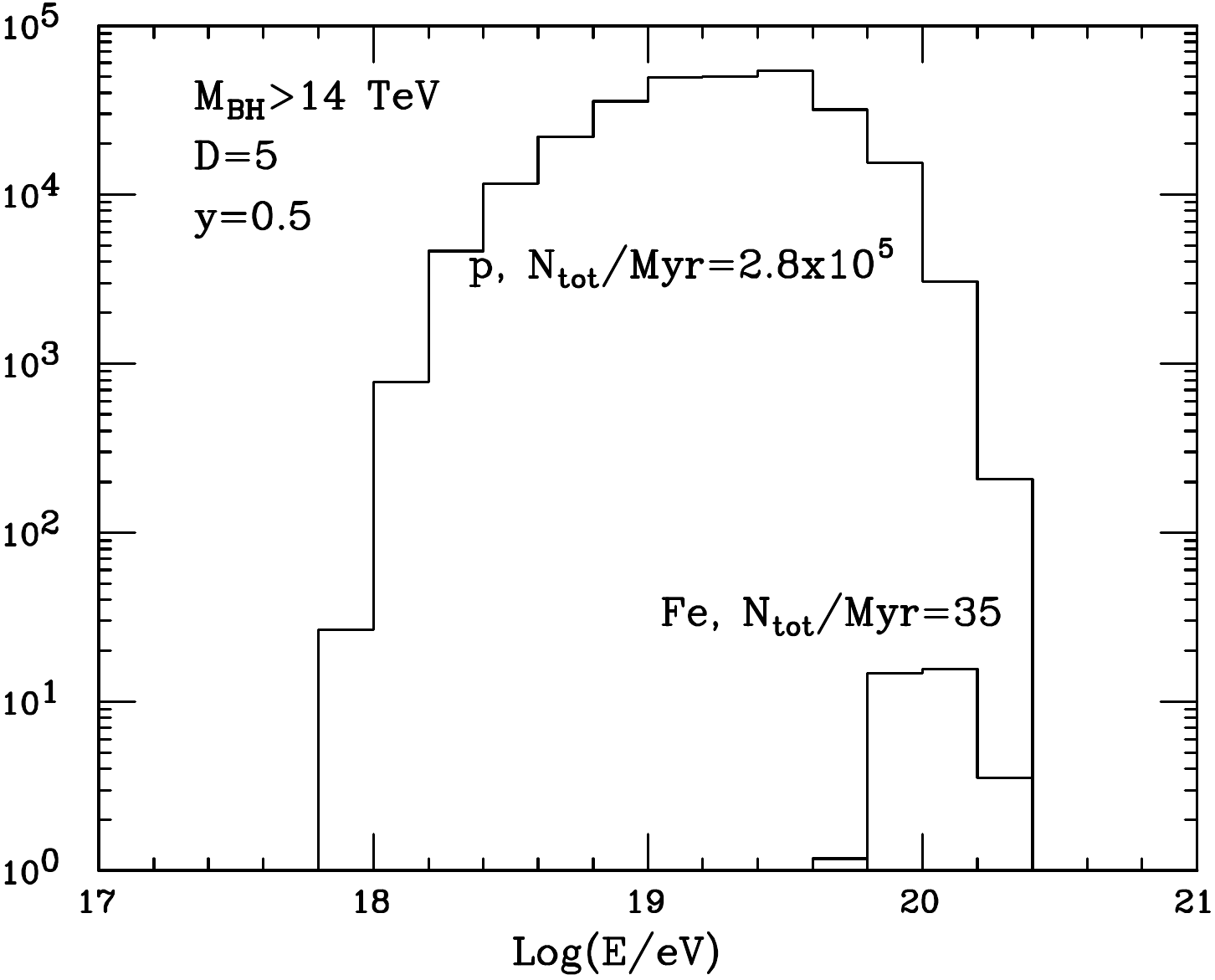}
\end{center}
\vskip -1cm
\ccaption{*}{\label{fig:wdrate05} \it Black hole production rates
  by cosmic rays impinging on a 5400~km radius white dwarf in the
  extreme case of  $M>14$~\tev\ and with
  the inelasticity parameter  $y=0.5$. 
  The upper curve corresponds to a pure proton cosmic-ray
  composition, the lower one to pure Fe.}
\end{figure}

For the discussion of slow-down and stopping of the black hole inside
the white dwarf, we also present here some relevant kinematical
distributions. 
Figure~\ref{fig:gamma} shows the distributions of the
Lorentz factor, $\gamma=E/M$, of the produced black hole, in the case
of proton (left) and of Fe (right) primaries. We present in
table~\ref{tab:wdeff} the rate-suppression factors due to requiring
that the produced black hole has a Lorentz factor
$\gamma<3M_{min}/m_p$, and a mass $M<M_{min}+1~\tev$, the criteria we
used in section~\ref{sec:WDstop}, to determine the stopping power of
white dwarfs. Notice that in all cases these efficiencies are large
enough to ensure abundant rates of produced black holes. 

\begin{figure}
\begin{center}
\includegraphics[width=0.48\textwidth,clip]{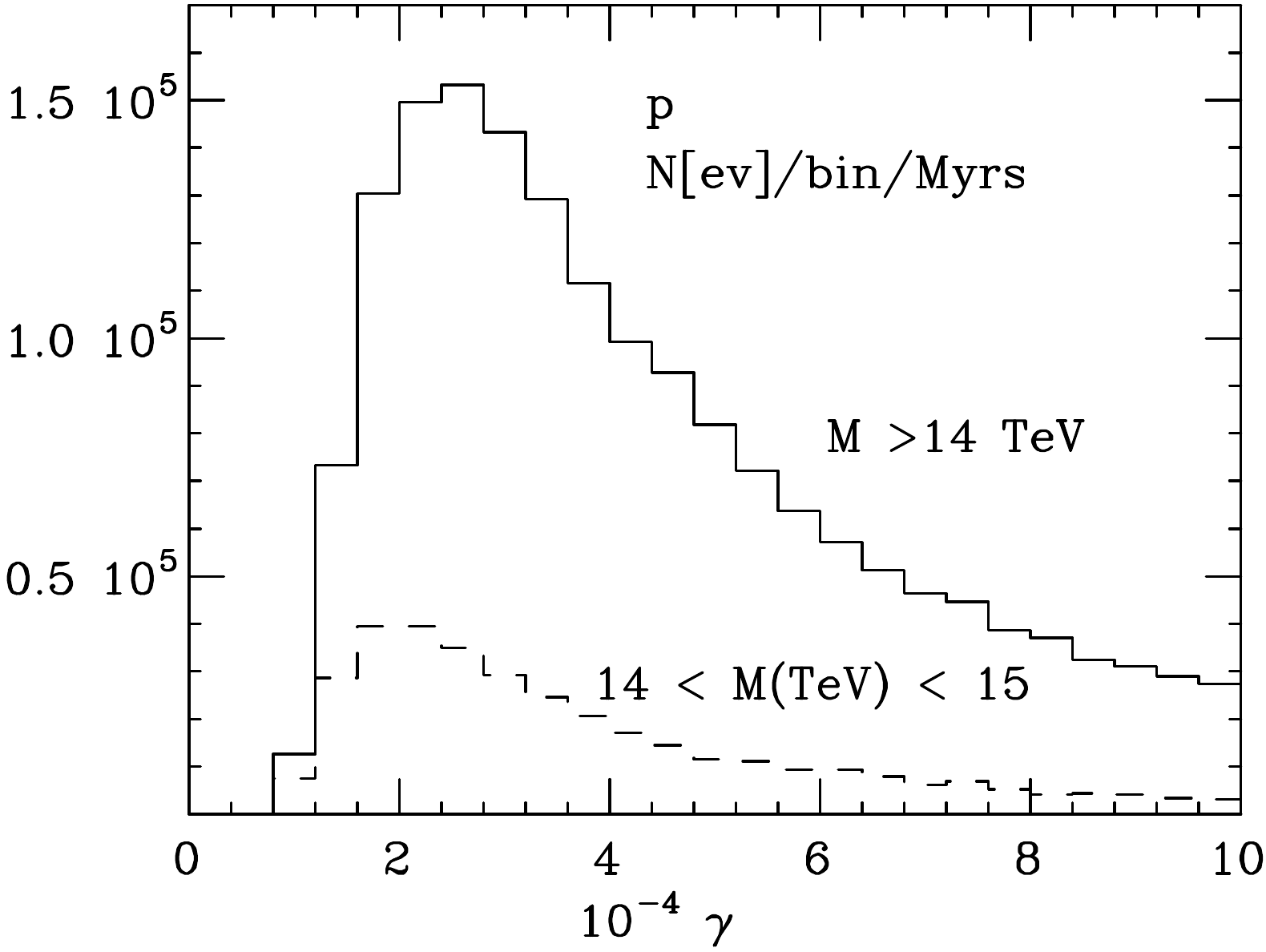}
\hfill
\includegraphics[width=0.48\textwidth,clip]{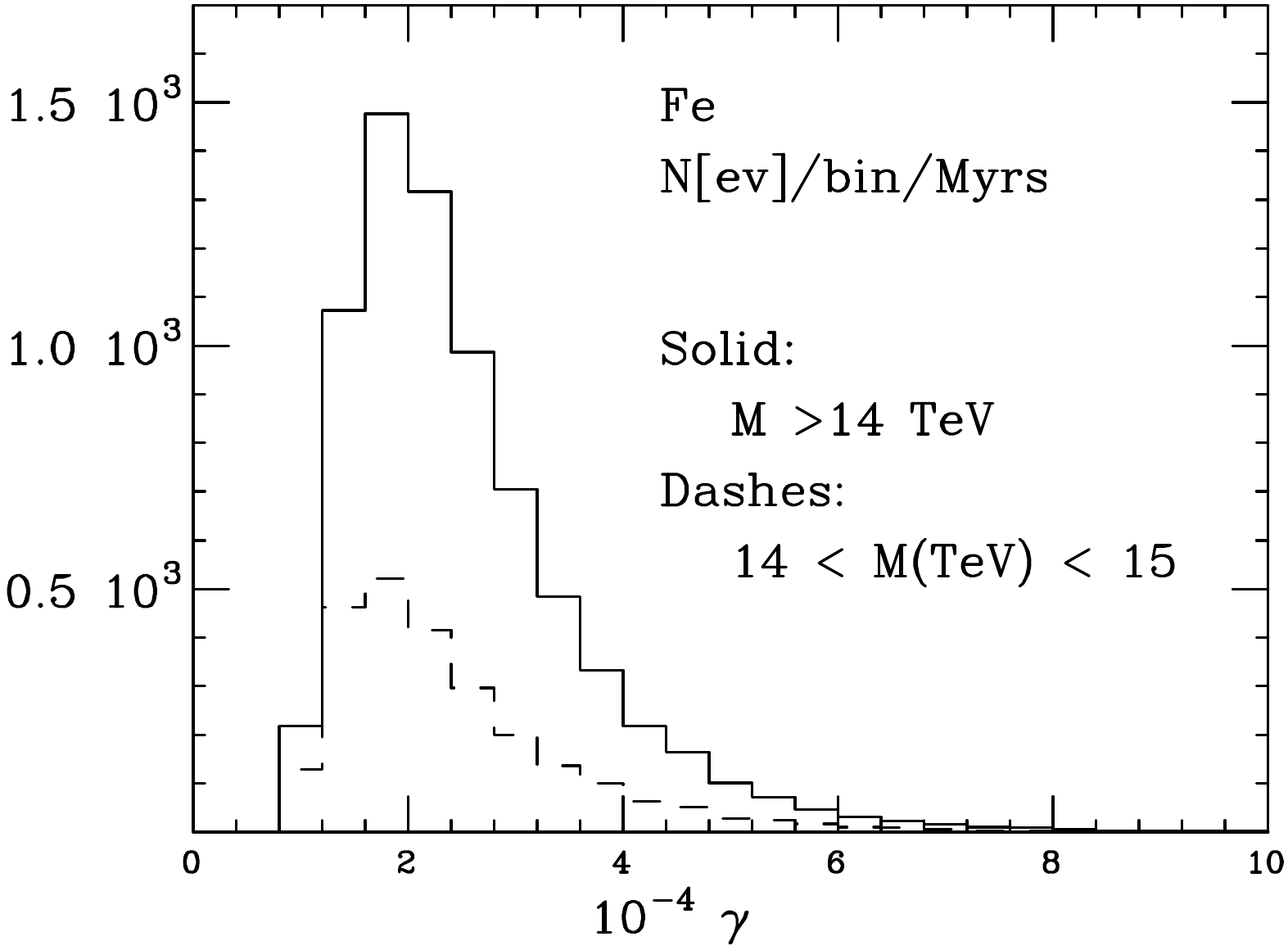}
\end{center}
\vskip -1cm 
\ccaption{*}{\label{fig:gamma} \it The Lorentz factors
$\gamma$, in units of $10^4$, for black holes with $M>14$~\tev
produced by proton (left) and Fe (right) cosmic rays. The lower curves
represent the distributions relative to the events with $14<M(\tev)<15$.}
\end{figure}

{\renewcommand{\arraystretch}{1.1}
\begin{table}
\begin{center}
\ccaption{*}{\label{tab:wdeff}
\it Efficiency factors $\epsilon_p$ (100\% proton flux) and
$\epsilon_{Fe}$ (100\% Fe flux) for the production of black holes with
$M<M_{min}+1~\tev$ and $\gamma<3M_{min}/m_p$. 
$M_D=M_{min}/3$, and inelasticity $y=M_{min}/14~TeV$ for all columns
except the last one, which has $y=0.5$.}
\vskip 2mm
\begin{tabular}{l|llllll}
\hline \hline
$M$(TeV)   &  7     &    8  & 10 & 12 & 14 & 14 ($y=0.5$)   \\ \hline
$\epsilon_p$
  &   $6.9 \times 10^{-2}$
  &   $8.3 \times 10^{-2}$
  &   $9.8 \times 10^{-2}$
  &   0.10
  &   0.10
  &   $4.4 \times 10^{-2}$
\\
$\epsilon_{Fe}$
  &   0.28   
  &   0.32
  &   0.35
  &   0.34
  &   0.32
  &   0.47  
\\
\hline \hline

\end{tabular}
\end{center}
\end{table}}

The production rates on a neutron star (neglecting the magnetic screening)
can be obtained from the white
dwarf's ones by rescaling by the surface area. Assuming a 10~km
radius, the proton rates in Table~\ref{tab:wdrate-p} are reduced by a factor of
$3.4\times 10^{-6}$, leading to the numbers in
Table~\ref{tab:nsrate}. The distributions as a function of the
cosmic ray energy have the same shape as those in the white dwarf
cases, fig.~\ref{fig:wdrate}. 
{\renewcommand{\arraystretch}{1.1}
\begin{table}
\begin{center}
\ccaption{*}{\label{tab:nsrate} \it 
Black hole production rates, per million years, induced by proton cosmic rays
impinging on a $R=10$~km neutron star. 
$M_D=M_{min}/3$ and $y=\rm{max}(0.5,M_{min}/14~\tev)$.  }
\vskip 2mm
\begin{tabular}{l||lllllll}
\hline \hline
$M_{min}$   &  $D=5$ & $D=6$ & $D=7$ & $D=8$ &    $D=9$   &   $D=10$    &
$D=11$
 \\ \hline 
3~\tev
& $1.3\times 10^4$ & $2.5\times 10^4$ & $4.0\times 10^4$ & 
$5.6\times 10^4$ & $ 7.4 \times 10^4$ & $ 9.2\times 10^4$ & $1.1\times
10^5$ \\
4~\tev
& $2.2\times 10^3$ & $4.5\times 10^3$ & $7.0\times 10^3$ &
$9.9\times10^3$ & $1.3\times 10^4$ & $1.6\times10^4$ & $1.9\times10^4$ \\
5~\tev
& 570 & 1100 & 1800 & 2500 & 3300 & 4100 & 5000 \\
6~\tev
& 190 & 380 & 590 & 830 & 1100 & 140 & 1600 \\
7~\tev
& 72 & 146 & 231& 323 & 422 & 526 & 633 \\   
8~\tev
& 47 & 99 & 161 & 229 & 301 & 378 & 457 \\
10~\tev
& 23 & 52 & 88  & 129 & 172 & 218 & 265 \\
12~\tev
& 13 & 31 & 54  & 80  & 109 & 139 & 171 \\
14~\tev
& 8  & 20 & 36  & 54  & 74  & 95  & 118 \\
\hline \hline
\end{tabular}
\end{center}
\end{table}}
 The results for $D=5,8$ are summarized in fig.~\ref{fig:nsrate}, where
 the corresponding rates for Fe cosmic rays are also shown. 
\begin{figure}
\begin{center}
\includegraphics[width=0.68\textwidth,clip]{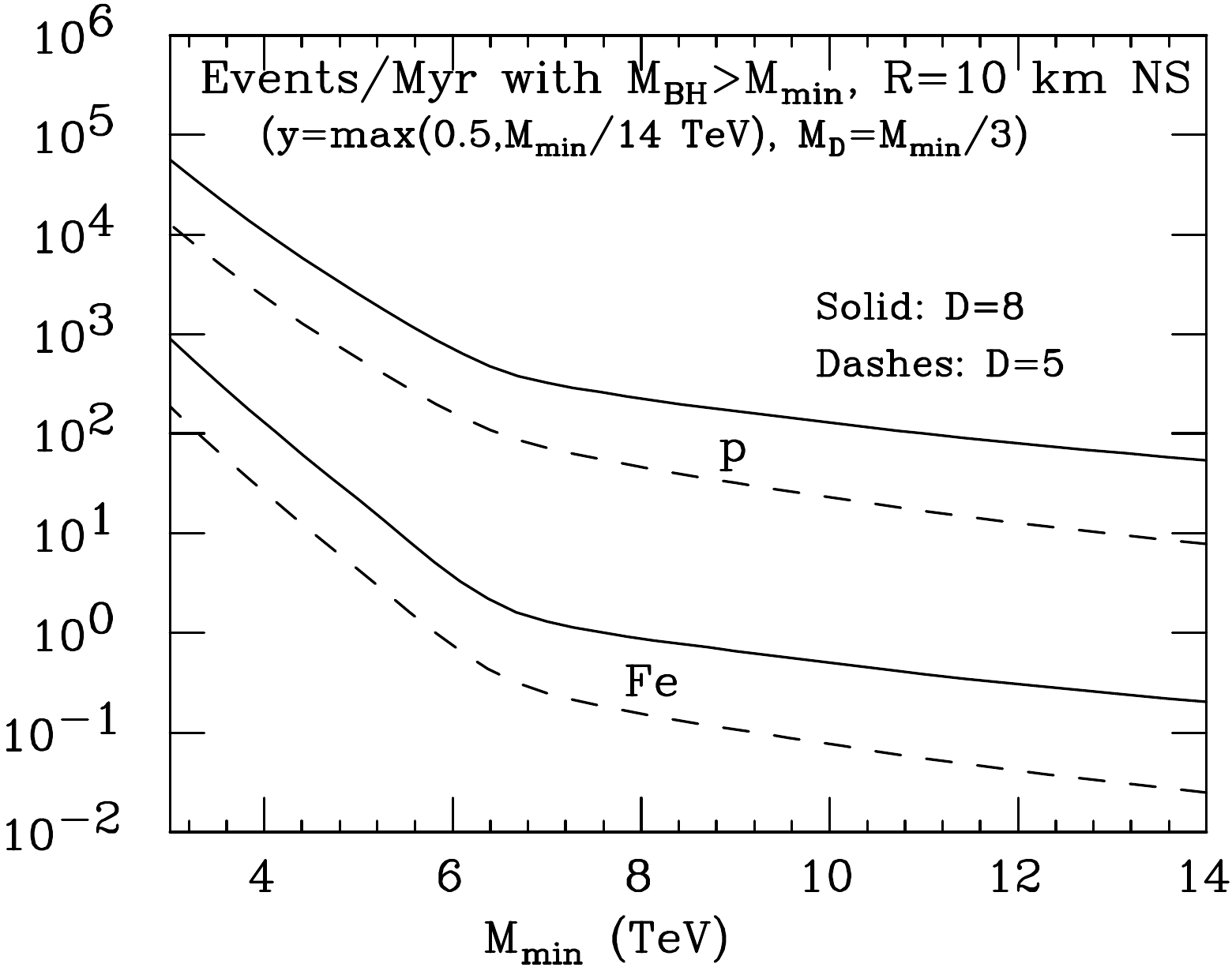}
\end{center}
\vskip -1cm
\ccaption{*}{\label{fig:nsrate} \it Black hole production rates
  by cosmic rays impinging on a 10~km radius neutron star, for
  the inelasticity parameter $y=max(0.5,M_{min}/14~\tev)$ and with
  $M_D=M_{min}/3$.  
  The upper curves correspond to a pure proton cosmic-ray
  composition, the lower ones to pure Fe.}
\end{figure}

 We conclude our discussion of production properties
by showing in fig.~\ref{fig:xdist} the $x$ spectrum of the partons
engaged in the production of black holes with $M>14~\tev$, 
for various cosmic ray
components and different inelasticity assumptions. Notice that the
bulk of the production is always obtained for $x\lsim 0.6$, namely the
region where the knowledge of the PDFs is accurate to better than 
10\%~\cite{Pumplin:2002vw}.

\begin{figure}
\begin{center}
\includegraphics[width=0.48\textwidth,clip]{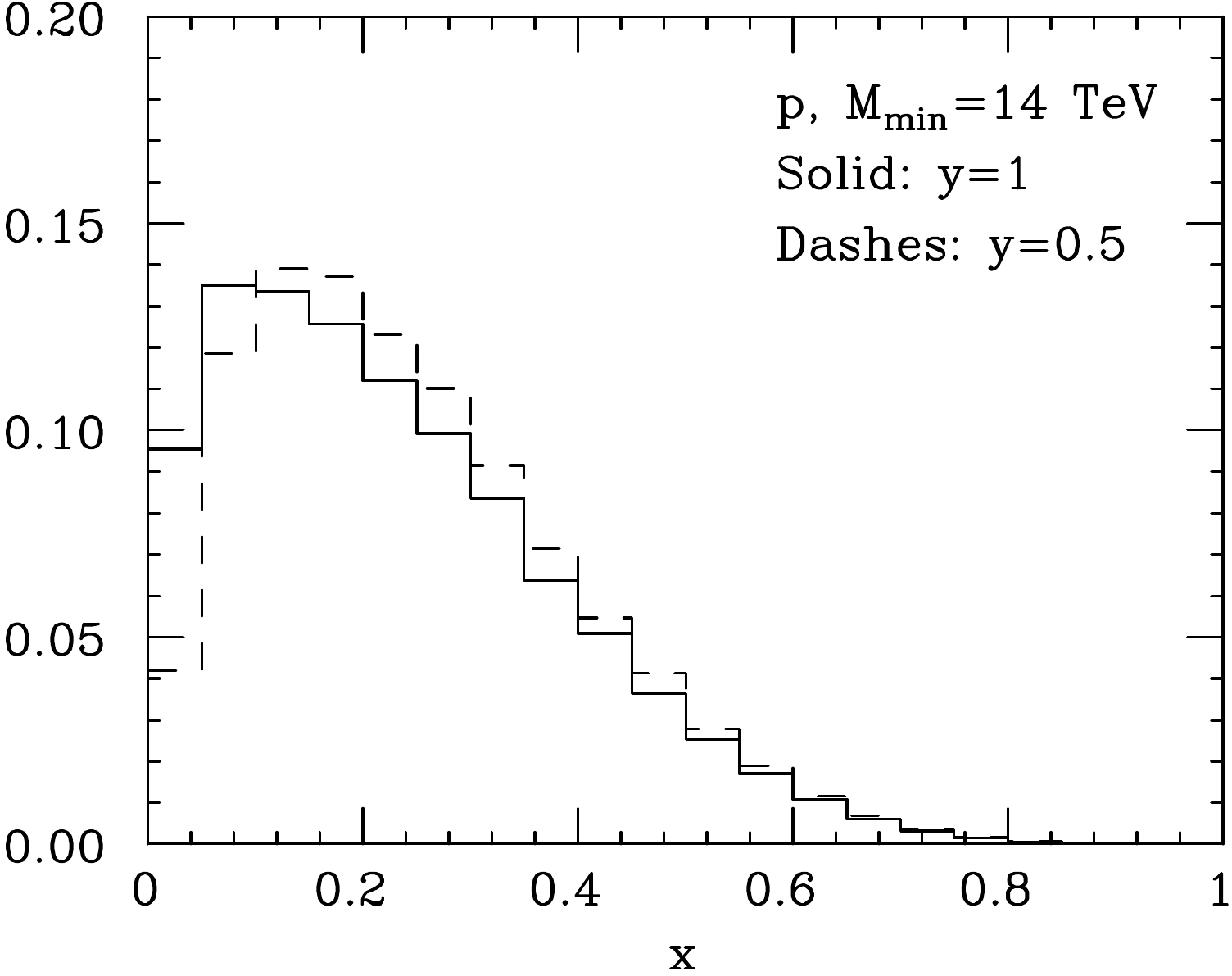}
\hfill
\includegraphics[width=0.48\textwidth,clip]{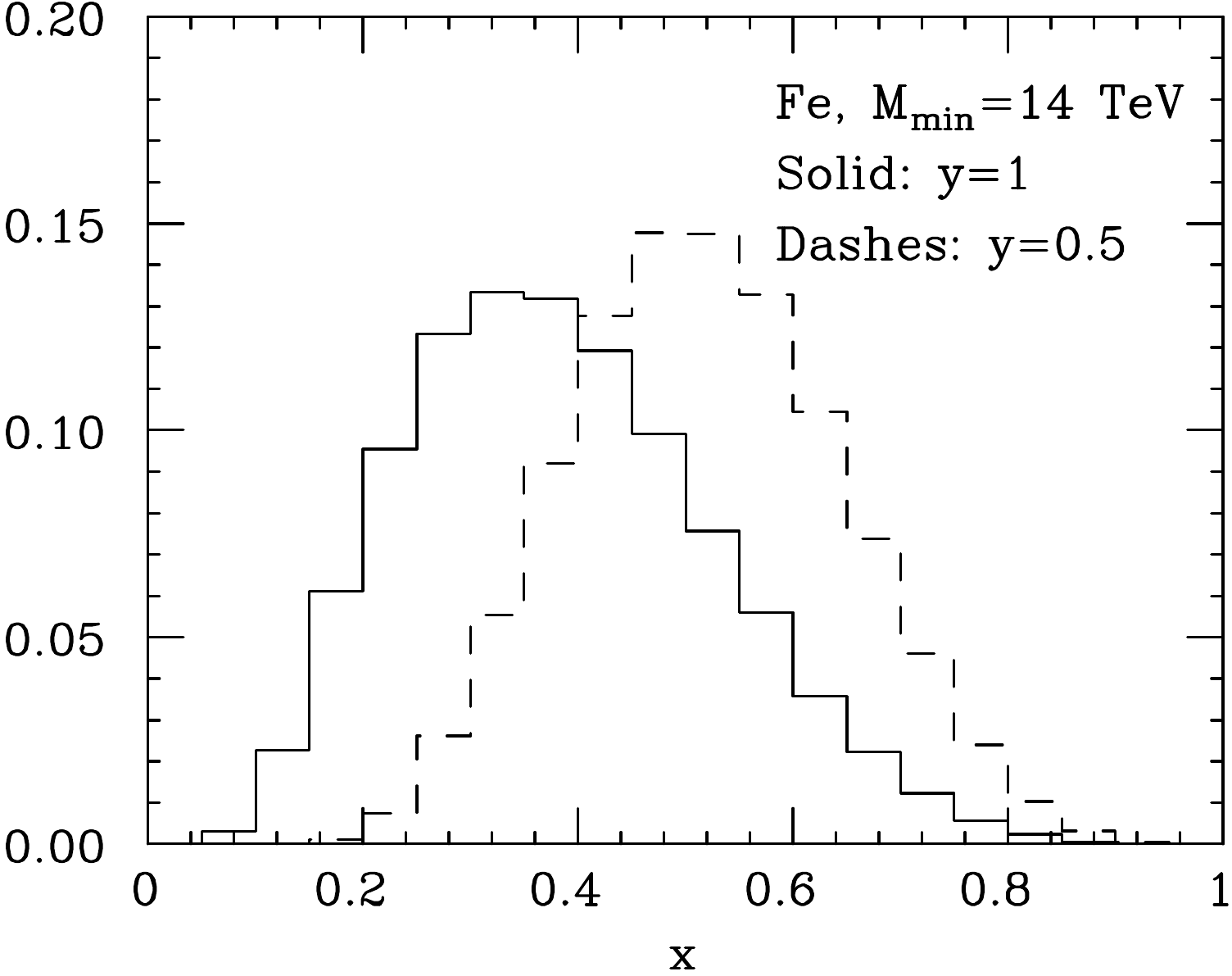}
\end{center}
\vskip -1cm 
\ccaption{*}{\label{fig:xdist} \it The normalized spectra of the
  partonic momentum fraction $x$, for proton (left) and Fe (right)
  cosmic rays, and for $y=1$ and $y=0.5$. $M_{min}=14$~TeV.}
\end{figure}

\subsection{Black hole production by cosmic neutrinos}
High-energy neutrinos have a black hole production rate much bigger
than nuclear cosmic rays of the same
energy (see
e.g.~\cite{Giddings:2001ih,Feng:2001ib,Anchordoqui:2001cg,Kowalski:2002gb, 
Ringwald:2001vk,Tyler:2000gt,Dutta:2002ca}).   
Two reasons justify this
statement. The first one is that
all the energy carried by the neutrino can be used for the production
of the black hole (contrary to protons or nuclei, where the partons
carry a small fraction of the primary energy). The second reason is
that the fraction of neutrino interactions that lead to black hole
production is much bigger than the similar fraction of nuclear interactions:
most collisions with a star
atmosphere result in generic strong-interaction processes, and the
rate of black hole production is proportional to the ratio of the black
hole cross section to the total inelastic cross section.  In the case
of neutrinos, black hole production only competes against the total
electroweak cross section~\cite{Dutta:2003fv}, 
which turns out to be of the same size as
the black hole cross section, when neutrino energies exceed
$10^{17}$~eV.

While no experimental evidence is as yet available for the existence
of such high-energy neutrinos, any modeling of cosmic ray production
and evolution predicts their presence, with rates that are consistent
with the current non-observation. Neutrinos can be produced directly at
the source of the highest energy cosmic rays, where they emerge as
decay products of the charged pions produced during the acceleration
of the primary charged cosmic
rays~\cite{Waxman:1998yy,Bahcall:1999yr}.  And they also would appear in the decay of pions produced by the collision of protons
with energy above the GZK cutoff with the cosmic microwave background:
$\gamma p \to \Delta^+ \to n\pi^+$ (the so-called {\it cosmogenic
neutrino flux}~\cite{Beresinsky:1969qj,Stecker:1978ah,Hill:1983xs}).
The observation of cosmic rays above the GZK cutoff, and the
assumption that even only a tiny fraction of these are protons,
automatically leads to the presence of neutrinos in the region above
$10^{14}$~eV. Several groups have extracted predictions for the
cosmogenic neutrino flux, most
recently~\cite{Engel:2001hd,Fodor:2003ph} and
\cite{Anchordoqui:2007fi}. The latter based their analysis on the
Auger study of the highest-energy cosmic ray
composition~\cite{Unger:2007mc}, and considered a broad range of
single and mixed compositions, including the extreme case of a Fe-only
spectrum at injection, in which case  the spectrum of protons dissociated from
the Fe nuclei would be soft enough to strongly suppress pion
photoproduction, and thus the neutrino flux.  For our calculations we
use the following parameterization of the neutrino flux: 
\be  \label{eq:nuflux} 
\frac{d\Phi_\nu(E_\nu)}{dE_{\nu}} = 10^{-7}
\,\left(\frac{\gev}{E_\nu}\right)^2 \;
\mathrm{m}^{-2}\mathrm{s}^{-1}\mathrm{sr}^{-1}\mathrm{GeV}^{-1}\; .
\ee 
This provides a lower limit to the acceptable fits of the worse-case, Fe-only,
scenarios considered in~\cite{Anchordoqui:2007fi}, in the region
$10^{17} \lsim E_\nu(\ev)\lsim 10^{19}$. We stress that this is a very
conservative assumption, and does not include the contribution from
neutrinos originating directly at the cosmic ray sources. A recent
evaluation of this contribution from AGN and
GRBs~\cite{Anchordoqui:2007tn}, for example, leads to a flux between
10 and 100 times larger than~(\ref{eq:nuflux}), over and beyond the range 
$10^{17} \lsim E_\nu(\ev)\lsim 10^{20}$.\footnote{See
also~\cite{Li:2007ps}, where the decay of muons produced in the
interactions of $\pi^0$ photons with the CMB photons leads to a flux
of comparable size.} We also notice that the flux in~(\ref{eq:nuflux})
is over three orders of magnitude below the Waxman-Bahcall upper
bound~\cite{Waxman:1998yy,Bahcall:1999yr}.

The corresponding black hole production
rate for a 10~km-radius neutron star is shown in fig.~\ref{fig:nu}, 
for $D=5$, $M_{min}=14$~\tev\ and $y=0.5$, as a function of the neutrino
energy.  The rates as a function of $D$ are given in
table~\ref{tab:nsrate-nu}.  Accumulation over more than  several hundred million
years of the life of a neutron star would lead to immense rates, even
if the cosmic neutrino flux turned out to be suppressed by several
orders of magnitude relative to our most conservative assumption. 
The direct detection of high-energy neutrinos in the next generation
of neutrino telescopes~\cite{Halzen:2002pg}
 will make it possible to strengthen these
estimates.

\begin{figure}
\begin{center}
\includegraphics[width=0.68\textwidth,clip]{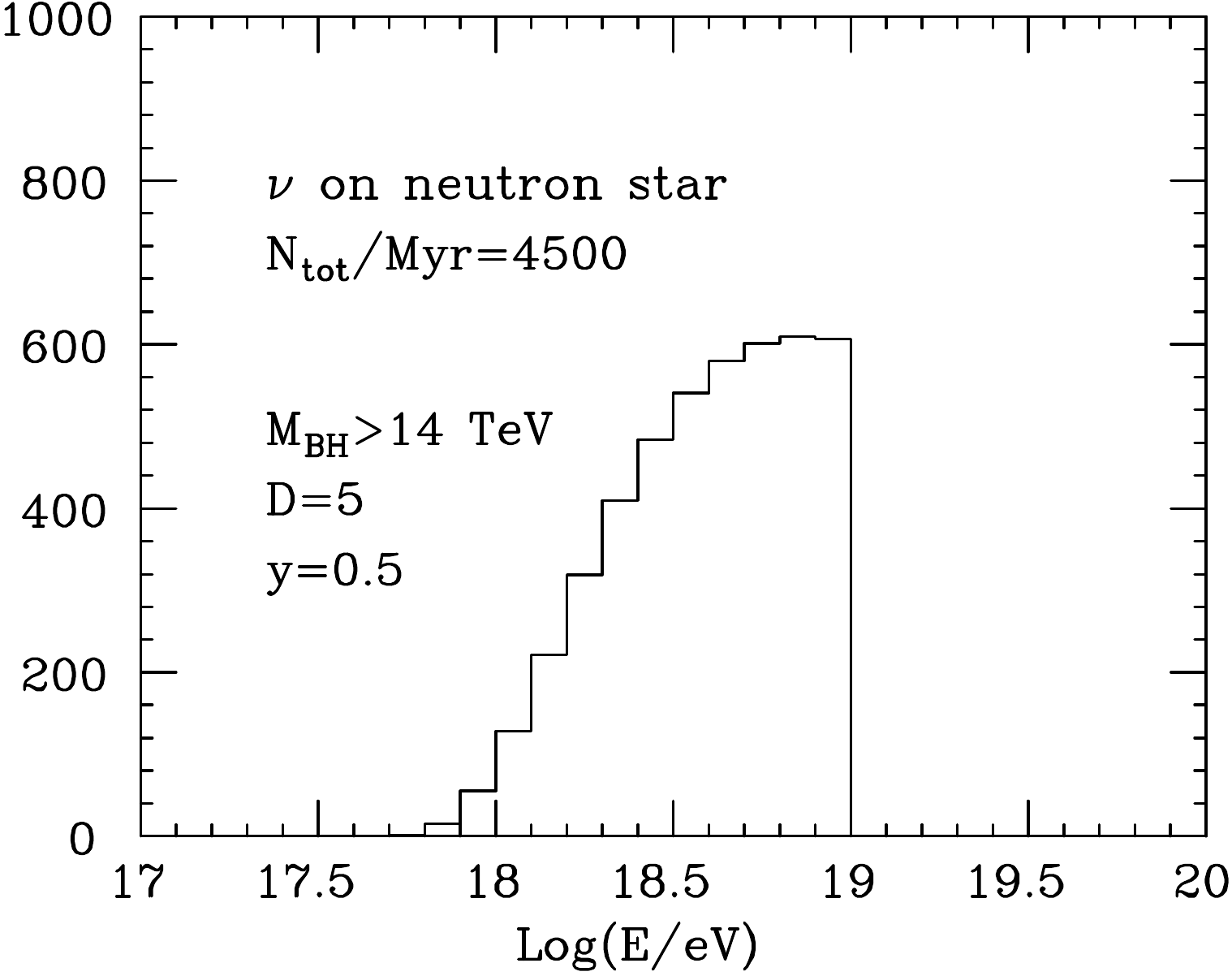}
\end{center}
\vskip -1cm 
\ccaption{*}{\label{fig:nu} \it Production rates for black holes of
  mass $M>14$~\tev, created
  by cosmic neutrinos impinging on a 10~km radius neutron star. The rates
  correspond to number of events, in one million years, in each energy
  bin.  $D=5$ and $y=0.5$.}
\end{figure}

{\renewcommand{\arraystretch}{1.1}
\begin{table}
\begin{center}
\ccaption{*}{\label{tab:nsrate-nu} \it 
Black hole production rates, per million years, induced by neutrino cosmic rays
impinging on a $R=10$~km neutron star. 
$M_{min}=14~\tev$, $M_D=M_{min}/3$ and $y=0.5$.}
\vskip 2mm
\begin{tabular}{l|lllllll}
\hline \hline
$D=$   &  5 & 6 & 7 & 8 &    9   &   10    & 11 \\ \hline 
$N=$
&  $4.5\times 10^3$
&  $1.1\times 10^4$
&  $2.0\times 10^4$
&  $3.0\times 10^4$
&  $4.0\times 10^4$
&  $5.1\times 10^4$
&  $6.2\times 10^4$
\\
\hline \hline
\end{tabular}
\end{center}
\end{table}}

\section{LHC production of gravitationally bound black holes}
\label{app:boundBH}
In this Appendix we estimate the number of LHC-produced black holes that could in these hypothetical scenarios become gravitationally trapped by the Earth.  This in particular addresses questions about whether there could be multi-black hole effects.
For a black hole to get trapped, and start its accretion, its
speed should not exceed the escape velocity from Earth, namely $v_E
\sim 11$~km s$^{-1} \sim 3.7 \times 10^{-5} c$.  In
the central LHC collisions, the black hole acquires a speed along the
beam axis because of an imbalance in the longitudinal momenta of the
colliding partons, and a transverse speed because of the bremstrahlung
emitted from the initial state. These velocities are typically much
larger than $v_E$. One should however also account for the slow-down
caused by the interactions with matter as the black hole crosses the
Earth. Such a slow-down will increase the chances that a black hole
will be captured in the Earth's gravitational field.
In this Appendix we study in detail the velocity
spectrum resulting at production, and convolute it with the stopping
power of the Earth, to obtain an estimate of the trapping probability,
as a function of the black hole mass.

We start by analyzing the slow-down due to accretion.
As pointed out in the main
text, the Earth's density does not provide enough material to
stop a highly relativistic black hole, such as those produced by
cosmic rays.  Indeed, the Earth's column density
$\delta_E=1.1 \times 10^{10}\gr/\cm^2 \sim 2\gev/\mathrm{TeV}^{-2}$ 
leads, for $v\sim 1$,
to the
accretion of at most a few GeV. This number will however increase
significantly at low velocity, where the accretion per unit length traveled  goes like
$1/v$, as shown in eq.~(\ref{Mnrevol}). 
Therefore some slow-down will typically arise for non-relativistic
black holes produced at the LHC. Repeating
the analysis of the slow-down in the non-relativistic regime given in
section~\ref{sec:WDstop}, 
we derive the following relation for the maximum
velocity at production, $v_{max}$, that can be stopped
 before the black hole exits the Earth:
\be
v_{max} = \frac{2\pi\, k_D \, \hat{b}_{min}^2}
{(D-3)M_D^3} \, \left(\frac{M_D}{k_D M}
\right)^{(D-5)/(D-3)} \; \delta_E \; .
\ee
For a given mass, $v_{max}$ is the largest in $D=11$. We give some
reference values for $v_{max}$ in table~\ref{tab:vmax}, using the largest allowed value for $M_D$ corresponding to a given mass, $M_D=M/3$. Notice that
these velocities can be significantly larger than the escape
velocity.
{\renewcommand{\arraystretch}{1.1}
\begin{table}
\begin{center}
\ccaption{*}{\label{tab:vmax} \it Maximum velocities at production for
  gravitational trapping.}
\vskip 2mm
\begin{tabular}{l|lllll}
\hline \hline
$M$(TeV)                       &  4  & 6   & 8   & 10   & 12   \\ \hline 
$v_{max}\times 10^3$, $D=8$    & 9.1 & 2.7 & 1.1 & 0.58 & 0.34 \\
$v_{max}\times 10^3$, $D=11$   & 15  & 4.5 & 1.9 & 0.96 & 0.56 \\
\hline \hline
\end{tabular}
\end{center}
\end{table}}
Black holes pointing away from the center of the Earth will travel
across a smaller column density, and their velocity should therefore
be smaller than $v_{max} \; \ell/D$, where $\ell$ is the length of the
path inside the Earth, and $D$ is the Earth's diameter. This condition
can be written as
\be
v < v_{max} \cos\theta = v_{max} \frac{v_z}{v} \; ,
\ee
where $\theta$ is the angle with respect to the vertical axis $\hat
z$. The stopping condition becomes:
\be
v^2 = y^2 + v_\perp^2 < v_{max} \, v_\perp \, \cos\phi \; ,
\ee
where $y$ is the black hole rapidity (equal to its longitudinal
velocity in the non-relativistic limit), $v_\perp$ its velocity
in the plane transverse to the beam direction, and $\phi$ is the
angle between the velocity direction in the transverse plane 
and $\hat{z}$. 

For $y\ll 1$, $d\sigma/dy$ is approximately flat, and independent of
the black hole transverse momentum, since at small velocity that
longitudinal  and transverse dynamics decouple. For a given value of
$v_\perp<v_{max}$, 
\be
y<y_0=\sqrt{v_{max}v_\perp \cos\phi -v_\perp^2}\ ,
\ee
and
the fraction of events that satisfy the stopping
condition is therefore given by: 
\be
\frac{1}{2\pi} \int_{-\phi_v}^{\phi_v} \; d\phi \, \int_{-y_0}^{y_0}
\frac{1}{\sigma}\frac{d\sigma}{dy}\vert_{y=0} \; dy \, 
= \,
\frac{2}{\pi\sigma}\frac{d\sigma}{dy}\vert_{y=0} \; v_{max} \;
\sqrt{\frac{v_\perp}{v_{max}}} \; \int_{0}^{\phi_v} \; d\phi \; (\cos\phi -
\frac{v_\perp}{v_{max}} )^{1/2}\ ,
\ee
where $\phi_v= \arccos(v_\perp/v_{max})$.
To excellent approximation, this can be written as:
\be
\frac{2}{\pi\sigma}\frac{d\sigma}{dy}\vert_{y=0} \; v_{max} \;
\sqrt{\frac{v_\perp}{v_{max}}} \; 1.2 \times (1 -
\frac{v_\perp}{v_{max}} )\ 
\ee
The normalized
rapidity spectra of black holes at $y=0$, as a function of $M$,
is shown in fig.~\ref{fig:dsigdy}. 
\begin{figure}
\begin{center}
\includegraphics[width=0.68\textwidth,clip]{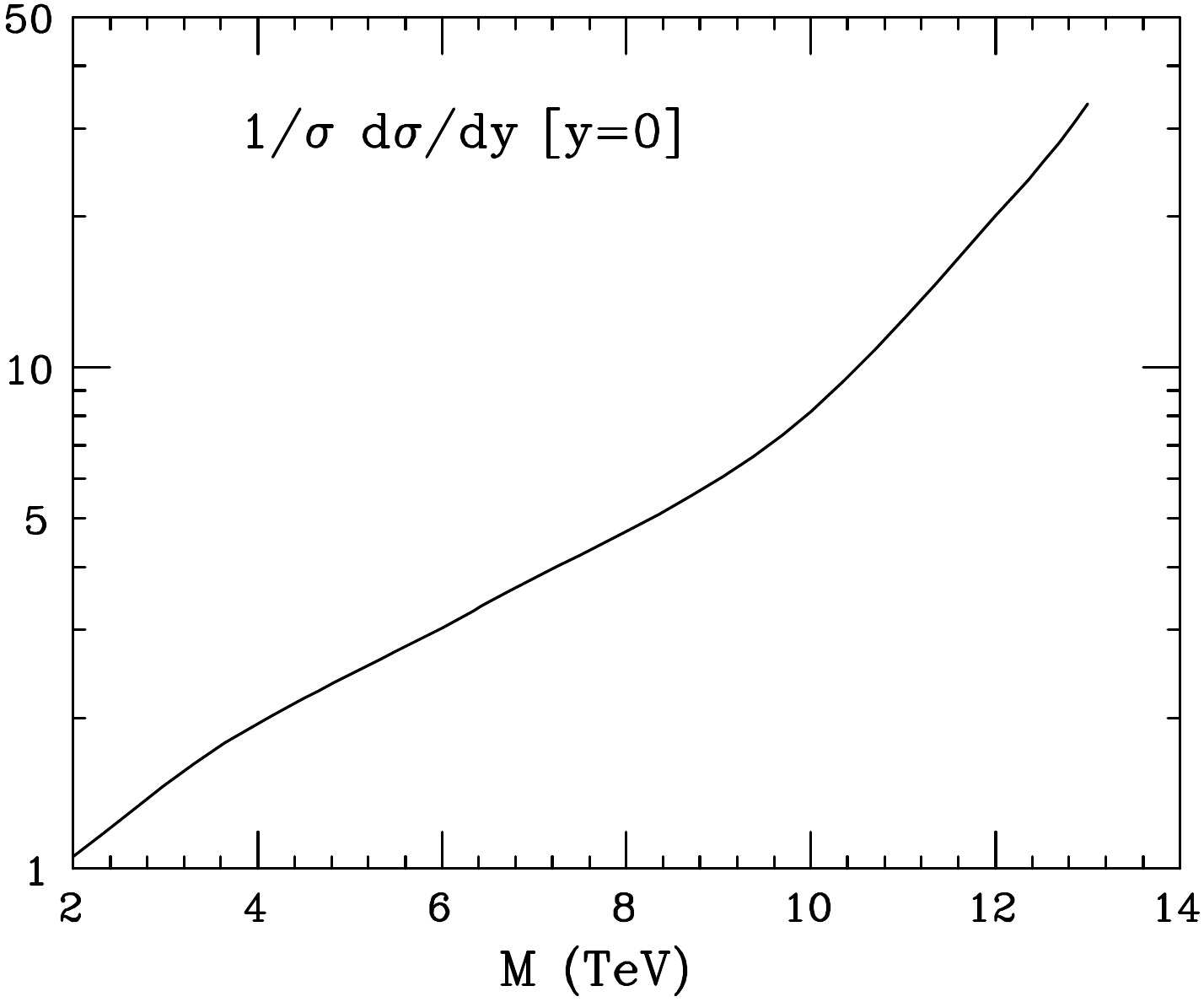}
\end{center}
\vskip -1cm
\ccaption{*}{\label{fig:dsigdy} \it Rapidity spectrum at $y=0$ for
  production at the LHC.}
\end{figure}

Convolution of this probability with the $v_\perp$ spectrum obtained,
as a function of the black hole mass, from the Herwig Monte
Carlo\cite{Corcella:2000bw}, 
leads to the stopping probabilities shown in 
 table~\ref{tab:BM}.
{\renewcommand{\arraystretch}{1.1}
\begin{table}
\begin{center}
\ccaption{*}{\label{tab:BM} \it Stopping probabilities
 for typical $M$ values.}
\vskip 2mm
\begin{tabular}{l|lllll}
\hline \hline
$M$(TeV)   &  4 & 6 & 8 & 10 & 12 \\
\hline
$P\times 10^4$, $D=8$   & 5.7 & 1.2 & 0.37 & 0.20 & 0.24  \\
$P\times 10^4$, $D=11$  & 14  & 3.4 & 1.2  & 0.71 & 0.92 \\
\hline \hline
\end{tabular}
\end{center}
\end{table} }
We then convolute these trapping probabilities with
the black hole production rates derived assuming inelasticity in the
realistic range from $y=0.5$
to $y=0.7$
(we trust that the coincidence of notation here of $y$ for the
rapidity and for the inelasticity will not be a source of
confusion!). The results are shown in fig.~\ref{fig:ntrapped}, for the
standard integrated luminosity of 1000~fb$^{-1}$. 
\begin{figure}
\begin{center}
\includegraphics[width=0.68\textwidth,clip]{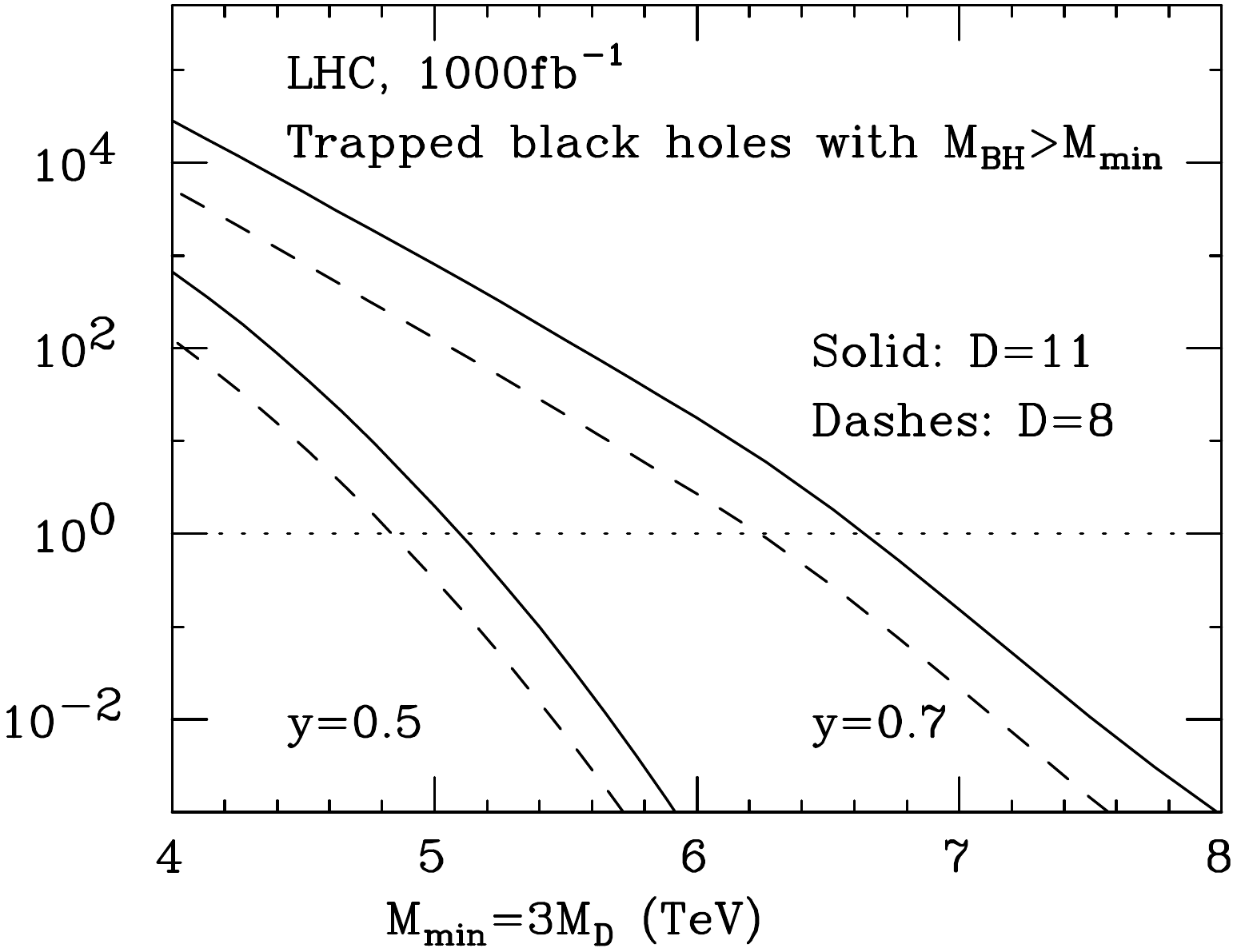}
\end{center}
\vskip -1cm
\ccaption{*}{\label{fig:ntrapped} 
\it Number of trapped black holes, for inelasticity $y=0.5$ and $y=0.7$,  as a
function of the black hole mass $M$, and for an integrated luminosity
of 1000~fb$^{-1}$.}
\end{figure}
As soon as $M\gsim 7$~TeV the expected number of trapped black holes
falls below 1 even with the looser inelasticity value of
$y=0.7$. We notice that mass values where the build up of multiple black holes could significantly exceed the value one are firmly excluded for $8\le D \le 11$
by the neutron stars, and for $D\le 7$ by the white dwarfs.

Some of the hypothesized black holes from LHC could reach the Sun or
the Moon. In the case of the Sun, its core is the only part that has a
significant stopping power. The density in the core, whose radius is
about $R_{core}\sim 0.2 R_{\sun} \sim 1.4\times 10^5~\km$, reaches
$150~\gr/\cm^3$, giving a column density hundreds of times greater than
the Earth, allowing to stop black holes proportionally faster.  On the
other hand, the geometric probability that an LHC-produced black hole
reaches the Sun's core is only about $2.2\times 10^{-7}$, a number
that by itself is much smaller than the probability of trapping inside
the Earth. The Moon has the same geometrical suppression as the Sun,
but a much higher suppression due to the limited stopping power, and
the smaller escape velocity.  We also note that the parameter
controlling macroscopic accretion, $d_0c_s$, is for the Sun
approximately four times its value on Earth (with $\rho=150$~gr/cm$^3$
and $c_s=500$~km/s), 
implying accretion time
scales that are four times longer.

\section{Synchrotron losses in  magnetic fields}
\label{app:synch}

In this Appendix we briefly summarize the limitations on cosmic ray penetration to stellar surfaces resulting from synchrotron radiation losses.  

We assume a dipole magnetic field, with polar intensity  $B_p$
defined by:
\be
\vec{B}(\theta,r)\;=\; B_p \,\left(\frac{R_0}{r}\right)^3 \,
     \frac{3\cos\theta \, \hat{r} \, - \,\hat{z}}{2} \;
\ee
where $R_0$ is the star radius, $\hat{z}$ points along the field axis,
and $\theta$ is the angle with respect to $\hat{z}$.
We consider an incident nucleus with
charge $Z$ and mass $A$, with radial momentum $p$ at an angle
$\theta$ from the direction of the magnetic axis.
One easily finds the Larmor radius as a function of $r$,
\be
r_L(r,\theta)=\frac{2p}{ZeB_p\sin\theta } \; \left({r\over R_0}\right)^3\ .
\ee
 To ensure that the cosmic ray reaches the star's surface, 
we require
that $r_L(r,\theta)>r$ for all values of the distance $r$. 
If we consider the trajectories subject to
the largest field ($\sin\theta=1$), this corresponds to requiring:
$p > Z\,e\,B_p \, R_0/2$, or:
\be
p \gsim 0.75\times 10^{17}~\ev~\frac{R_0}{5000~\km}\; \frac{ZB_p}{10^6\,\gauss} 
= 1.5\times 10^{17}~\ev~\frac{R_0}{10~\km}\; \frac{ZB_p}{10^9\,\gauss}
\; .
\ee
The two expressions correspond to typical radii of white dwarfs and
neutrons stars, and to reference 
magnetic fields chosen to be
in the range of actual measured values~\cite{Kawka,Lorimer:2005bw}.
For the energies we are dealing with, in the range of $10^{18}\ev$ or more,
we see therefore 
that magnetic fields of 1MG (WD) or 1000MG (NS) correspond to
relatively small deflections (and even smaller at lower fields), which
are compatible with our criterion. 

While typical magnetic fields do not seem to induce sufficient
deflection in the trajectory, they may nevertheless lead to great
energy losses due to synchrotron radiation emission. 
The synchrotron power loss  for relativistic cosmic rays  
corresponding to the above Larmor radius is given by:
\be
\frac{dE}{dt} = - \frac{1}{6} \, \frac{(Ze)^4 B_p^2 \sin^2\theta \,
  R_0^6}{A^4 m_p^4 \,r^6} \; E^2 \; .
\ee
Integrating along a cosmic ray trajectory
from $r=\infty$, where $E=E_\infty$, to $r=R_0$,
where $E=E_{R_0}$, gives:
\be
E_{R_0} = E_{max} \frac{E_{\infty}}{E_\infty+E_{max}} \; ,
\ee
where 
\be
E_{max} = \frac{
30 A^4 m_p^4}{(Ze)^4(\sin\theta B_p)^2 R_0} \,  \; .
\ee
Notice that $E_{R_0}$ is always smaller than $E_{max}$, regardless of
the initial cosmic ray energy. 
This is therefore a maximum energy
that can be retained by a cosmic ray impinging on the  star.
For initial energies above $E_{max}$, the higher the
energy, the more is radiated off, ending up always with the same
limiting energy. Introducing numbers appropriate for a neutron star, we get:
\be
E_{max} \approx 1.8\times 10^{17} 
\ev{A^4\over Z^4} {10\km\over R_0}
\left({10^8 G\over B_p \sin \theta}\right)^2\ . 
\ee
In a strong field,
cosmic rays of the highest energies are therefore allowed to penetrate only
within an angle $\theta(E)$ from the northern/southern hemispheres of the
 star, with:
\be \label{eq:polar}
\sin\theta(E) <  \sin\theta_{max} = 
\sqrt{\left({1.8\times 10^{17}\ev\over E}\right)\left(
  {10\km\over R_0}\right)} {A^2\over Z^2} \left({10^8 G\over
  B_p}\right) 
\, .
\ee
Notice that the reduction factor is less harmful for nuclei, since $A/Z\sim 2$.
The lowest magnetic fields that have been measured for neutron stars
are of the order of 100~MG~\cite{Lorimer:2005bw}, 
and this is therefore the lowest field
that we can assume for our study. In the case of proton cosmic rays,
penetration to the star's surface with energies in the range of
$5\times 10^{18}\ev$, as required for efficient production of
$M>14~\tev$ black holes (see fig.~\ref{fig:wdrate}), would require
$\theta\lsim 0.2$. 
Combining the reduction in the star surface area
that can be reached (just the polar caps, of area $\sim \pi (\theta_{max}
R_0)^2$) with the limited angular acceptance (cosmics with
$\theta<\theta_{max}$ from the polar zenith), we obtain an approximate
rate suppression of order $\theta_{max}^4$, and thus of order
$10^{-3}$ for the case considered. These numbers are too small to
allow sufficient rate for all cases, and specifically those
at the highest black hole masses.

\section{Production on background objects}
\label{app:background}

Given the self-screening behavior resulting from neutron-star magnetic
fields, one is naturally led to consider production of neutral black
holes on background objects, such as on a binary companion of the
star, or on the interstellar medium.  We here outline some such
considerations.

\subsection{On lifetimes and solid angles in X-ray binaries}
\label{app:binaries}

We would like to understand the range of possible values for the full
coverage equivalent (FCE), defined in (\ref{FCEdef}), for neutron star
binary systems.  A number of such systems have been well-studied and
modeled.  Thus, deviations from the FCEs given below would require
substantial revision of our theoretical description of the formation
and evolution of these systems.

Specifically, neutron stars are often found in binaries tight enough
for mass transfer to occur from the donor to the NS. These X-ray
binaries (see \cite{Verbunt} for an overview) are well studied
systems, and are also known to exist in other galaxies.  
The range of possible donor masses, $M_d$, is
large\cite{Verbunt}, from $M_d=0.01\Msun$ Helium white dwarfs in
ultracompact binaries to $30\Msun$ stars around accreting X-ray
pulsars.  The closest possible distance is when the donor star fills
its Roche lobe (e.g. the tidal radius), yielding the largest value for
$f=\Delta\Omega/4\pi=(1-\cos\theta)/2$, where\cite{Eggleton}
$\tan\theta=R_{RL}/a=0.49[0.6+q^{-2/3}\ln(1+q^{1/3})]^{-1}$ is only a
function of the mass ratio $q=M_d/M_{NS}$. For $M_d=0.01-10\Msun$ (and
a fixed NS mass of $M_{NS}=1.4\Msun$), we find $f=0.002-0.06$.
 
  Massive X-ray binaries (see \cite{Bildsten:1997vw} for an overview)
have donor masses $M_d>5\Msun$ and range up to $50\Msun$. The longest
time that such a system can live as a mass transferring system (and
thereby visible in X-rays) is when the donor always underfills the
Roche Lobe, and the NS accretes some of the stellar wind leaving the
companion. Such a system lives for at most the hydrogen burning main
sequence lifetime~\cite{Hansen}, $T_{MS}\approx 10 {\rm Gyr}
(\Msun/M)^{2.5} $ of the massive star, which is $<10^8$ years for
$M_d>5\Msun$. Such a star has $f=0.05$, yielding $\approx 5$ Myr of
FCE.  More massive main sequence companions would have slightly larger
values of $f$, but their much shorter main sequence lifetimes make
them less constraining.

 A well-known class is that of NSs accreting from a $M\approx \Msun$
 red giant that result in millisecond radio pulsars orbiting the
 remaining He WD core. While the Roche-lobe filling mass transfer is
 occurring, $f\approx 0.03$, and the lifetime in this phase is set by
 the nuclear evolution rate of the red giant. Those that start
 Roche-lobe filling at orbital periods less than 10 days can have a
 mass transfer lifetime of $10^9$ years \cite{Verbunt}, giving
 $\approx 30$ Myr of FCE.
 
The best examples are traditional low-mass X-ray binaries (those with
$M_d<\Msun$) that have shorter $10^8-10^9 $ year lifetimes set by the
mass transfer rates of $\dot M\approx 10^{-8}-10^{-9} \Msun\ {\rm
yr^{-1}}$. However, at very short orbital periods, the donors become
so low in mass, $M_d\approx 0.05 \Msun (f=0.006)$, as to be brown
dwarfs (e.g. SAX J1808.4-3658 \cite{Bildsten:2001gu}) with mass transfer
rates of $10^{-11} \Msun \ {\rm yr^{-1}}$, consistent with that expected from
gravitational wave losses. These are certainly mass-transferring for
over a Gyr, giving $6$ Myr of FCE

 The lowest mass companions are in ultracompact binaries where a He
  white dwarf donates material to a NS at a rate set by gravitational
  wave losses (see \cite{Deloye:2003tb} for an overview). The most
  constraining of these systems are those at the longest orbital
  period of 40 minutes, where $T=10^9 {\rm yr}$ and $M_d\approx
  0.01\Msun (f=0.002)$ and $\approx 2 \ {\rm Myr}$ of FCE.
 
 In closing, the exposures of known systems are in the range of $2-30$
Myr of FCE, and one can be very confident in those scenarios in the
2-6 Myr range~\cite{Bildstencomm}.  Even the NSs accreting from red
giants that give 30 Myr are relatively robust, and plausibly could be
used in improving bounds.

 \subsection{Production on the interstellar medium}

We study here the possibility that black holes are produced by the
collision of cosmic rays with the interstellar medium (ISM). This is
not meant to be a fully robust study, but to provide an indication of
further directions that could be undertaken to produce additional
constraints, largely complementary to ours.
We first
estimate the column density that cosmic rays travel through as they
reach a star. We consider stars in the disk as embedded in a disk of
ISM of height $h$ and radius $L$, where reasonable values are given by
$h\sim 6\times 10^2$ light years, the disk width, and
$L\sim60h$. The average ISM density $n_{ISM}$ is about 1
proton/cm$^3$.  The average column density for cosmic rays pointing
toward the star is given by $n_{ISM} \, (h/2) \, \log(2L/h) \sim
1.4\times 10^{-4}$ protons/(100\,mb). This means an interaction
probability of about $1.4\times 10^{-4}$ for each nucleon in the
cosmic rays.  When the cosmic ray is pointing directly toward the
star, and when it produces a black hole, this will end up hitting the
star. The star's magnetic field has no influence, and therefore both
neutron stars and white dwarfs of arbitrary magnetic field can be
considered as targets. The number of black holes is obtained by applying
this reduction factor to the rates calculated for the production
of black holes via cosmic rays directly hitting a star. This
suppression is too large  to give acceptable rates on neutron stars. In
the case of white dwarfs, and for $D=5$, $M_{min}=14$~TeV and $y=1$,
this means a rate of over 30/Myrs for a 10\% proton composition
(see table~\ref{tab:wdrate-p}), and of about 1/Myrs for 100\% Fe
(see table~\ref{tab:wdrate-Fe}). Even in the latter, most
conservative, case, this means 100 black holes produced over
100Myrs. Any white dwarf with mass in the range 1--1.2 solar
masses, independently of the magnetic field, will absorb and stop such
black holes, which will then catalyze its decay on time scales short
as compared to observed ($\gsim$ Gyr) lifetimes. A white dwarf like
Sirius-B, for example, with a mass of exactly one solar mass and an
age of about 120~Myrs~\cite{Liebert:2005aw}, would not have
escaped destruction by $D=5,6$ black holes up to minimum mass
14~TeV, or by $D=7$ black holes up to around 10~TeV (see
fig.~\ref{fig:wdstop}).

 Finally, we point out the possible use of massive weakly-interacting
dark matter as a target for black hole production by cosmic
rays~\cite{Draggiotis:2008jz}. While lack of direct experimental
evidence for it makes it insufficient today for our purposes, the expected
densities~\cite{Kamionkowski:1997xg} of about 0.3~GeV/cm$^3$ 
could provide sufficient to generate large numbers of black holes to
be absorbed by white dwarfs. The much lower $\gamma$ factor due to
production on such a heavy target would extend the range of capture of even the
highest mass black holes to lighter white dwarfs, and extend the
stopping potential to black holes significantly heavier than those accessible
at the LHC.

\section{Useful conventions, conversion factors, and reference quantities}

Conversion factors:
\ba
&& 1~\yr = 3.2\times 10^7 s\\
&& 1~\gr = 0.56 \times 10^{24}~\gev \rightarrow  1~\gev =1.78\times
10^{-24}~\gr \\
&& \quad \rightarrow 1~\tev = 1.78\times
10^{-21}~\gr \\
&& 1~\fm^{-1}=197~\mev \rightarrow 1~\gev = 5.1\times 10^{13}~\cm^{-1} \\
&& \quad \rightarrow 1~TeV^{-1} = 1.97\times 10^{-17}~\cm \\
&&  1~K = 8.6\times 10^{-5}~\ev = 4.5\times 10^{-8}~\mathrm{\AA}^{-1} 
\ea

Fundamental constants:
\ba
&&M_4 = 1/\sqrt{8\pi G_N} = 2.4 \times 10^{18} \gev = 2.4\times 10^{15}\tev= 4.3 \times
10^{-6} \gr \\
&& m_p = 1.84\times 10^3 m_e = 9.4\times 10^2~\mev
\ea

Astronomical quantities: 
\ba
&& M_{E}= 6.0\times 10^{27}~\gr  =
3.3\times 10^{51}~\gev  \\
&&R_E = 6.4 \times 10^8 ~\cm \\
&&\rho_E = 5.5 ~\gr/\cm^3\\
&&\langle A_E\rangle = 4\times 10^1 \\
&& v_E = 1.12 \times 10^6 ~\cm/s = 3.7 \times 10^{-5} \\
&& M_{NS} \approx 1.5 \Msun = 3\times 10^{33} ~\gr = 2\times 10^{57}~\gev \\
&& R_{NS} \approx 10 ~\km \\
&& \rho_{NS} \gsim 2\times 10^{14}~\gr/\cm^{3}\approx {0.1m_p\over \fm^3}  \sim 10^{-3}~\gev^{4}
\ea

\end{document}